\documentclass[10pt, journal]{IEEEtran}

\usepackage{tcolorbox}
\usepackage{amsmath,amsfonts}
\usepackage{algorithm}
\usepackage{algpseudocode}
\usepackage{array}
\usepackage{amssymb}
\usepackage{graphicx}
\usepackage[strings]{underscore}
\usepackage{cite}
\usepackage{xcolor}
\definecolor{lightred}{rgb}{1,0.9,0.9}
\definecolor{lightgreen}{rgb}{0.9,1,0.9}
\definecolor{darkblue}{rgb}{0.4, 0.6, 1}
\definecolor{darkred}{rgb}{0.6, 0.0, 0.0}
\definecolor{lightgray}{gray}{0.9} 
\usepackage{colortbl,booktabs}
\usepackage{tabularx}
\usepackage{threeparttable}
\usepackage{multirow}
\usepackage{pifont}
\usepackage{enumitem}

\usepackage{hyperref}
\usepackage[numbered]{bookmark}
\hypersetup{
	colorlinks=true,       
	linkcolor=darkred,       
	filecolor=blue,       
	urlcolor=black,     
	citecolor=blue,   
}

\definecolor{darkgreen}{RGB}{0,128,0}

\newcommand{\subtitleone}[1]{{\textsc{#1}}}

\newcommand{\subtitletwo}[1]{\textit{#1}}

\begin{document}

    \title{A Survey of Wireless Sensing Security from a Role-Based View: Victim, Weapon, and Shield}
	
    \author{Ruixu~Geng\textsuperscript{*}, ~
    Jianyang~Wang\textsuperscript{*},~
    Yuqin~Yuan,~
    Fengquan~Zhan,~
    Tianyu~Zhang,~
    Rui~Zhang,~
    Pengcheng~Huang,~\\ Dongheng~Zhang,~\IEEEmembership{Member,~IEEE}, 
    Jinbo~Chen,~\IEEEmembership{Member,~IEEE},
    Yang~Hu,~\IEEEmembership{Member,~IEEE},
    Yan~Chen\textsuperscript{\textdagger},~\IEEEmembership{Senior~Member,~IEEE}

    \thanks{This work has been submitted to the IEEE for possible publication. Copyright may be transferred without notice, after which this version may no longer be accessible. (*These authors contributed equally to this work. \textdagger Corresponding author: Yan Chen.)}
    \thanks{Ruixu Geng, Jianyang Wang, Yuqin Yuan, Fengquan Zhan, Tianyu Zhang, Rui Zhang, Pengcheng Huang, Dongheng Zhang, Jinbo Chen, Yan Chen are with the School of Cyber Science and Technology, University of Science and Technology of China, Hefei 230026, China}
    \thanks{Yang Hu is with the School of Information Science and Technology, University of Science and Technology of China, Hefei 230026, China }
    
    }

	\markboth{IEEE Communications Surveys \& Tutorials, VOL. XX, 202X}
	{Geng \MakeLowercase{\textit{et al.}}: A Survey of Wireless Sensing Security from a Role-Based View: Victim, Weapon, and Shield}

	\maketitle

	\begin{abstract}
        Wireless sensing technology has become prevalent in healthcare, smart homes, and autonomous driving due to its non-contact operation, penetration capabilities, and cost-effectiveness. As its applications expand, the technology faces mounting security challenges: sensing systems can be attack targets, signals can be weaponized, or signals can function as security shields. Despite these security concerns significantly impacting the technology's development, a systematic review remains lacking. This paper presents the first comprehensive survey of wireless sensing security through a role-based perspective. Analyzing over 200 publications from 2020-2024, we propose a novel classification framework that systematically categorizes existing research into three main classes: (1) wireless systems as victims of attacks, (2) wireless signals as weapons for attacks, and (3) wireless signals as shields for security applications. This role-based classification method is not only intuitive and easy to understand, but also reflects the essential connection between wireless signals and security issues. Through systematic literature review and quantitative analysis, this paper outlines a panoramic view of wireless sensing security, revealing key technological trends and innovation opportunities, thereby helping to promote the development of this field. Project page: \url{https://github.com/Intelligent-Perception-Lab/Awesome-WS-Security}.
	\end{abstract}
	
	
	\IEEEpeerreviewmaketitle
	
	\section{Introduction} \label{sec:intro}

    \begin{figure*}[!t]
    \centering
    \includegraphics[width=0.85\textwidth]{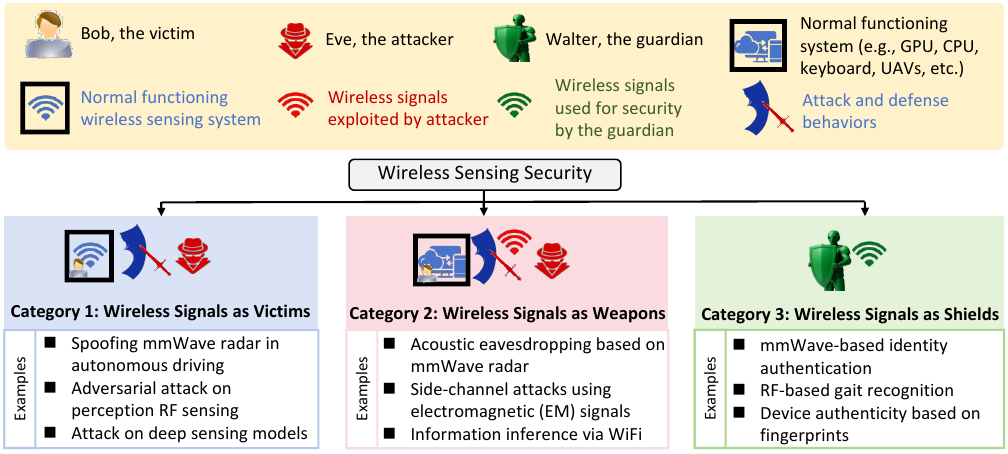}
    \caption{Wireless sensing security can be categorized into three classes based on the role wireless signals play: Victims, Weapons and Shields.}
    \label{fig:introduction}
    \end{figure*}
    
    \subsection{Motivation}
    In recent years, wireless sensing technology has gained widespread attention and application in both academia and industry. This emerging sensing paradigm utilizes the interaction of wireless signals with the environment and objects to acquire information, including but not limited to WiFi~\cite{jiConstruct3DHand2023,pallaproluBegDiffractRF2023,zhangMultipersonPassiveWifi2023,renPersonReidentification3D2023,yangMultipleWiFiAccess2024,zhangRLocRobustIndoor2023}, acoustic signals~\cite{fuPushingLimitsAcoustic2024,fuAdaptiveMetasurfaceBasedAcoustic2024,bureauThreedimensionalUltrasoundMatrix2023,maoAIMAcousticImaging2018}, millimeter-wave (mmWave) radar~\cite{zhuMaliciousAttacksMultiSensor2024,laiEnablingVisualRecognition2024,zhangRFSearchSearchingUnconscious2023,songRFURLUnsupervisedRepresentation2022,karimian-sichaniAntennaArrayWaveform2024,xu3DHighResolutionImaging2024,schenkelRadarEnabledMillimeterWaveSensing2024}, Bluetooth~\cite{iannizzottoPerspectivePassiveHuman2022,basalamahSensingCrowdsUsing2016,zhangBluetoothLowEnergy2014}, and LoRa~\cite{geLoGaitLoRaSensing2023,zhangExploringLoRaLongrange2020}. Compared to traditional optical or contact sensors, wireless sensing technology offers unique advantages: \subtitletwo{(1) Non-contact and non-invasive:} Wireless sensing enables non-contact, non-invasive sensing with significantly less threat to privacy information than CMOS sensors, which is particularly important in healthcare monitoring~\cite{wangContactlessRadarHeart2024,chenContactlessElectrocardiogramMonitoring2024,wang2024rf} and privacy-sensitive scenarios~\cite{chenMMCameraImagingModality2022}; \subtitletwo{(2) Superior and harmless penetration:} The longer wavelength of wireless signals provides penetration capability, allowing operation in non-line-of-sight (NLOS) conditions, thus enabling more flexible all-weather, all-time applications in complex environments such as indoor spaces~\cite{liHighresolutionHandheldMillimeterwave2024,gengDREAMPCDDeepReconstruction2024}; \subtitletwo{(3) Integration with ubiquitous communication devices:} The integration of communication and sensing allows many wireless sensing systems to utilize existing communication infrastructure, such as WiFi routers, enabling constant sensing while reducing deployment costs.
 
	However, as wireless sensing technology becomes more prevalent, security issues are increasingly prominent. Specifically, attackers can exploit vulnerabilities in wireless systems to compromise the accuracy of sensing systems through interference or deception of wireless signals. A typical case in autonomous driving scenarios is where attackers might spoof radar echo signals, causing vehicles to misjudge their surroundings and potentially leading to serious safety incidents~\cite{chenMetaWaveAttackingMmWave2023,huntMadRadarBlackBoxPhysical2024}. Moreover, wireless signals themselves may leak privacy: for instance, by analyzing reflected signals from mmWave radar, attackers might infer occupants' activity patterns~\cite{yuRFPoseOTRFbased3D2023,yuRFGANRFBasedHuman2023}, physiological states~\cite{gongEnablingOrientationFreeMmwaveBased2024,ranContactlessBloodPressure2022}, or even conversation content within a room~\cite{xuMmEarPushLimit2024}. Some attackers might also use wireless sensing systems as tools to attack other systems. For example, exploiting the penetration capability of mmWave radar, attackers could potentially eavesdrop on conversations behind walls or monitor others' behavior~\cite{chenOverviewHumanPose2024}.
 
    These security issues are wide-ranging, encompassing personal privacy, property security, public safety, etc. Unresolved security issues significantly hinder the technology's adoption in sensitive domains like healthcare, finance, and public security, while undermining public trust. Therefore, addressing these security challenges is essential for both advancing the field academically and enabling the widespread practical implementation of wireless sensing technology.

	Unfortunately, despite the growing significance of wireless sensing security, a systematic review and classification of related research remains absent in academia. This gap prevents researchers from establishing connections across security studies and developing a comprehensive understanding of the field. The lack of a thorough review creates multiple challenges: newcomers struggle to grasp research trends, established researchers face difficulties in positioning their work, and practitioners may overlook critical security risks during system implementation. Thus, a systematic review of wireless sensing security is imperative.
    
    \subsection{Challenges}
    However, conducting a comprehensive review of wireless sensing security is not straightforward. The key to the review is finding a reasonable classification method, which should satisfy three critical requirements:
    
    \textit{Challenge 1: Intuitive, logical, and essential classification.} The classification should be easily comprehensible while revealing fundamental insights. Device-based classifications (e.g., WiFi, microphone, mmWave radar) are intuitive but fail to capture cross-device security patterns and technical implications. An effective framework should focus on core security principles rather than surface-level technical details.

    \textit{Challenge 2: Comprehensive coverage.} The classification should encompass the field's breadth and adaptability. Wireless sensing security spans multiple domains including cyber-physical security, metamaterials, backdoor attacks, adversarial samples, authentication, and autonomous driving. Simple threat-based classifications (e.g., spoofing, jamming) often miss cross-domain attacks and security applications. Furthermore, the framework should accommodate both existing research and future developments.

    \textit{Challenge 3: Clear boundaries between categories.} Each research work should fit distinctly into a single category to establish a clear cognitive framework. However, the multifaceted nature of wireless sensing security complicates categorization. Method-based classifications (e.g., signal processing vs. deep learning) often lead to overlaps, as many approaches combine multiple techniques. The framework should minimize such ambiguities to provide clear research positioning.

	\begin{figure*}[!t]
    \centering
    \includegraphics[width=0.95\textwidth]{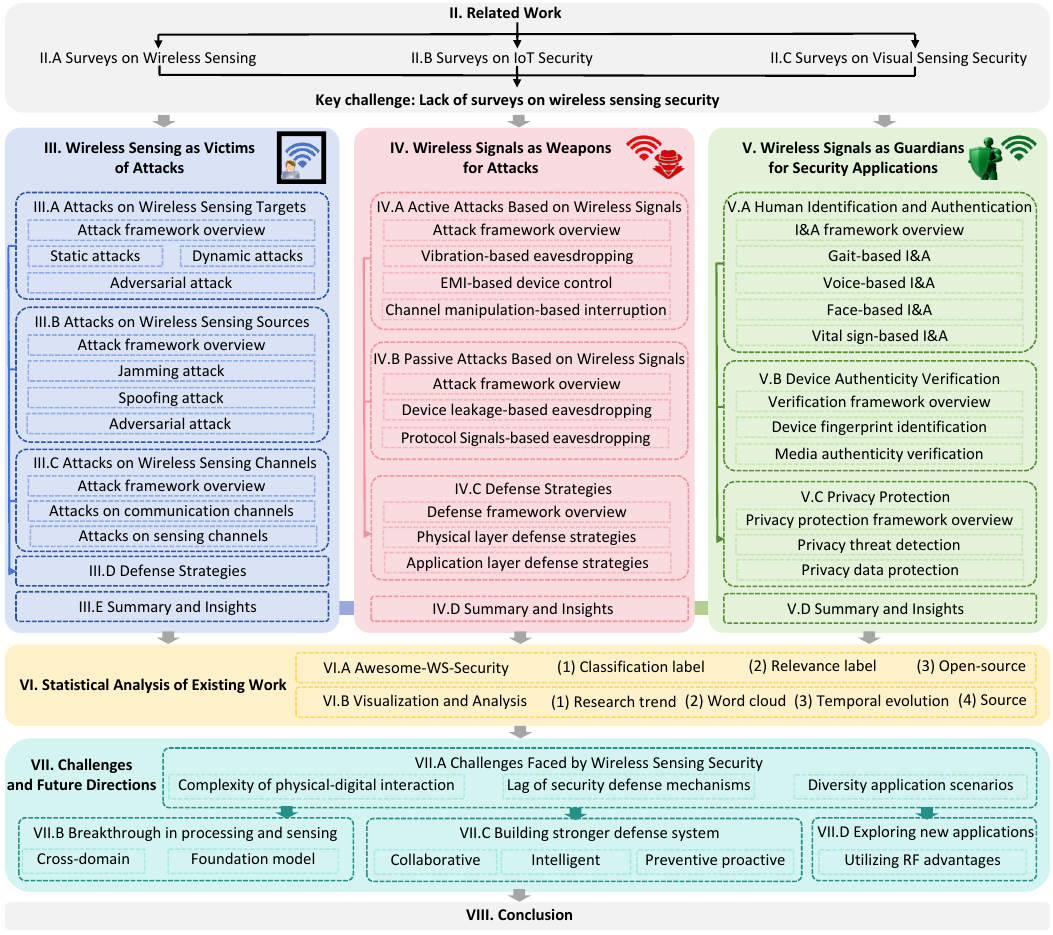}
    \caption{Overall structure of the paper.}
    \label{fig:structure}
    \end{figure*}

    \subsection{Our Classification Method}
    
    To find such a classification method, we summarized over 200 related articles. We discover that almost all wireless sensing security-related research can be categorized into three classes based on the role wireless signals play in security scenarios (as shown in Fig.~\ref{fig:introduction}):
    
    \noindent \textbf{Category 1: Wireless systems as attack targets.} For research in this category, wireless systems composed of wireless signals are ``innocent victims''. The focus of \textit{Category 1} is on how to attack these wireless systems, rendering them inoperable. For example, attacks on automotive mmWave Frequency-Modulated Continuous Wave (FMCW radar)~\cite{reddyvennamMmSpoofResilientSpoofing2023,huntMadRadarBlackBoxPhysical2024,sunWhoControlPractical2021,nashimotoLowcostDistancespoofingAttack2021}, attacks on human activity recognition (HAR) systems~\cite{xieUniversalTargetedAdversarial2023}, and attacks on WiFi-based detection systems~\cite{zhouRIStealthPracticalCovert2024} all belong to this category. Based on different attack targets, we further subdivide this category into three types: attacks on the signal source, attacks on the channel~\cite{weiMetasurfaceenabledSmartWireless2023}, and attacks on the target. Additionally, research on how to protect these wireless systems to defend against attacks ~\cite{liuExploitingFineGrainedChannel2023,mengSecurFiSecureWireless2023} also belongs to this category.
    
    \noindent \textbf{Category 2: Wireless signals as attack tools.} For research in this category, wireless signals are the attacker's ``evil accomplices''. The research focus is on how to use wireless signals to attack other normally functioning physical systems, thereby stealing private information from physical systems or even rendering them inoperable. For example, using mmWave radar to eavesdrop on mobile phone microphones\cite{basakMmSpySpyingPhone2022, fengMmEavesdropperSignalAugmentationbased2023, huMILLIEARMillimeterwaveAcoustic2022, huMmEchoMmWavebasedAcoustic2023, shiPrivacyLeakageSpeechinduced2023, wangMmEveEavesdroppingSmartphones2022, wangMmPhoneAcousticEavesdropping2022, wangWavesdropperWallWord2022, xuMmEarPushLimit2024, zhangAmbiEarMmWaveBased2022} or CPU information~\cite{liSpiralspyExploringStealthy2022}, WiFi-based eavesdropping on POS terminal keyboard inputs~\cite{chenSilentThiefPassword2024}, and inferring GPU-related information through electromagnetic leakage~\cite{zhanGraphicsPeepingUnit2022} all belong to this category. Based on whether the attacking device actively emits wireless signals, we further subdivide this category into two types: active attacks and passive attacks. Furthermore, corresponding countermeasures~\cite{staatIRShieldCountermeasureAdversarial2022 ,asaadSecureActivePassive2022} also belongs to this category.

    \noindent \textbf{Category 3: Wireless signals as guardians of security applications.} For research in this category, wireless signals act as ``righteous guardians'' dedicated to completing security tasks. The research focus is on how to use wireless signals as a ``shield'' to implement security applications. This category can be further divided into three main types: (1) Biometric Authentication: This includes gait-based recognition~\cite{renPersonReidentification3D2023,liangDCSGaitClassLevelDomain2024}, voiceprint-based identification~\cite{liuWavoIDRobustSecure2023}, face recognition~\cite{xuMaskDoesNot2022}, and physiological signal-based authentication~\cite{wangHeartPrintExploringHeartbeatBased2022}.
    (2) Device Authenticity Verification: This encompasses device fingerprinting~\cite{fengFingerprintingIoTDevices2023} and media authenticity verification~\cite{fangNowhereHideDetecting2023}.
    (3) Privacy Protection: This includes privacy threat detection~\cite{qiuRadar2PassiveSpy2023,dengDrDefenderProactive2024} and privacy data protection~\cite{liuExploitingFineGrainedChannel2023}.
    
    This role-based classification effectively addresses the aforementioned challenges. It provides an intuitive framework aligned with cognitive patterns, enabling quick comprehension of the field. The classification captures fundamental relationships between wireless signals and security: wireless signals as potential targets, attack vectors, or security tools. This essential perspective unifies diverse research areas into a coherent framework. Moreover, it enables clear categorization based on core contributions, regardless of the technical approaches employed.
    
    Based on the above classification method, we have systematically reviewed and analyzed over 200 papers from the past five years, establishing the \textit{Awesome-WS-Security} database. The database includes metadata such as publication year, venue, and category for each paper.
    
    Finally, based on the preceding review, statistics, and analysis, we deeply discussed the challenges and future trends in wireless sensing security. This survey serves as both an entry point for newcomers to grasp the field's landscape and a reference for researchers to identify promising directions based on their interests in wireless signals' roles as targets, weapons, or shields.

    \subsection{Contribution}
    In summary, the main contributions of this paper include:
    \begin{enumerate}
    \item We provide a systematic review for the wireless sensing security field. To our knowledge, this is the first comprehensive survey focusing on wireless sensing security, filling the gap of a lack of systematic reviews in this field.
    \item We propose an innovative classification framework based on the roles of wireless signals. This framework divides wireless sensing security research into 3 main categories, 10 secondary classifications, and 24 tertiary classifications. It is not only intuitive and easy to understand but also reflects the essential connection between wireless signals and security issues. This classification method successfully addresses the challenges of logical classification, comprehensive coverage, and avoiding overlap, providing a clear approach for understanding and organizing research in this field.

    \item We systematically analyze the development patterns and trends in wireless sensing security research. Through analysis of over 200 papers, we have identified evolution patterns in different research directions, recognized hot topics and underdeveloped areas, providing new perspectives and ideas to promote the development of this field.
    \end{enumerate}
    
    As shown in Fig.~\ref{fig:structure}, the rest of this paper is arranged as follows: Section \ref{sec:relatedWork} reviews related survey work; Sections \ref{sec:roleTarget}, \ref{sec:roleTool}, and \ref{sec:roleGuardian} discuss in detail the current research status of wireless signals as attack targets, attack tools, and guardians of security applications, respectively; Section \ref{sec:roleStat} provides statistical analysis of existing work; Section \ref{sec:discussion} explores current challenges and future research directions; finally, Section \ref{sec:conclusion} concludes the paper.

    \section{Related Work} \label{sec:relatedWork}
    
        \begin{table*}[!t]
    \caption{Comparison with Related Survey Works}
    \label{tab:comparison}
    \renewcommand{\arraystretch}{1.3}
    \centering
    \begin{tabular}{p{1.8cm}|p{3.8cm}|p{3.8cm}|p{3.8cm}||p{3cm}}
    \toprule
    \textbf{Dimension} & \textbf{Wireless Sensing Surveys} & \textbf{IoT Security Surveys} & \textbf{Visual Sensing Security Surveys} & \textbf{This Paper} \\
    \midrule
    Focus & Sensing algorithms & Network security & Visual system security & \textbf{Wireless sensing security} \\
    \midrule
    Literature & \textit{mmWave:}\cite{wangReviewActiveMillimeter2019,mafukidzeScatteringCentersPoint2022,applebyMillimeterWaveSubmillimeterWaveImaging2007,han4DMillimeterWaveRadar2023,zhouMMWRadarBasedTechnologies2020,abduApplicationDeepLearning2021,yaoRadarPerceptionAutonomous2023,zhangSurveyMmWaveBasedHuman2023,shastriReviewMillimeterWave2022}; \textit{WiFi:}\cite{maWiFiSensingChannel2020,chenCrossDomainWiFiSensing2023,liuWirelessSensingHuman2019,heWiFiVisionSensing2020,makkiSurveyWiFiPositioning2015,roySurveyUbiquitousWiFibased2022,geContactlessWiFiSensing2022,ahmadWiFiBasedHumanSensing2024,yangSenseFiLibraryBenchmark2023,tanCommodityWiFiSensing2022};\textit{Acoustic:}\cite{caiUbiquitousAcousticSensing2022,baiAcousticbasedSensingApplications2020,gorshkovScientificApplicationsDistributed2022,zhuDistributedAcousticSensing2022,rahmanReviewDistributedAcoustic2024,huPhysiologicalAcousticSensing2014} & \textit{Architecture:}\cite{ahemdIoTSecurityLayered2017,gouConstructionStrategiesIoT2013,riahiSystemicApproachIoT2013}; \textit{Threats:}\cite{khanIoTSecurityReview2018,neshenkoDemystifyingIoTSecurity2019,hassijaSurveyIoTSecurity2019}; \textit{Defense:}\cite{xuSecurityIoTSystems2014,zhangIoTSecurityOngoing2014,siwakotiAdvancesIoTSecurity2023}; \textit{Technologies:}\cite{al-garadiSurveyMachineDeep2020,hussainMachineLearningIoT2020,xiaoIoTSecurityTechniques2018,alwahediMachineLearningTechniques2024};\textit{Dataset:}\cite{alexComprehensiveSurveyIoT2023,kaurInternetThingsIoT2023} & \textit{Adversarial:}\cite{akhtarAdvancesAdversarialAttacks2021,dhamijaHowDefendSecure2024}; \textit{Physical:}\cite{guesmiPhysicalAdversarialAttacks2023}; \textit{System:} (Autonomous)\cite{gaoAutonomousDrivingSecurity2022,phamSurveySecurityAttacks2021,singandhupeReviewSLAMTechniques2019}; (Visual)\cite{winklerSecurityPrivacyProtection2014}; (Video)\cite{vennamAttacksPreventiveMeasures2021,zhaoReviewComputerVision2021} & \textbf{First wireless sensing security survey} \\
    \midrule
    Classification & By scenarios/technology & By architecture/threats & By attacks/scenarios & \textbf{Role-based view} \\
    \midrule
    Contribution & Wireless sensing advances & Low-level IoT security & Key insights in security analysis & \textbf{Systematic review of wireless sensing security}  \\
    \midrule
    Limitation & Lacks security analysis & Ignores sensing security & Not for wireless sensing & -- \\
    \bottomrule
    \end{tabular}
    \end{table*}
    
    Despite the increasing significance of wireless sensing security, no comprehensive survey has specifically addressed this field. Related surveys fall into three categories: wireless sensing surveys, IoT security surveys, and visual sensing security surveys (Tab.~\ref{tab:comparison}). While each category offers valuable insights, none fully captures the unique challenges of wireless sensing security.
    
    \subsection{Surveys on Wireless Sensing}
    
    Recent wireless sensing surveys have covered various sensors but with limited focus on security aspects:
    \textit{(1) mmWave radar sensing surveys} encompass multiple research levels~\cite{kongSurveyMmWaveRadarBased2024}. At the signal processing level, Wang et al. \cite{wangReviewActiveMillimeter2019} reviewed high-resolution imaging technologies including antenna array design and beamforming, while Mafukidze et al. \cite{mafukidzeScatteringCentersPoint2022} examined radar point cloud generation methods. Appleby et al. \cite{applebyMillimeterWaveSubmillimeterWaveImaging2007} analyzed mmWave imaging applications in security. At the application level, autonomous driving and human sensing represent two major directions. For autonomous driving, Han et al. \cite{han4DMillimeterWaveRadar2023} analyzed 4D mmWave radar signal processing and resolution enhancement, Zhou et al. \cite{zhouMMWRadarBasedTechnologies2020} investigated weather adaptability, Abdu et al. \cite{abduApplicationDeepLearning2021} examined deep learning applications, and Yao et al. \cite{yaoRadarPerceptionAutonomous2023} introduced a data representation-based classification framework. In human sensing, Zhang et al. \cite{zhangSurveyMmWaveBasedHuman2023} and Shastri et al. \cite{shastriReviewMillimeterWave2022} covered applications ranging from vital sign detection to behavior recognition.
    \textit{(2) WiFi-based sensing surveys} demonstrate significant diversity. Ma et al. \cite{maWiFiSensingChannel2020} and Chen et al. \cite{chenCrossDomainWiFiSensing2023} reviewed Channel State Information (CSI)-based sensing technologies comprehensively. For specific applications, Liu et al. \cite{liuWirelessSensingHuman2019} and He et al. \cite{heWiFiVisionSensing2020} focused on human activity sensing, Makki et al. \cite{makkiSurveyWiFiPositioning2015} and Roy et al. \cite{roySurveyUbiquitousWiFibased2022} examined indoor positioning, while Ge et al. \cite{geContactlessWiFiSensing2022} investigated healthcare monitoring. With deep learning advancement, Ahmad et al. \cite{ahmadWiFiBasedHumanSensing2024} and Yang et al. \cite{yangSenseFiLibraryBenchmark2023} surveyed deep learning applications in WiFi sensing. Tan et al. \cite{tanCommodityWiFiSensing2022} reviewed commercial WiFi sensing development over the past decade.
    \textit{(3) Acoustic sensing surveys} have provided valuable insights across multiple domains. In IoT devices, \cite{caiUbiquitousAcousticSensing2022} and \cite{baiAcousticbasedSensingApplications2020} outlined acoustic sensing frameworks on commercial devices. In distributed acoustic sensing, Gorshkov et al. \cite{gorshkovScientificApplicationsDistributed2022}, Zhu et al. \cite{zhuDistributedAcousticSensing2022}, and ReviewDistributedAcoustic \cite{rahmanReviewDistributedAcoustic2024} examined applications in scientific research, infrastructure monitoring, and railway monitoring respectively. Hu et al. \cite{huPhysiologicalAcousticSensing2014} focused on accelerometer-based physiological acoustic sensing.
    
    These surveys comprehensively cover basic principles, key algorithms, and typical applications. However, they primarily focus on sensing technology development, with limited examination of security aspects.
    
    \subsection{Surveys on IoT Security}
    
    IoT security surveys provide critical insights for wireless sensing security. In terms of \subtitleone{layered security architecture}, Ahmed et al. \cite{ahemdIoTSecurityLayered2017} and Gou et al. \cite{gouConstructionStrategiesIoT2013} proposed layered security frameworks from the perception layer to the application layer, while Riahi et al. \cite{riahiSystemicApproachIoT2013} first proposed a systemic approach to IoT security. Regarding \textit{security threat classification}, Khan et al. \cite{khanIoTSecurityReview2018} and Neshenko et al. \cite{neshenkoDemystifyingIoTSecurity2019} systematically reviewed various security threats faced by IoT systems, while Hassija et al. \cite{hassijaSurveyIoTSecurity2019} focused on security challenges in different application scenarios. In terms of \textit{defense mechanisms}, Xu et al. \cite{xuSecurityIoTSystems2014} and Zhang et al. \cite{zhangIoTSecurityOngoing2014} analyzed the technical challenges faced by IoT security, while Siwakoti et al. \cite{siwakotiAdvancesIoTSecurity2023} comprehensively reviewed the latest advances in IoT security. In \subtitleone{emerging technologies}, with the development of machine learning technology, Artificial intelligence (AI)-based IoT security solutions have become a research hotspot. Al-Garadi et al. \cite{al-garadiSurveyMachineDeep2020}, Hussain et al. \cite{hussainMachineLearningIoT2020}, and Xiao et al. \cite{xiaoIoTSecurityTechniques2018} systematically reviewed the application of machine learning in IoT security. Recently, Alwahedi et al. \cite{alwahediMachineLearningTechniques2024} further explored the potential of generative AI and large language models in IoT security. Additionally, Alex et al. \cite{alexComprehensiveSurveyIoT2023} and Kaur et al. \cite{kaurInternetThingsIoT2023} systematically classified and analyzed \textit{IoT security datasets}, providing important references for subsequent research.
    
    However, these surveys focus primarily on communication security rather than sensing security, emphasizing device and network-level security over physical-layer signal propagation challenges.
    
    \subsection{Surveys on Visual Sensing Security}
    
    Visual sensing security surveys examine security challenges at three key levels, providing valuable methodological references for wireless sensing security: \subtitleone{(1) Adversarial sample attacks}, Akhtar et al. \cite{akhtarAdvancesAdversarialAttacks2021} systematically summarized the latest advances in adversarial attacks in computer vision, while Dhamija et al. \cite{dhamijaHowDefendSecure2024} focused on defense strategies for deep learning models. \subtitleone{(2) Physical world attacks}, Guesmi et al. \cite{guesmiPhysicalAdversarialAttacks2023} conducted comprehensive survey on physical adversarial attacks in camera-based intelligent systems. \subtitleone{(3) System-level security},  Gao et al. \cite{gaoAutonomousDrivingSecurity2022} and Pham et al. \cite{phamSurveySecurityAttacks2021} analyzed security threats from the perspectives of on-board perception and vehicle networking respectively, while Singandhupe et al. \cite{singandhupeReviewSLAMTechniques2019} focused on security issues in SLAM systems.  Winkler et al. \cite{winklerSecurityPrivacyProtection2014} systematically analyzed issues such as data security, node security, and user privacy protection. Vennam et al. \cite{vennamAttacksPreventiveMeasures2021} and Zhao et al. \cite{zhaoReviewComputerVision2021} conducted surveys from the perspectives of attack-defense confrontation and computer vision methods, respectively.
    
    While these surveys focus on optical systems, they offer valuable insights for wireless sensing security. The adversarial attack framework proposed by ~\cite{akhtarAdvancesAdversarialAttacks2021} informed wireless signal perturbation analysis, while the physical attack analysis developed by~\cite{guesmiPhysicalAdversarialAttacks2023} addressed vulnerabilities in wireless signal propagation. However, these works do not address wireless sensing security's specific requirements, highlighting the need for a dedicated framework.
    
    \section{Wireless Sensing System as Targets of Attacks} \label{sec:roleTarget}
    \begin{table*}[!t]
\centering
\caption{Summary of Attacks and Defense Methods for Wireless Sensing Systems}
\label{tab:summaryVictim}
\resizebox{\textwidth}{!}{
\begin{tabular}{|p{1cm}|p{2.5cm}|p{5cm}|p{4cm}|p{4.2cm}|p{5cm}|}
\hline
\textbf{Category} & \textbf{Subcategory} & \textbf{Representative Work} & \textbf{Core Principle} & \textbf{Challenges} & \textbf{Insights} \\
\hline

\multirow{3}{*}[-5em]{\rotatebox[origin=c]{90}{\textbf{\makebox[0em]{Attacks on Targets}}}}

 & Static Attacks 
 & \begin{itemize}[leftmargin=*]
\item Single Modality~\cite{chenMetaWaveAttackingMmWave2023},~\cite{zhuTileMaskPassiveReflectionbasedAttack2023}
\item Multi-Modality~\cite{zhuMaliciousAttacksMultiSensor2024}
\end{itemize}
 & Reducing SNR at the receiver through absorption or reflection
 & \begin{itemize}[leftmargin=*]
\item Accurate placement
\item Shape design
\item Material characteristics
\end{itemize}
 & \multirow{3}{*}{\begin{minipage}[t]{\linewidth}
 \begin{itemize}
\item Intelligent and precise attacks
\item Cross-modality joint attacks
\item Diversity of attack techniques
\item Practicality of adversarial samples
\item Application of new materials and manufacturing technologies
\item Co-evolution of offensive and defensive methods
\end{itemize}
\end{minipage}}
 \\
\cline{2-5}

 & Dynamic Attacks 
 & \begin{itemize}[leftmargin=*]
\item Vibration modulation~\cite{rueggVibrationRotationMillimeterWave2007}
\item Rotation modulation~\cite{zhangMicroDopplerEffectsRemoved2021}
\end{itemize}
 & Modulating signals passively in the time domain to deceive sensing systems
 & \begin{itemize}[leftmargin=*]
\item Accuracy of Signal Modeling 
\item Temporal Modulation Complexity
\end{itemize}
 &
 \\
\cline{2-5}
 & Adversarial Example Attacks 
 & \begin{itemize}[leftmargin=*]
\item Human Activity Recognition~\cite{xieUniversalTargetedAttacks2022},~\cite{xieUniversalTargetedAdversarial2023}
\item WiFi Localization~\cite{wangAdversarialDeepLearning2022}
\end{itemize}
 & Deceiving models through adversarial samples for misclassification
 & \begin{itemize}[leftmargin=*]
\item Stealth of perturbations
\item Robustness of adversarial samples
\end{itemize}
 &
 \\

\hline
\multirow{3}{*}[-8em]{\rotatebox[origin=c]{90}{\textbf{\makebox[0em]{Attacks on Sources}}}}
 & Jamming Attacks
 & \begin{itemize}[leftmargin=*]
\item mmWave-based jamming~\cite{yan2016can}
\item Acoustic-based jamming~\cite{yan2016can}
\item Transduction Attacks~\cite{gaoKITEExploringPractical2023}
\end{itemize}
 & Lowering SNR by introducing high-power interference signals, preventing useful signal parsing
 & \begin{itemize}[leftmargin=*]
\item High-precision frequency generators
\item High-power transmitters
\item Matched modulation methods
\item Directional jamming
\end{itemize}
 & \multirow{3}{*}{\begin{minipage}[t]{\linewidth}
 \begin{itemize}
\item Specialization and complexity of attack methods
\item Miniaturization and portability of tools
\item High-precision hardware design
\item Multi-modal joint attacks
\item Diversity of attack techniques
\item More covert attack methods
\item Improved robustness of attack effects
\item Intelligent attack strategies
\end{itemize}
\end{minipage}}
 \\
\cline{2-5}
 & Spoofing Attacks  
 & \begin{itemize}[leftmargin=*]
\item Range Spoofing~\cite{reddyvennamMmSpoofResilientSpoofing2023},~\cite{huntMadRadarBlackBoxPhysical2024},~\cite{shenoyRFprotectPrivacyDevicefree2022}
\item Velocity Spoofing~\cite{reddyvennamMmSpoofResilientSpoofing2023},~\cite{huntMadRadarBlackBoxPhysical2024}
\item Angle Spoofing~\cite{shenoyRFprotectPrivacyDevicefree2022}
\item Time/Frequency Spoofing~\cite{zhangDolphinAttackInaudibleVoice2017},~\cite{yangRemoteAttacksSpeech2023}
\end{itemize}
 & Crafting false signals to deceive the target system into recognizing them as legitimate
 & \begin{itemize}[leftmargin=*]
\item Accuracy of parameter estimation
\item Real-time requirements
\item Authenticity of spoofing signals
\item Stability of signal transmission
\end{itemize}
 & 
 \\
\cline{2-5}
 & Adversarial Example Attacks 
 & \begin{itemize}[leftmargin=*]
\item mmWave-based~\cite{duPracticalDeceptiveJamming2022}
\item WiFi-based~\cite{liuPhysicalWorldAttackWiFibased2022},~\cite{zhouWiAdvPracticalRobust2022},~\cite{wangAdversarialExamplesWiFi2024},~\cite{zhaoExplanationGuidedBackdoorAttacks2024}
\item Acoustic Sensing~\cite{chenWhoRealBob2021},~\cite{shiAudiodomainPositionindependentBackdoor2022},~\cite{liInaudibleAdversarialPerturbation2024}
\item Transduction Attacks~\cite{jiPoltergeistAcousticAdversarial2021}
\end{itemize}
 & Introducing subtle perturbations into input signals to deceive models for misclassification
 & \begin{itemize}[leftmargin=*]
\item Stealth of perturbations
\item Stability in physical environments
\item Independence from time
\item Black-box characteristics
\end{itemize}
 & 
 \\

 \hline
 \multirow{2}{*}[-4em]{\rotatebox[origin=c]{90}{\textbf{\makebox[0em]{Attacks on Channels}}}}
 & Communication Channels
 & \begin{itemize}[leftmargin=*]
\item Data Tampering~\cite{weiMetasurfaceenabledSmartWireless2023}
\item Traffic Manipulation~\cite{staatMirrorMirrorWall2022},~\cite{niuReconfigurableIntelligentSurfaceAssisted2024}
\end{itemize}
 & Altering the propagation properties of communication signals to disrupt communication systems
 & \begin{itemize}[leftmargin=*]
\item Stealth of attacks
\item Real-time requirements
\end{itemize}
 & \multirow{3}{*}{\begin{minipage}[t]{\linewidth}
 \begin{itemize}
\item Joint communication and sensing attacks
\item Use of novel techniques like RIS
\item More covert methods
\item Enhanced robustness
\item Modular attack tools
\item Integration with AI
\end{itemize}
\end{minipage}}
 \\
\cline{2-5}
 & Sensing Channels  
 & \begin{itemize}[leftmargin=*]
\item Intrusion detection jamming~\cite{staatIRShieldCountermeasureAdversarial2022},~\cite{zhouRIStealthPracticalCovert2024}
\item Imaging jamming~\cite{xuNovelApproachRadar2023},~\cite{liJammingISARImaging2023}
\end{itemize}
 & Altering signal propagation properties in sensing channels to disrupt sensing systems
 & \begin{itemize}[leftmargin=*]
\item Compatibility
\item Independence
\item High-speed modulation
\end{itemize}
 & 
 \\

\hline
 \multirow{2}{*}[-4em]{\rotatebox[origin=c]{90}{\textbf{\makebox[0em]{Defense Strategies}}}}
 & Active Defense 
 & \begin{itemize}[leftmargin=*]
\item Parameter randomization~\cite{nallaboluFrequencyDomainSpoofingAttack2021a},~\cite{sunWhoControlPractical2021}
\item Waveform fingerprinting~\cite{sunWhoControlPractical2021}
\item Channel Characteristic modeling~\cite{zhangEarArrayDefendingDolphinAttack2021}
\end{itemize}
 & Proactively preventing, detecting, and responding to attacks
 & \begin{itemize}[leftmargin=*]
\item Real-time requirements
\item False alarms and missed detections
\end{itemize}
 & \multirow{3}{*}{\begin{minipage}[t]{\linewidth}
 \begin{itemize}
\item Security design with hardware collaboration
\item Cross-layer defense strategies
\item Lightweight mechanisms
\item Intelligent and adaptive defenses
\item Integration with authentication mechanisms
\end{itemize}
\end{minipage}}
 \\
 \cline{2-5}
 & Passive Defense 
 & \begin{itemize}[leftmargin=*]
\item Signal obfuscation~\cite{wangPriSensePrivacyPreservingWireless2024}
\item Physical layer encryption~\cite{liProtegoSecuringWireless2022},~\cite{luoMIMOCryptMultiUserPrivacyPreserving2024}
\end{itemize}
 & Designing inherent system security features to resist or mitigate attacks without real-time intervention
 & \begin{itemize}[leftmargin=*]
\item Compatibility
\item System complexity
\end{itemize}
 & 
 \\
\hline
\end{tabular}
}
\end{table*}

    In our role-based classification framework, we first examine scenarios where wireless sensing systems serve as targets of attacks. This represents a fundamental security concern as these systems become increasingly deployed in healthcare, smart homes, and autonomous driving applications.
    Wireless sensing systems comprise three key components: sensing targets, sources, and channels, each susceptible to distinct attack vectors.
    
    This section proposes a systematic classification framework for these attacks (as shown in Tab.~\ref{tab:summaryVictim}). Specifically, attacks on sensing targets can be categorized into static attacks, dynamic attacks, and adversarial attacks. Attacks targeting sensing sources include jamming, spoofing, and adversarial attacks, while attacks on sensing channels operate in both communication and sensing domains. 

    \subsection{Attacks on Wireless Sensing Targets}
    \subsubsection{Overview}
\begin{figure}[!t]
    \centering
    \includegraphics[width=\linewidth]{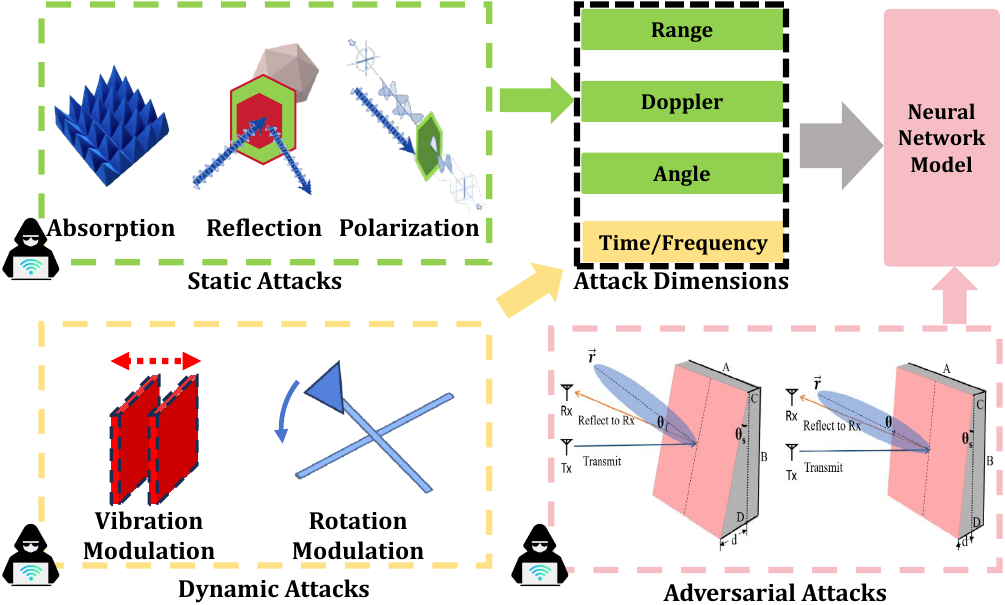}
    \caption{A framework diagram for conducting attacks from the target's perspective.}
    \label{fig:Fig3Targets}
    \end{figure}

    Attacks on sensing targets involve manipulating or altering the characteristics of the sensed target through three main approaches, as shown in Fig.~\ref{fig:Fig3Targets}.
    
    \begin{itemize}
    \item Static Attacks: Signals can not be modulated over time.
    \item Dynamic Attacks: Signals can be modulated over time.
    \item Adversarial Example Attacks: Employ specially crafted adversarial examples to attack models.
    \end{itemize}
    
    These methods are applicable to various wireless sensing systems, such as radar imaging~\cite{liHighresolutionHandheldMillimeterwave2024}, gesture recognition~\cite{liDomainIndependentRealTimeGesture2023}, fall detection~\cite{liSCLFallReliableFall2024}, and vital sign monitoring~\cite{haContactlessSeismocardiographyDeep2020,chenContactlessElectrocardiogramMonitoring2024}.

    \subsubsection{Static Attacks}
    Static attacks employ strategically placed material tags to interfere with sensing systems without temporal modulation. Key scenarios include:

    \paragraph{Single Modality}
    MetaWave~\cite{chenMetaWaveAttackingMmWave2023} introduced Meta-material Tags, including absorption tags, reflection tags, and polarization tags, which interfere with mmWave radar by altering its echo properties. To enhance the effectiveness of these attacks, the authors employed a simulator based on Shooting and Bouncing Rays (SBR) principles to create digital replicas of attack scenarios and optimized the design and deployment parameters using~\cite{chenWhoRealBob2021}, achieving robust mmWave attacks. Similarly, Zhu et al.~\cite{zhuTileMaskPassiveReflectionbasedAttack2023} developed TileMask, a passive reflection-based method that generates adversarial signals to deceive DNN-based detection models, rendering them unable to detect targets.

    \paragraph{Multi-Modality}
    Building upon research in single-modality attacks~\cite{chenMetaWaveAttackingMmWave2023,zhuTileMaskPassiveReflectionbasedAttack2023}, Zhu et al. innovatively designed composite adversarial objects. By combining existing attack methods for mmWave radar, LiDAR, and cameras, they executed passive reflection-based joint attacks on multi-sensor fusion systems~\cite{zhuMaliciousAttacksMultiSensor2024}.

    \paragraph{Key Technical Challenges and Solutions}
    Current static attacks on targets are primarily focused on the autonomous driving domain due to the rapid development of this technology and its reliance on advanced sensing techniques. The evolution of attack methods faces the following key challenges:
    \subtitletwo{(1) Accurate Placement:}
    Optimizing tag positions and angles using digital models for maximum disruption~\cite{chenMetaWaveAttackingMmWave2023}.
    \subtitletwo{(2) Shape Design:}
    Tag shapes significantly influence reflected signal RCS, necessitating optimized tag designs for effective interference~\cite{chenMetaWaveAttackingMmWave2023,zhuTileMaskPassiveReflectionbasedAttack2023}.
    \subtitletwo{(3) Material Characteristics:}
    Tags require specific electromagnetic properties, which pose manufacturing challenges. Advancements in 3D printing have made it easier to produce such materials~\cite{chenMetaWaveAttackingMmWave2023}.

    \subsubsection{Dynamic Attacks}
    Dynamic attacks utilize vibration or rotational mechanisms to modulate signals over time, resulting in more destructive interference with sensing systems. These attacks require higher technical expertise compared to static attacks. Dynamic attacks can be classified into vibration and rotation-based interference:
    \paragraph{Vibration Modulation}
    In 2007, Ruegg et al. proposed a dynamic attack system targeting SAR imaging using vibration devices~\cite{rueggVibrationRotationMillimeterWave2007}.

    \paragraph{Rotation Modulation}
    Rotating components on rigid bodies (e.g., ship antennas, airplane propellers) can cause phase modulation of received signals due to Doppler effects. This leads to smearing effects in Doppler spectrums, degrading imaging performance~\cite{zhangMicroDopplerEffectsRemoved2021}.

    \paragraph{Key Technical Challenges and Solutions}
    Dynamic attacks face two primary challenges:
    \subtitletwo{(1) Accuracy of Signal Modeling:}
    Dynamic interference must mimic real target signal characteristics, requiring a deep understanding of target features and sensing principles~\cite{rueggVibrationRotationMillimeterWave2007}.
    \subtitletwo{(2) Temporal Modulation Complexity:}
    Dynamic interference is more complex than static methods, as vibration or rotation frequencies must align with target signals~\cite{zhangMicroDopplerEffectsRemoved2021}.

    \subsubsection{Adversarial Example Attacks}
    The rise of deep learning in digital media has led to the emergence of adversarial example attacks. While deep learning achieves remarkable performance in tasks like image recognition~\cite{heDeepResidualLearning2016}, speech recognition~\cite{wang2021unispeech}, and language modeling~\cite{devlinBERTPretrainingDeep2019}, these models are vulnerable to adversarial inputs~\cite{szegedy2013intriguing}. Such attacks, which can disrupt image classification~\cite{moosavi2017universal} and speech recognition~\cite{qin2019imperceptible} through subtle perturbations, have now been extended to wireless sensing systems where attackers inject specific perturbations to compromise sensing models. Two key tasks are affected:
    
    \paragraph{Human Activity Recognition (HAR)}
    Xie et al. proposed a universal attack method for HAR systems by generating pre-computed perturbations to deceive systems with a 95\% success rate~\cite{xieUniversalTargetedAdversarial2023}. They later improved this by developing an iterative algorithm that produces perturbations capable of generalizing across activity samples without additional training, achieving over 90\% success in black-box scenarios~\cite{xieUniversalTargetedAdversarial2023}.

    \paragraph{WiFi Localization}
    Neural network-based WiFi indoor localization systems~\cite{wangLearningDomainInvariantModel2024} are also vulnerable to adversarial examples. Wang et al. examined the impact of six adversarial attack methods in both black-box and white-box scenarios and proposed adversarial training strategies to enhance model robustness~\cite{wangAdversarialDeepLearning2022}.

    \paragraph{Key Technical Challenges and Solutions}
    Adversarial example attacks face two major challenges:
    \subtitleone{(1) Stealth of Perturbations:}
    Perturbations should remain imperceptible to users~\cite{moosavi2017universal,qin2019imperceptible}. For example, adversarial samples can be added to a dataset. These samples are carefully crafted inputs designed to subtly alter the data in a way that misleads the sensing or recognition models~\cite{xieUniversalTargetedAdversarial2023}.
    \subtitleone{(2) Robustness of Adversarial Samples:}
    Perturbations also should be effective under various conditions, including changes in angles, distances, and environmental factors~\cite{wangAdversarialDeepLearning2022}.

    \subsubsection{Development Trends}
    
    Future attack methods are advancing toward greater intelligence and precision, utilizing AI-driven strategies to enhance stealth and robustness. Multi-modal attacks combining various sensing technologies pose comprehensive threats, while the integration of static, dynamic, and adversarial techniques diversifies attack strategies. Emerging technologies like 3D printing further accelerate the design and production of disruptive objects. This ongoing evolution of offensive strategies drives continuous innovation in defense mechanisms, fostering a dynamic adversarial landscape.

    \subsection{Attacks on Wireless Sensing Sources}
    \subsubsection{Overview}
\begin{figure*}[!t]
    \centering
    \includegraphics[width=\linewidth]{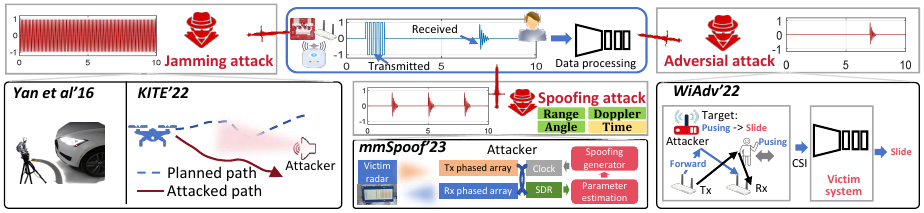}
    \caption{Overview of attacks on wireless sensing sources. The top row illustrates three primary attack types: jamming attacks that overwhelm the receiver with high-power interference, spoofing attacks that inject false signals to deceive signal processing, and adversarial attacks that introduce subtle perturbations.}
    \label{fig:attackSource}
    \end{figure*}

    Attacks on wireless sensing sources aim to disrupt wireless sensing systems by interfering with the signal source. These attacks may involve transmitting jamming signals, fabricating false signals, or introducing subtle perturbations into the signal. Based on existing interference methods, such attacks can be categorized into three types: jamming attacks, spoofing attacks, and adversarial attacks, as shown in Fig.~\ref{fig:attackSource}.
    
    \begin{itemize}
    \item Jamming Attack: The attacker transmits interference signals to the victim device, reducing its signal-to-noise ratio (SNR), preventing accurate reception and decoding of signals.
    \item Spoofing Attack: The attacker generates false signals, deceiving the target system into interpreting them as genuine. This results in errors in recognizing parameters like distance, velocity, and angle of the target.
    \item Adversarial Attack: The attacker injects subtle perturbations into the received signals, deceiving deep learning models into incorrect classification or recognition.
    \end{itemize}
    
    \subsubsection{Jamming Attack} 
    A jamming attack involves emitting high-power interference signals within the victim’s frequency band, causing saturation in the system's ADC components and rendering it non-functional. This straightforward technique is often employed in electronic warfare, military conflicts, and autonomous driving. Depending on the sensing technology, jamming attacks can be classified as mmWave-based, acoustic-based, or transduction attacks.

    \paragraph{MmWave-Based Jamming}
    Yan et al.~\cite{yan2016can} investigated jamming attacks on mmWave radar systems in autonomous driving. They designed single-frequency signals and swept-frequency signals to exploit vulnerabilities in SNR-threshold-based detection mechanisms, where signals must exceed background noise to be detected. This effectively disrupted the radar system.

    \paragraph{Acoustic-based Jamming}
    Yan et al.~\cite{yan2016can} also studied attacks on ultrasonic systems used in autonomous vehicles. Using ultrasonic transducers, the attackers continuously emitted ultrasonic waves matching the sensors’ operating frequency, drowning out the target signal and preventing object detection. These attacks were validated on real vehicles with parking assist features, demonstrating significant disruption to both indoor and outdoor ultrasonic sensors.

    \paragraph{Transduction Attacks}
    Acoustic sensors can also be exploited for cross-modality transduction attacks. Research shows that inertial sensors are highly vulnerable to acoustic transduction attacks, where malicious acoustic signals induce measurement errors. Gao et al.~\cite{gaoKITEExploringPractical2023} generated stable false signals by controlling the frequency and phase of acoustic signals, causing resonance interference. This affected multiple degrees of freedom, enabling attackers to manipulate the speed and direction of target devices and successfully execute Acoustic Transduction Attacks on moving targets.

    \paragraph{Key Technical Challenges and Solutions}
    The primary technical challenges of jamming attacks lie in the hardware platform and algorithms, encompassing the following four key aspects:
    \subtitleone{(1) High-Precision Frequency Generators:}
    Interference signals must match the victim's frequency band. Devices require wideband coverage or fast tuning capabilities.
    \subtitleone{(2) High-Power Emitters:}
    Interference signals must have sufficient power to exceed the target echo signal at the radar receiver input.
    \subtitleone{(3) Matching Modulation Schemes:}
    The modulation of interference signals should align with the radar's modulation characteristics for maximal disruption.
    \subtitleone{(4) Directional Jamming:}
    The directionality of interference signals must align with the radar's main beam. For phased-array radars, rapid beam scanning capabilities are required.

    \subsubsection{Spoofing Attack}
    Spoofing attacks mislead wireless sensing systems by generating false signals. Strategies include modulating reflected signals or actively generating spoofing signals to deceive the system. Spoofing can target various dimensions: distance, velocity, angle, and time/frequency (Duality in Fourier Theory).
    
    \paragraph{Range Spoofing}
    Distance spoofing in wireless sensing systems~\cite{deriberollesCharacterizingRadarNetwork2020,nallaboluFrequencyDomainSpoofingAttack2021,leuGhostPeakPractical2022} involves fast-time modulation of reflected signals, achieved in two ways: one approach is to use a reflective array to capture the signals transmitted by the victim radar, modulate them by applying specific frequency shifts, and then reflect the signals back. For example, mmSpoof~\cite{reddyvennamMmSpoofResilientSpoofing2023} capture the victim radar's signal and reflect it back with a specific frequency shift. mmSpoof demonstrated effective deception at distances up to 100 meters, even in dynamic scenarios. RF-protect~\cite{shenoyRFprotectPrivacyDevicefree2022}, though designed for privacy protection, operates similarly in distance spoofing. Another approach involves generating false waveforms to deceive the radar, provided that the parameters of the victim radar are accurately estimated beforehand. Techniques like MadRadar~\cite{huntMadRadarBlackBoxPhysical2024} estimate the victim radar's parameters in real time to generate spoofing waveforms, enabling precise deception.
    
    \paragraph{Velocity Spoofing}
    Velocity spoofing is achieved by modulating signals over slow time scales, while distance spoofing is implemented via fast-time modulation. Since the principles behind velocity and range spoofing are similar, these two techniques often co-occur but can also be executed independently. The methods described in mmSpoof~\cite{reddyvennamMmSpoofResilientSpoofing2023} and MadRadar~\cite{huntMadRadarBlackBoxPhysical2024} demonstrate this dual capability. The former achieves independent spoofing of distance and velocity by modulating the victim radar’s transmitted signals and adding fixed offsets. In contrast, the latter generates fake reflected signals to implement these attacks.
    
    \paragraph{Angle Spoofing}
    Angle spoofing, which targets Direction of Arrival (DoA) estimation, presents unique challenges compared to distance and velocity spoofing. The difficulty stems from DoA estimation's dependence on phase differences between signals at radar antennas, which cannot be directly manipulated through conventional signal modification methods. Nevertheless, RF-protect~\cite{shenoyRFprotectPrivacyDevicefree2022} demonstrates successful angle spoofing through strategic radar positioning in home privacy protection scenarios. This hardware-software integrated system injects fake human reflections into sensing data, adopting principles from online privacy protection methods to confuse eavesdroppers' tracking capabilities.

    \paragraph{Time/Frequency Spoofing}
    Time and frequency domains present additional attack vectors. RF-protect~\cite{shenoyRFprotectPrivacyDevicefree2022} achieves time-domain spoofing by generating fake respiratory waveforms through phase-shifter manipulation. In acoustic sensing systems, DolphinAttack~\cite{zhangDolphinAttackInaudibleVoice2017} exploits ultrasonic waves above 20 kHz to modulate voice commands, enabling covert control of voice assistants. Building on this, SINGATTACK~\cite{yangRemoteAttacksSpeech2023} overcomes proximity limitations by injecting modulated current into shared power grids, causing switching power supplies to generate human-like audio signals detectable by speech recognition systems.

    \paragraph{Key Technical Challenges and Solutions}
    Successful spoofing attacks face four key technical challenges:
    \subtitleone{(1) Accuracy of Parameter Estimation:} Operating under black-box conditions requires accurate estimation of victim system parameters. MadRadar~\cite{huntMadRadarBlackBoxPhysical2024} demonstrates this capability by estimating radar parameters like pulse slope and cycle time in real-time without prior knowledge.
    \subtitleone{(2) Real-time requirements:} Attacks must operate in real-time with precise synchronization. mmSpoof~\cite{reddyvennamMmSpoofResilientSpoofing2023} and MadRadar~\cite{huntMadRadarBlackBoxPhysical2024} achieve this in radar systems, while DolphinAttack~\cite{zhangDolphinAttackInaudibleVoice2017} maintains real-time operation in voice assistant attacks.
    \subtitleone{(3) Authenticity of Spoofing Signals:} Signals must convincingly mimic legitimate inputs to deceive detection algorithms, as demonstrated in~\cite{reddyvennamMmSpoofResilientSpoofing2023,shenoyRFprotectPrivacyDevicefree2022,huntMadRadarBlackBoxPhysical2024,zhangDolphinAttackInaudibleVoice2017,yangRemoteAttacksSpeech2023}.
    \subtitleone{(4) Stability of Signal Transmission:} Environmental factors such as interference, multipath effects, and attenuation can degrade spoofing signal effectiveness.

    \subsubsection{Adversarial Attack}
    Adversarial attacks exploit algorithmic vulnerabilities within the victim's sensing system by injecting subtle perturbations into the received signals to mislead the decision-making models~\cite{chenWhoRealBob2021,duPracticalDeceptiveJamming2022,shiAudiodomainPositionindependentBackdoor2022,liInaudibleAdversarialPerturbation2024}. While adversarial attacks share similarities with spoofing attacks in their goal to deceive wireless sensing systems, the two differ in their scope. Spoofing attacks primarily target the physical and signal layers, whereas adversarial attacks focus on the algorithmic layer. Subsequently, we provide an overview of adversarial attacks across three key technologies: mmWave-based sensing, WiFi-based sensing, and acoustic sensing.

    \paragraph{mmWave-based}
    Liu et al. proposed an adversarial attack method targeting High-Resolution Range Profiles (HRRP). By identifying ``vulnerable positions'' within HRRP samples, they injected interference pulses into specific range bins to mimic the echo distribution of target scattering centers~\cite{duPracticalDeceptiveJamming2022}. This approach leverages differential evolution (DE) and other non-gradient optimization methods, making it independent of the gradients or structure of the recognition network and thus enhancing its applicability to real-world scenarios.

    \paragraph{WiFi-based}
    Liu et al.~\cite{liuPhysicalWorldAttackWiFibased2022} discovered that inducing Channel State Information (CSI) losses via interference signals could disrupt WiFi-based behavior recognition (WBR) systems. They proposed two methods for generating physical adversarial samples capable of conducting untargeted and targeted attacks. By controlling the frequency and duration of interference signals to mimic normal CSI losses, the attack maintains a high level of stealth.
    Zhou et al.~\cite{zhouWiAdvPracticalRobust2022} introduced the WiAdv framework, which uses full-duplex devices to simulate dynamic multipath effects caused by user gestures. By adjusting signal phases and features, WiAdv generates adversarial signals. Given the non-differentiable nature of WiFi signal preprocessing modules, WiAdv employs black-box attack strategies, including ``Constant Attack'' and ``Greedy Attack.'' On the Widar3.0 gesture recognition system, WiAdv achieved over 70\% attack success rates while maintaining robustness across different environments.
    Wang et al.~\cite{wangAdversarialExamplesWiFi2024} proposed Phy-Adv, the first physical adversarial attack targeting deep learning-based WiFi fingerprinting localization (FL) models. By designing physical attenuation loss and differentiable simulation modules, they generated adversarial interference signals feasible in real-world settings, misleading DFLM results across multiple DNN models. Moreover, RMBN (Robust Mean-Based Normalization) significantly enhanced model robustness against such attacks.
    Zhao et al.~\cite{zhaoExplanationGuidedBackdoorAttacks2024} investigated security vulnerabilities in RF fingerprinting systems using deep neural networks (DNNs). They developed a model-agnostic backdoor attack method utilizing explainable machine learning techniques (e.g., Local Interpretable Model-agnostic Explanations, LIME) and autoencoders~\cite{wuNetworkPruningExplicit2024} to generate effective backdoor triggers without access to model gradients or training processes. Experiments conducted on four different datasets and various DNN architectures showed over 95\% attack success rates, with triggers remaining effective in both the time and frequency domains.

    \paragraph{Acoustic-based}
    Chen et al.~\cite{chenWhoRealBob2021} proposed FAKEBOB, a black-box adversarial attack method towards Speech Recognition Systems (SRS). By framing adversarial sample generation as an optimization problem, they combined threshold estimation algorithms with Natural Evolution Strategies (NES) to execute efficient attacks. 
    Shi et al.~\cite{shiAudiodomainPositionindependentBackdoor2022} introduced a position-independent and imperceptible audio backdoor attack. They embedded triggers at random positions in audio samples during training, simulating environmental sounds to create highly stealthy triggers. To enhance stability in real-world environments, physical distortion simulation was incorporated during training. 
    Li et al.~\cite{liInaudibleAdversarialPerturbation2024} proposed VRIFLE, an inaudible adversarial perturbation attack delivered via ultrasound. Exploiting the inaudibility of ultrasonic frequencies to human ears, attackers injected adversarial perturbations during user speech to manipulate Automatic Speech Recognition (ASR) systems in real time. To address distortion and noise challenges during physical signal transmission, the authors developed an ultrasonic conversion model, enhancing the stability of perturbations in real-world environments. 

    \paragraph{Transduction Attacks}
    Adversarial attacks can also exploit acoustic sensors for transduction attacks. Ji et al.~\cite{jiPoltergeistAcousticAdversarial2021} introduced an acoustic transduction attack targeting inertial sensors embedded in visual sensors. Modern visual sensors often include inertial sensors for image stabilization, mitigating blurring caused by camera movement. However, attackers used carefully crafted sound signals to manipulate the inertial sensor output, triggering unnecessary motion compensation. This resulted in blurred images, leading to misclassification of objects and affecting safety-critical decisions.

    \paragraph{Key Technical Challenges and Solutions}
    Researchers face numerous technical challenges when implementing adversarial attacks. These challenges can be categorized into four primary areas: stealth of perturbations, stability of signal transmission in physical environments, independence from time and location, and black-box characteristics.
    \subtitleone{(1) Stealth of Perturbations:}
    To execute adversarial attacks without being detected, perturbations must be sufficiently inconspicuous. For instance, injecting interference pulses into specific range bins rather than the entire range profile can enhance stealth~\cite{duPracticalDeceptiveJamming2022}. VRIFLE uses ultrasonic signals to transmit inaudible perturbations~\cite{liInaudibleAdversarialPerturbation2024}, while audio backdoor attacks disguise triggers by mimicking environmental sounds~\cite{shiAudiodomainPositionindependentBackdoor2022}.
    \subtitleone{(2)  Stability in Physical Environments:}
    Signals often degrade during physical transmission due to noise, attenuation, or other environmental factors, increasing the difficulty of adversarial attacks. For example, WiAdv improves the adaptability of attack signals in real-world scenarios by dynamically simulating multipath effects~\cite{zhouWiAdvPracticalRobust2022}.
    \subtitleone{(3) Independence from Time:}
    In real-time audio streams, attackers cannot control the precise insertion position of triggers. Therefore, triggers must be designed to remain effective regardless of their placement. Position-independent audio backdoor attacks achieve robustness by randomizing trigger positions, enabling effectiveness across varying time contexts~\cite{zhouWiAdvPracticalRobust2022}.
    \subtitleone{(4) Black-Box Characteristics:}
    In black-box scenarios, attackers do not have access to the target model’s structure or gradient information. As a result, non-gradient or gradient-free attack methods are required. FAKEBOB uses score threshold estimation and NES to perform efficient black-box attacks~\cite{chenWhoRealBob2021}.

    \subsubsection{Development Trends}

    Attacks on wireless sensing sources are expected to become more specialized and complex, focusing on enhanced stealth, robustness, and adaptability. Advancements will likely enable undetectable perturbations, compact and portable attack tools, and high-precision interference or spoofing signals. Multi-modal joint attacks combining radar, WiFi, and acoustic technologies will further expand their scope and impact. These evolving threats highlight the urgent need for intelligent, adaptive, and cross-modal defenses to ensure the resilience of sensing systems in dynamic and challenging environments.

    \subsection{Attacks on Wireless Sensing Channels}
    \subsubsection{Overview}
    Channel attacks manipulate the propagation characteristics of signals, altering the channel response. For example, attackers can use Reconfigurable Intelligent Surfaces (RIS)~\cite{zhangPracticalPassiveIndoor2024} to adjust the reflection and refraction properties of signals, thereby influencing the perception results at the receiver. By controlling the propagation environment, attackers can execute attacks between the signal source and the receiver. Originally developed for communication technologies, RIS has been introduced into sensing systems and is considered a key enabler for integrated sensing and communication (ISAC). We categorize channel attacks into two applications:
    \begin{itemize}
    \item Attacks on Communication Channels: Alter the propagation characteristics of communication channels to tamper with transmitted information or manipulate traffic.
    \item Attacks on Sensing Channels: Modify the propagation properties of sensing channels to interfere with tasks such as target detection or imaging.
    \end{itemize}
    
    \subsubsection{Attacks on Communication Channels}
    By adjusting the phase shifts and reflective properties of RIS, attackers can manipulate channel characteristics, making it difficult for the receiver to identify genuine signals. Two typical scenarios are discussed: data tampering and traffic manipulation.

    \paragraph{Data Tampering}
    Wei et al.~\cite{weiMetasurfaceenabledSmartWireless2023} demonstrated an attack prototype using RIS in the WiFi band to eavesdrop and tamper with wireless communication content. Their findings highlight significant security risks in future intelligent wireless environments, where metasurfaces could enable stealthy and undetectable attacks.
    
    \paragraph{Traffic Manipulation}
    Staat et al.\cite{staatMirrorMirrorWall2022} proposed an Environmental Reconfiguration Attack (ERA) leveraging RIS's fast software-controlled properties. This novel, low-cost, and low-complexity interference method uses RIS to reflect legitimate signals and disrupt the receiver without emitting additional interference signals. Niu et al.\cite{niuReconfigurableIntelligentSurfaceAssisted2024} further proposed an alignment elimination interference scheme that induces self-interference at the receiver using legitimate signals, significantly degrading communication quality without requiring extra interference emissions.

    \paragraph{Key Technical Challenges and Solutions}
    \subtitleone{(1) Stealth of Attacks:}
    Stealth is crucial for preventing detection by communication systems and ensuring prolonged attacks. RIS can dynamically generate false signals or eavesdrop on target signals.
    \subtitleone{(2) Real-Time Requirements:}
    Communication systems demand real-time responses, necessitating efficient algorithms for rapid signal path switching. For example,~\cite{staatMirrorMirrorWall2022} implemented a fast channel switching algorithm to enable real-time interference.
    
    \subsubsection{Attacks on Sensing Channels}
    As a cutting-edge ISAC technology, RIS can be used not only for communication channel attacks but also for attacks on sensing systems. Two typical attack scenarios are discussed:

    \paragraph{Intrusion Detection Jamming}
    Although IRShield\cite{staatIRShieldCountermeasureAdversarial2022} was designed for defense, it effectively targets attacker channels by suppressing malicious motion detection, reducing their detection success rate to below 5\%. Zhou et al. proposed RIStealth, an RIS-based attack method that disables WiFi intrusion detection systems from detecting moving targets\cite{zhouRIStealthPracticalCovert2024}. RIStealth uses RIS's programmable phase reflection properties to modulate the WiFi sensing channel, disrupting the environmental change signals extracted by Intrusion Detection Systems (IDS).

    \paragraph{Imaging Jamming}
    Xu et al.\cite{xuNovelApproachRadar2023} proposed an effective deception interference method that creates false targets and misleading signals in the range-Doppler domain through periodic phase encoding, targeting mmWave radar sensing systems. Similarly, Li et al.\cite{liJammingISARImaging2023} proposed an interference strategy against Inverse Synthetic Aperture Radar (ISAR) systems.

    \paragraph{Key Technical Challenges and Solutions}
    \subtitleone{(1) Compatibility:}
    RIS, as a key technology for ISAC, is already deployed in specialized indoor scenarios to enhance communication performance~\cite{weiMetasurfaceenabledSmartWireless2023,dongGPSMirrorExpandingAccurate2023}. This makes channel-based attacks more feasible. On integrated sensing-communication platforms, RIS can be used to attack sensing systems, demonstrating strong compatibility with communication systems.
    \subtitleone{(2) Independence:} 
    Attacks on sensing channels do not directly occupy wireless channels or affect communication performance. For example, IRShield's interference strategy~\cite{staatIRShieldCountermeasureAdversarial2022} does not impact communication channels.
    \subtitleone{(3) High-speed Modulation:}
    For RIS to achieve specific deception effects in sensing systems, it must maintain high modulation rates during attacks to ensure effectiveness. For example, to deceive range detection, RIS must modulate at MHz frequency level~\cite{liJammingISARImaging2023}.
    
    \subsubsection{Development Trends}

    Future channel attacks are expected to grow in complexity, with joint communication and sensing attacks potentially enhancing attack range and impact, especially in ISAC systems. Emerging technologies like RIS will provide stealthier and more efficient means of attack, achieving undetectable interference through dynamic signal manipulation. Attack strategies will also become more robust, adapting to diverse environments and system conditions. Modular attack tool designs will lower the barriers to implementation, enhancing deployment flexibility. Additionally, integrating AI will enable attackers to optimize strategies intelligently, supporting real-time decision-making and automated attacks. These trends pose severe challenges to existing defense mechanisms.

    \subsection{Defense Strategies}
    \subsubsection{Overview}
    This work explores attacks on wireless sensing systems from the perspectives of targets, sources, and channels. To ensure the security and reliability of such systems, researchers have proposed various defense strategies, broadly classified into active defense and passive defense:
    \begin{itemize}
    \item Active Defense: Prevents attacks through real-time monitoring and intervention.
    \item Passive Defense: Employs inherent security features to resist attacks, ensuring reliable system operation even under attack.
    \end{itemize}

    \subsubsection{Active Defense}
    Active defense involves proactive measures to prevent, detect, and counter attacks. These measures often include real-time monitoring, disrupting attackers’ actions, or preemptively stopping attacks. Key active defense strategies include parameter randomization, waveform fingerprinting, and channel characteristic modeling.

    \paragraph{Parameter Randomization}
    Nallabolu et al.\cite{nallaboluFrequencyDomainSpoofingAttack2021a} proposed a hybrid Chirp FMCW radar defense method. By transmitting Chirp signals with varying slopes and analyzing the received signals, this approach distinguishes real targets from spoofed ones. Even if attackers alter the frequency shift, the system effectively differentiates real and fake targets, significantly enhancing FMCW radar's anti-spoofing capability. Similarly, Sun et al.\cite{sunWhoControlPractical2021} proposed a challenge-response mechanism to protect wireless sensing systems. Radar systems randomly alter waveform parameters (e.g., frequency, phase, or modulation) for each transmission (challenge) and validate reflected signals based on these parameters (response). If the reflected signal matches the random parameters, the target is deemed legitimate; otherwise, a spoofing attack is suspected.

    \paragraph{Waveform Fingerprinting}
    Sun et al.~\cite{sunWhoControlPractical2021} also utilized radar signal characteristics such as spectrum and phase noise to record legitimate waveform fingerprints, which are compared with received signals to detect potential attacks.
    
    \paragraph{Channel Characteristic Modeling}
    Zhang et al.\cite{zhangDolphinAttackInaudibleVoice2017} proposed a defense strategy to detect and identify the direction of DolphinAttack attackers without additional hardware or modifications\cite{zhangEarArrayDefendingDolphinAttack2021}. By analyzing the channel characteristics and attenuation rates of ultrasonic signals, their method distinguishes audible from ultrasonic waves, effectively defending against DolphinAttacks.
    
    \paragraph{Key Technical Challenges and Development Trends}
    \subtitleone{(1) Real-Time Requirements:}
    Active defense often increases system complexity and cost due to additional computation requirements. Efficient algorithms and hardware acceleration techniques are needed to meet real-time demands~\cite{zhangEarArrayDefendingDolphinAttack2021}.
    \subtitleone{(2) False Alarms and Missed Detections:} 
    Balancing false alarm and missed detection rates is critical to avoid disrupting normal communications. For instance, Nallabolu’s hybrid Chirp FMCW radar ensures high accuracy when differentiating real and spoofed targets~\cite{nallaboluFrequencyDomainSpoofingAttack2021a}.

    \subsubsection{Passive Defense}
    Passive defense strategies focus on designing inherent system security features to resist or mitigate attacks without real-time intervention. Key passive defense techniques include signal obfuscation and physical layer encryption.

    \paragraph{Signal Obfuscation}
    Wang et al.~\cite{wangPriSensePrivacyPreservingWireless2024} proposed a privacy-preserving defense by carefully designing pilot signals to introduce artificial peaks at specific frequencies. This creates false variations in equivalent CSI, misleading attackers and resulting in inaccurate estimations of vital signs.
    
    \paragraph{Physical Layer (PHY) Encryption}
    Li et al.\cite{liProtegoSecuringWireless2022} introduced a RIS-based security strategy that shifts protection functions from transmitters to the environment. By deploying programmable metasurfaces as ``security barriers,'' legitimate receivers experience directional signal enhancement, while phase noise is introduced toward eavesdroppers to disrupt decoding. RIS, as a critical ISAC technology, can also defend against sensing system attacks. Luo et al.\cite{luoMIMOCryptMultiUserPrivacyPreserving2024} proposed MIMOCrypt, a physical layer encryption scheme for wireless sensing systems. It protects user privacy while supporting multi-user scenarios, maintaining both communication and sensing capabilities. MIMOCrypt uses multi-objective optimization and efficient decryption methods to balance security, sensing accuracy, and communication quality.

    \paragraph{Key Technical Challenges and Development Trends}
    \subtitleone{(1) Compatibility:}
    Passive defense mechanisms must align with existing communication standards and devices, avoiding interference with normal functionality. This requires balancing security and performance through careful system design and software-hardware integration~\cite{luoMIMOCryptMultiUserPrivacyPreserving2024}.
    \subtitleone{(2) System Complexity:} 
    Introducing new defense mechanisms may increase system complexity and cost, necessitating simplified designs~\cite{liProtegoSecuringWireless2022}.

    Future defense strategies will likely focus on integrating security design with hardware collaboration, cross-layer protection, lightweight and adaptive systems, and authentication mechanisms. By leveraging software-hardware co-design, multi-layered protection, and AI technologies, future defenses aim to enhance robustness. 

    \subsection{Summary and Insights}
    \subsubsection{Unified Attack Framework}
    \begin{figure}[!t]
    \centering
    \includegraphics[width=\linewidth]{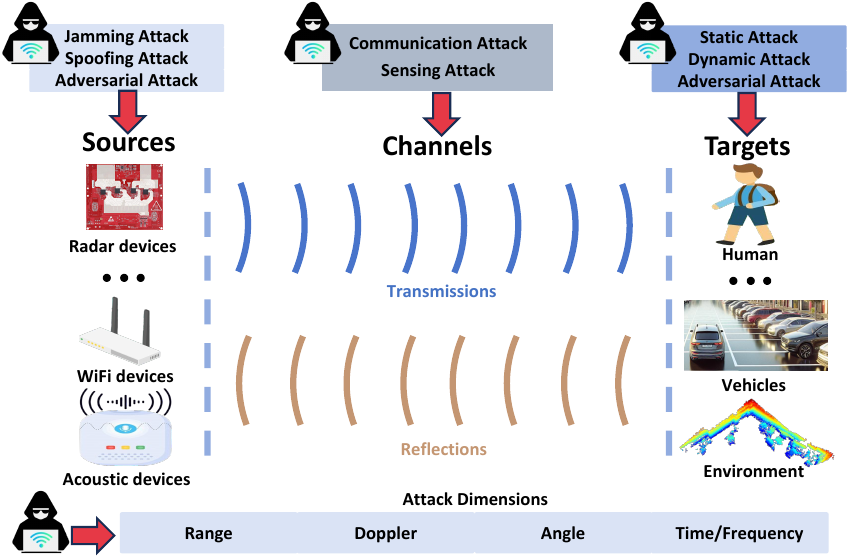}
    \caption{A unified attack framework for wireless sensing systems. Attackers can target three perspectives (source, target, channel) across four key dimensions (range, Doppler, angle, time/frequency).}
    \label{fig:ToAwoT}
    \end{figure}
    This paper proposes a classification framework based on targets, sources, and channels, providing a comprehensive perspective for studying attacks on wireless sensing systems. Representative studies are summarized from the perspectives of sensing technologies, tasks, and strategies. Furthermore, we propose four key attack dimensions: range, Doppler, angle (including elevation and azimuth), and time/frequency (leveraging Fourier duality). These dimensions complement the classifications of targets, sources, and channels, allowing attackers to combine these perspectives to mount multi-dimensional attacks.

    Studies demonstrate that single or combined attacks on these dimensions can threaten all RF sensing applications. For instance, human activity recognition (HAR)~\cite{xieRPMRFbasedPose2024,xieRPM20RFBased2024,yuMobiRFPosePortableRFBased2024} can be attacked from the range-angle dimension, gesture recognition from the range-Doppler-angle dimension, imaging from the range-Doppler-angle dimension, and vital sign sensing from the time dimension. Autonomous driving systems face threats across all dimensions. Since RF sensing tasks rely on these dimensions, an attack on any single dimension can disrupt the sensing task, encompassing all current attack methods within this unified framework, as illustrated in Fig.~\ref{fig:ToAwoT}.

    \subsubsection{Future Research Directions}
    This framework highlights the transferability of attack methods across different sensing systems, suggesting that techniques developed for one task can inspire defenses for others. From a dialectical perspective, studying attack methods for one task can inform the design of defenses for others, as countering an attack inherently strengthens the system.
    Key future directions for wireless sensing system attacks include: \subtitleone{(1) Co-Evolution of Attacks and Defenses:} Both will increasingly rely on intelligent techniques, such as AI and machine learning, for precision and adaptability. \subtitleone{(2) Cross-Technology and Cross-Modal Synergies:}Sensing system designs must consider broader multi-modal and multi-dimensional collaborative effects. \subtitleone{(3) Dual Evolution of New Technologies and Infrastructure:} Advances in attack and defense will depend on infrastructure upgrades, including software-hardware collaboration and high-precision systems, evolving toward modularity, intelligence, and collaboration.

    \section{Wireless Signals as Tools for Attacks} \label{sec:roleTool}
    \begin{table*}[!htbp]
\centering
\caption{Summary of Wireless Signal-based Attack and Defense Methods}
\label{tab:summaryTool}
\resizebox{\textwidth}{!}{
\begin{tabular}{|p{1cm}|p{2cm}|p{5cm}|p{3.7cm}|p{3cm}|p{6cm}|}
\hline
\textbf{Category} & \textbf{Subcategory} & \textbf{Representative Works} & \textbf{Core Principles} & \textbf{Main Challenges} & \textbf{Development Trends} \\
\hline
\multirow{3}{*}[-10em]{\rotatebox[origin=c]{90}{\textbf{\makebox[3em]{Active Attacks}}}} & Vibration Detection & 
\begin{itemize}[leftmargin=*]
\item Headphone eavesdropping\cite{xuMmEarPushLimit2024,wangMmPhoneAcousticEavesdropping2022}
\item Object vibration detection\cite{zhangAmbiEarMmWaveBased2022,wangWavesdropperWallWord2022,fengMmEavesdropperSignalAugmentationbased2023}
\item Phone eavesdropping\cite{basakMmSpySpyingPhone2022,wangMmEveEavesdroppingSmartphones2022}
\item Speaker eavesdropping\cite{huMILLIEARMillimeterwaveAcoustic2022,huMmEchoMmWavebasedAcoustic2023,weiAcousticEavesdroppingWireless2015,wangUWHearWallExtraction2020}
\item Covert channels\cite{liSpiralspyExploringStealthy2022}
\end{itemize} & 
Using wireless signals to detect and reconstruct vibration-induced information & 
\begin{itemize}[leftmargin=*]
\item Weak signal detection
\item Noise processing
\item Signal quality
\end{itemize} & 
\begin{itemize}[leftmargin=*]
\item Range: Short to long-range detection\cite{wangWavesdropperWallWord2022,fengMmEavesdropperSignalAugmentationbased2023}
\item Processing: Traditional to deep learning\cite{wangMmEveEavesdroppingSmartphones2022,xuMmEarPushLimit2024}
\item Scope: Single to multiple sources\cite{zhangAmbiEarMmWaveBased2022,liSpiralspyExploringStealthy2022}
\end{itemize} \\
\cline{2-6}
& EMI Attacks & 
\begin{itemize}[leftmargin=*]
\item Touchscreen control\cite{shanInvisibleFingerPractical2022,wangAnalyzingDefendingGhostTouch2024}
\item Mouse control\cite{songPuppetMousePracticalContactless2024}
\item Sensor interference\cite{jangParalyzingDronesEMI2023,gaoExploringPracticalAcoustic2024}
\item Process control\cite{gaoPracticalEMIAttacks2024}
\end{itemize} & 
Controlling devices through electromagnetic interference & 
\begin{itemize}[leftmargin=*]
\item Signal design
\item Attack control
\item Stealth
\end{itemize} & 
\begin{itemize}[leftmargin=*]
\item Theory: Empirical to theoretical\cite{wangAnalyzingDefendingGhostTouch2024}
\item Precision: Coarse to fine control\cite{songPuppetMousePracticalContactless2024}
\item Efficiency: High to low power\cite{gaoExploringPracticalAcoustic2024}
\end{itemize} \\
\cline{2-6}
& Channel Manipulation & 
\begin{itemize}[leftmargin=*]
\item Charging attacks\cite{kohlerBrokenwireWirelessDisruption2023}
\item Voice assistant attacks\cite{yanSurfingattackInteractiveHidden2020}
\item Motion tracking\cite{zhuTuAlexaWhen2018}
\item Screen theft\cite{liuScreenGleaningScreen2021}
\item Voice backdoors\cite{zhangInaudibleBackdoorAttack2024}
\end{itemize} & 
Exploiting protocol and channel vulnerabilities & 
\begin{itemize}[leftmargin=*]
\item Protocol exploitation
\item Signal synchronization
\end{itemize} & 
\begin{itemize}[leftmargin=*]
\item Precision: General to protocol-level\cite{kohlerBrokenwireWirelessDisruption2023}
\item Scope: Generic to specialized\cite{zhangInaudibleBackdoorAttack2024}
\item Stealth: Obvious to covert\cite{liuScreenGleaningScreen2021}
\end{itemize} \\
\hline
\multirow{3}{*}[-5em]{\rotatebox[origin=c]{90}{\textbf{\makebox[3em]{Passive Attacks}}}}  & Device Leakage & 
\begin{itemize}[leftmargin=*]
\item CPU monitoring\cite{chengMagAttackGuessingApplication2019,ningDeepMagSniffingMobile2018}
\item GPU monitoring\cite{zhanGraphicsPeepingUnit2022,maiaCanOneHear2022}
\item Fingerprint sensing\cite{niRecoveringFingerprintsDisplay2023}
\item Microphone detection\cite{rameshTickTockDetectingMicrophone2022,rameshYourMicLeaks2024}
\item Mining detection\cite{xiaoMagTracerDetectingGPU2023}
\end{itemize} & 
Analyzing electromagnetic emissions from devices & 
\begin{itemize}[leftmargin=*]
\item Signal extraction
\item Feature recognition
\item Attack practicality
\end{itemize} & 
\begin{itemize}[leftmargin=*]
\item Scope: Single to complex systems\cite{zhanGraphicsPeepingUnit2022,xiaoMagTracerDetectingGPU2023}
\item Precision: State to data reconstruction\cite{maiaCanOneHear2022,rameshYourMicLeaks2024}
\item Usage: Threats to monitoring\cite{xiaoMagTracerDetectingGPU2023}
\end{itemize} \\
\cline{2-6}
& Protocol-Defined Signals Analysis & 
\begin{itemize}[leftmargin=*]
\item IoT analysis\cite{zouIoTBeholderPrivacySnooping2023,classenAttacksWirelessCoexistence2022}
\item Location tracking\cite{abediNoncooperativeWifiLocalization2022,guWiFiLeaksExposingStationary2024,liRFTrackStealthyLocation2024}
\item Password inference\cite{chenSilentThiefPassword2024,wangMuKIFiMultiPersonKeystroke2024}
\end{itemize} & 
Inferring information from protocol-defined signals & 
\begin{itemize}[leftmargin=*]
\item Feature extraction
\item Data analysis
\item System efficiency
\end{itemize} & 
\begin{itemize}[leftmargin=*]
\item Features: Enhanced extraction\cite{zouIoTBeholderPrivacySnooping2023}
\item Analysis: Finer granularity\cite{chenSilentThiefPassword2024}
\item Intelligence: Data fusion\cite{liRFTrackStealthyLocation2024}
\end{itemize} \\
\hline
\multirow{2}{*}[-5em]{\rotatebox[origin=c]{90}{\textbf{\makebox[3em]{Defense}}}}  & Physical Layer & 
\begin{itemize}[leftmargin=*]
\item Reflecting surface
\item Beamforming\cite{asaadSecureActivePassive2022}
\item Collaborative defense\cite{liTwoWayAerialSecure2024}
\item Power allocation\cite{jiaSecureMultiantennaTransmission2022}
\end{itemize} & 
Controlling signal propagation and power & 
\begin{itemize}[leftmargin=*]
\item Channel uncertainty
\item System complexity
\end{itemize} & 
\begin{itemize}[leftmargin=*]
\item Integration: Multi-technology\cite{asaadSecureActivePassive2022}
\item Methods: Adaptive strategies\cite{jiaSecureMultiantennaTransmission2022}
\item AI integration
\end{itemize} \\
\cline{2-6}
& Application Layer & 
\begin{itemize}[leftmargin=*]
\item Adversarial defense\cite{yeScreenPerturbationAdversarial2023,testaPrivacyRealTimeSpeech2023}
\item Privacy protection\cite{corbettBystandARProtectingBystander2023}
\end{itemize} & 
Scenario-specific countermeasures & 
\begin{itemize}[leftmargin=*]
\item Signal complexity
\item Environmental factors
\end{itemize} & 
\begin{itemize}[leftmargin=*]
\item Strategy transfer
\item Deep learning integration
\item Multi-layer defense
\end{itemize} \\
\hline
\end{tabular}
}
\end{table*}
    Wireless signals, beyond their communication and sensing functions, have emerged as powerful attack tools. Compared to traditional cyberattacks, wireless signal-based attacks directly affect the physical layer. This section systematically analyzes these attack methods, categorizing them into (1) active attacks, (2) passive attacks, and (3) defense strategies (i.e., how to protect targeted systems from wireless signal attacks), as summarized in Tab.~\ref{tab:summaryTool}.

    \textbf{Active attacks} actively emit wireless signals to attack target systems, through vibration detection (e.g., headphone audio eavesdropping\cite{xuMmEarPushLimit2024}), electromagnetic interference (e.g., mouse control\cite{songPuppetMousePracticalContactless2024}), and channel manipulation (e.g., charging system disruption\cite{kohlerBrokenwireWirelessDisruption2023}).
    
    \textbf{Passive attacks} capture and analyze naturally existing signals to extract information covertly, e.g., GPU electromagnetic leakage\cite{zhanGraphicsPeepingUnit2022}, light reflection-based sound eavesdropping\cite{nassiLamphonePassiveSound2022}, and WiFi traffic patterns\cite{zouIoTBeholderPrivacySnooping2023}.
    
    \textbf{Defense strategies} serve as countermeasures against these attacks, implementing protections from physical layer beamforming\cite{staatIRShieldCountermeasureAdversarial2022} to application layer security\cite{testaPrivacyRealTimeSpeech2023}.

    \subsection{Active Attacks Based on Wireless Signals}

    \subsubsection{Attack Framework Overview} 
    Active wireless signal attacks exploit three main mechanisms:
    \subtitleone{(1) Vibration detection mechanism:} Wireless signals detect and capture minute surface vibrations, thereby extracting audio information from headphones and environmental objects for eavesdropping attacks\cite{xuMmEarPushLimit2024,shiPrivacyLeakageSpeechinduced2023}.
    \subtitleone{(2) EMI mechanism:} Wireless signals interfere with and control electronic devices through precisely designed electromagnetic interference, enabling unauthorized manipulation of touchscreens, mice, and sensors\cite{shanInvisibleFingerPractical2022,songPuppetMousePracticalContactless2024,jangParalyzingDronesEMI2023}.
    \subtitleone{(3) Channel manipulation mechanism:} Wireless signals exploit protocol vulnerabilities to disrupt device communications and compromise system operations\cite{kohlerBrokenwireWirelessDisruption2023}.

    \subsubsection{Vibration-based eavesdropping}
    
    Here we first introduces the principles of vibration detection, followed by a discussion of representative methods from an application perspective.
    
    \paragraph{Physical Principle}
    
    Acoustic waves and mechanical motion can generate micrometer or even nanometer-scale vibrations on the surface of objects. Although these vibrations are very small, they can cause significant changes in the phase of reflected wireless signals (especially mmWave)~\cite{guoMeasuringMicrometerLevelVibrations2023,basakMmSpySpyingPhone2022,wangMmPhoneAcousticEavesdropping2022}.

    \begin{figure}[!htbp]
    \centering
    \includegraphics[width=\linewidth]{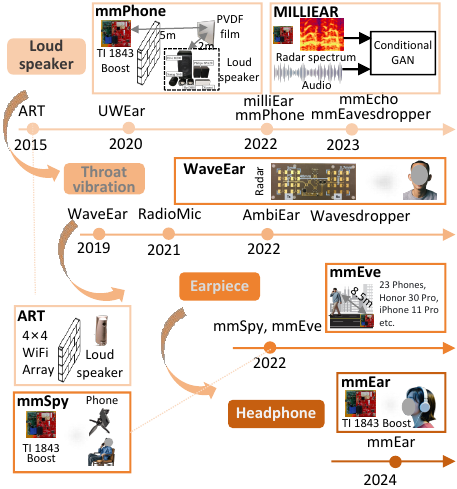}
    \caption{Evolution of acoustic eavesdropping techniques based on wireless signals (2015-2024).}
    \label{fig:acousticEvolution}
    \end{figure}
    
    \paragraph{Attack Application 1 - Sound Eavesdropping}
    
    Sound eavesdropping through vibration detection has become a significant attack vector. These attacks reconstruct voice information by capturing surface vibrations caused by sound waves. The field has evolved from targeting high-amplitude sources, such as loudspeakers, throat vibrations, to more subtle sources, including earpiece receivers and headphones (as shown in Fig.~\ref{fig:acousticEvolution}).
    \subtitleone{(1) Speaker Eavesdropping.}
    Early work like ART\cite{weiAcousticEavesdroppingWireless2015} focused on speaker-emitted sounds, detecting digits through RF reflection phase changes. UWHear\cite{wangUWHearWallExtraction2020} advanced this approach using ultra-wideband (UWB) RF reflections for multi-speaker recognition. mmPhone\cite{wangMmPhoneAcousticEavesdropping2022} introduced an innovative approach utilizing piezoelectric films as passive ``microphones'', achieving speech reconstruction at distances over 5 meters through denoising neural networks.
    MILLIEAR\cite{huMILLIEARMillimeterwaveAcoustic2022,huUnconstrainedVocabularyEavesdropping2024} combined high-resolution mmWave FMCW ranging with generative models for complete audio reconstruction across various conditions. Its successor, mmEcho\cite{huMmEchoMmWavebasedAcoustic2023}, enhanced signal processing for robust speech reconstruction. mmEavesdropper\cite{fengMmEavesdropperSignalAugmentationbased2023} improved attack directionality using beamforming and Chirp-Z transform, while developing theoretical models for signal distortion calibration. VibSpeech\cite{wangVibSpeechExploringPractical2024} demonstrated wideband speech reconstruction from band-limited vibration signals, achieving 8kHz reconstruction from sub-500Hz side-channel data.
    \subtitleone{(2) Throat Vibration Eavesdropping.} 
    Besides speakers, the human throat also produces relatively large vibrations during speech. 
    WaveEar\cite{xuWaveEarExploringMmWavebased2019} pioneered throat vibration detection using custom mmWave hardware. RadioMic\cite{ozturkRadioMicSoundSensing2021} extended this to multiple vibration sources. AmbiEar\cite{zhangAmbiEarMmWaveBased2022} achieved NLOS speech recognition through correlated object vibrations. The key insight of AmbiEar is that human voice causes correlated vibrations in surrounding objects, regardless of the speaker's position and posture. AmbiEar successfully overcame the challenges of low signal-to-noise ratio and signal distortion using component extraction and encoder-decoder networks.
    The latest Wavesdropper system\cite{wangWavesdropperWallWord2022} goes further, capable of recovering speech content by detecting the speaker's throat vibrations through walls. RF-Mic\cite{chenRFMicLiveVoice2023} introduced RFID-based speech eavesdropping via glasses, analyzing facial movements through a Conditional Denoising Autoencoder (CDAE) network. The system extracts facial movements, bone conduction, and air vibrations to construct a facial speech dynamics model for real-time eavesdropping.
    \subtitleone{(3) Mobile Phone Receiver Eavesdropping.} 
    Recent research has progressed from large-amplitude vibrations to challenging weak vibration detection in phone receivers and headphones, enabled by mmWave sensitivity to small displacements. mmSpy\cite{basakMmSpySpyingPhone2022} achieved 83-44\% accuracy in digit and keyword classification at 1-6 feet using 77GHz and 60GHz radar. mmEve\cite{wangMmEveEavesdroppingSmartphones2022} extended attack distances to 6-8 meters through I/Q plane optimization and GAN denoising, validating attacks across 23 smartphone models.
    \subtitleone{(4) Headphone Eavesdropping.} 
    Xu et al. extended the attack target to headphone devices, proposing the mmEar system\cite{xuMmEarPushLimit2024}. Compared to mobile phone receivers, headphone surface vibrations are even weaker, posing greater challenges for attacks. To address this, the authors proposed a weak vibration enhancement method and designed a deep denoising network to further improve the signal-to-noise ratio. The system also employs a pretrain-finetune conditional GAN model for cross-device attack generalization.
    
    \paragraph{Attack Application 2 - Vibration-Based Side Channel Attacks}
    
\begin{figure}
        \centering
        \includegraphics[width=1\linewidth]{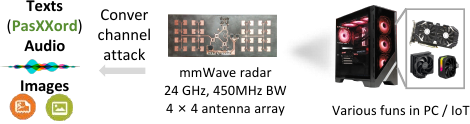}
        \caption{Framework diagram of SpiralSpy}
        \label{fig:SpiralSpyFramework}
    \end{figure}

    Vibration detection technology enables covert channels for data exfiltration from physically isolated systems by controlling mechanical components to generate specific vibration patterns. The SpiralSpy system\cite{liSpiralspyExploringStealthy2022} demonstrates this capability by modulating computer fan speeds, achieving 6bps transmission rates within 8 meters (Fig.~\ref{fig:SpiralSpyFramework}). Unlike passive sound eavesdropping, these active vibration control methods establish reliable communication channels with higher capacity, circumventing traditional network isolation and electromagnetic shielding measures.
    
    \paragraph{Key Technical Challenges and Development Trends}
    
    The fundamental challenge in vibration-based eavesdropping lies in recovering speech signals from minute surface vibrations captured by wireless signals. This process faces significant technical hurdles in \textit{vibration detection}, \textit{noise processing}, and \textit{signal reconstruction}.
    \subtitleone{(1) Weak Vibration Detection.} Extracting micron-level displacements from reflected signals remains a core challenge. The field has progressed through three key advances: First, signal representation evolved from basic amplitude detection to advanced methods, exemplified by mmEve's\cite{wangMmEveEavesdroppingSmartphones2022} IQ plane fitting optimization and mmEar's\cite{xuMmEarPushLimit2024} Feeble Vibration Enhancement technique. Second, spatial resolution improved through phased array MIMO technology, as demonstrated by Shi et al.\cite{shiPrivacyLeakageSpeechinduced2023}. Third, detection capabilities advanced from single to multi-target sensing, as shown in AmbiEar\cite{zhangAmbiEarMmWaveBased2022}.
    \subtitleone{(2) Environmental Noise Processing.} Solutions evolved from passive suppression to active elimination through two approaches: Signal processing and deep learning. The Wavesdropper system\cite{wangWavesdropperWallWord2022} introduced CEEMD-based suppression, while mmEavesdropper\cite{fengMmEavesdropperSignalAugmentationbased2023} improved SNR through beamforming. In deep learning, mmEve\cite{wangMmEveEavesdroppingSmartphones2022} pioneered GAN-based denoising, and mmEar\cite{xuMmEarPushLimit2024} enhanced generalization through pretrain-finetune frameworks.
    \subtitleone{(3) Progress advanced from feature recognition to full signal reconstruction.} The field has advanced from feature recognition to complete signal reconstruction. MILLIEAR\cite{huMILLIEARMillimeterwaveAcoustic2022} achieved unconstrained speech reconstruction using conditional GANs, further refined by mmEcho\cite{huMmEchoMmWavebasedAcoustic2023}. mmMIC\cite{fanMmMICMultimodalSpeech2023} introduced multi-modal fusion combining lip and vocal cord data.
    
    These developments point to three major future directions: enhanced multi-dimensional detection capabilities, advanced signal processing integrating deep learning and multi-modal fusion, and expanded application scenarios. Particularly promising is the exploitation of wireless signals' unique properties for challenging scenarios, such as through-wall sensing and NLOS applications\cite{wenNonlineofsightSparseAperture2024,gengRecentAdvancesNonlinesight2021,liDeepNonlineofsightImaging2024,liNlostNonlineofsightImaging2023,gengPassiveNonLineSightImaging2022,zhangPassiveNonlinesightImaging2023}, where traditional optical sensors face limitations.

    \subsubsection{Device Control Based on EMI}

\begin{figure}[!t]
    \centering
    \includegraphics[width=\linewidth]{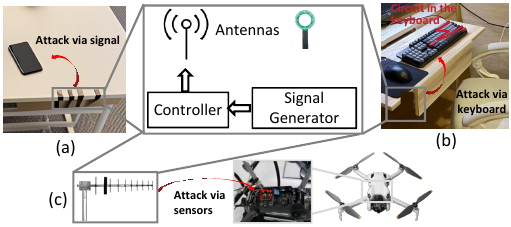}
    \caption{Examples of active attacks based on EMI. All these attacks are based on the same basic framework: generating specific waveforms through a signal generator, modulating them via a controller, and transmitting them through an antenna.}
    \label{fig:emiAttack}
    \end{figure}

    EMI attacks represent a powerful method for compromising device security through targeted electromagnetic emissions. As shown in Fig.~\ref{fig:emiAttack}, these attacks can be implemented through a common framework of signal generation, modulation, and transmission to achieve various levels of device control.
    
    \paragraph{Physical Principle}
    EMI attacks exploit electromagnetic coupling to induce interfering voltages and currents in target devices, affecting their operation at three levels: (1) Signal-level interference affects signal processing, such as inducing false touch inputs on capacitive screens\cite{shanInvisibleFingerPractical2022}; (2) Circuit-level interference causes component malfunctions, demonstrated in keyboard scanning circuit attacks\cite{jiangGhosttypeLimitsUsing2024}; and (3) System-level interference disrupts critical components like IMU sensors in drones, leading to control failure\cite{jangParalyzingDronesEMI2023}.
    
    \paragraph{Work Review}
    
    In recent years, researchers have demonstrated theeffectiveness of EMI attacks in multiple application scenarios.
    \subtitleone{(1) Input Device Control.}
    Early work by Maruyama et al.\cite{maruyamaTapGhostCompilation2019} established techniques framework for touchscreen attacks. The Invisible Finger system\cite{shanInvisibleFingerPractical2022} advanced this field by conducting systematic analysis of touchscreen electromagnetic response mechanisms, establishing minimum electric field requirements and optimal frequencies for phantom touch generation. The system incorporated touchscreen localization and touch injection feedback modules, enabling multi-touch simulation and complex gesture injection.
    Subsequent research expanded both scope and precision. The PuppetMouse system\cite{songPuppetMousePracticalContactless2024} achieved precise computer mouse control with 4ms response latency through optimized electromagnetic interference parameters. Wang et al.\cite{wangAnalyzingDefendingGhostTouch2024} refined GhostTouch attacks to achieve 14.6×19.2 pixel injection precision. Recent advances by Gao et al.\cite{gaoPracticalEMIAttacks2024} demonstrated capabilities beyond touch injection, including the suppression of legitimate user inputs to prevent termination of malicious programs.
    \subtitleone{(2) Sensor Interference.}
    Early research primarily focused on attacking individual sensors, such as attack on image stabilizers\cite{jiPoltergeistAcousticAdversarial2021}. Jang et al.\cite{jangParalyzingDronesEMI2023} proposed an EMI attack targeting Unmanned Aerial Vehicle (UAV) Inertial Measurement Unit (IMU) sensors. Compared to traditional acoustic resonance-based attacks, this method directly acts on the sensor's communication channel, offering higher efficiency and greater difficulty in defense. The authors' research indicates that this attack method can effectively disrupt a UAV's attitude control, leading to flight accidents. The latest research has expanded to multi-degree-of-freedom systems, such as the KITE attack framework proposed by Gao et al.\cite{gaoExploringPracticalAcoustic2024}, which maintains stable injection effects even with frequency offsets. This attack is characterized by its ability to bypass existing defense mechanisms while maintaining a high attack success rate.
    
    \paragraph{Key Technical Challenges and Development Trends}
    
    Device control based on EMI faces three main technical challenges:
    \subtitleone{(1) Precise design of interference signals.}
    Modern electronic devices generally employ anti-interference designs, requiring attack signals to be sufficiently precise to produce the intended effects. Researchers address this issue by combining theoretical modeling with experimental optimization. For instance, Wang et al.\cite{wangAnalyzingDefendingGhostTouch2024} achieved high-precision touch point control by systematically modeling the capacitive characteristics of touchscreens.
    \subtitleone{(2) Controllability of attack effects.}
    Unlike simple interference, precise device control requires good controllability in both time and space. Gao et al.\cite{gaoPracticalEMIAttacks2024} achieved selective cancellation of specific touch commands without affecting other normal operations by precisely controlling the timing and intensity of EMI signals.
    \subtitleone{(3) Stealthiness of attacks.}
    To avoid detection and defense, attacks must be as covert as possible. Researchers enhance attack stealthiness by optimizing signal waveforms, reducing energy consumption, and simulating normal behavior patterns. For example, the KITE attack framework designed by Gao et al.\cite{gaoExploringPracticalAcoustic2024} achieves stable attack effects while minimizing power consumption.
    
    Based on the above analysis, these attack methods are transitioning from the proof-of-concept stage to practical applications. Future research directions focus on further overcoming the aforementioned challenges from three aspects: theoretical foundations, control precision, and efficiency optimization. Additionally, these attacks also pose new challenges to device security design.

    \subsubsection{Communication Interruption Based on Channel Manipulation}

    Channel manipulation attacks represent a sophisticated class of wireless security threats that exploit both physical propagation characteristics and protocol-level vulnerabilities. Strictly speaking, this attack is more closely related to wireless communication security and have a relatively minor connection to wireless sensing security. However, considering that these methods indeed use injected wireless signals to attack normally functioning systems, they still fall within the scope of this paper.
    
    \paragraph{Physical Principle}
    
    Channel manipulation attacks operate at two distinct levels: At the physical level, attackers modify channel characteristics through targeted signal injection, as demonstrated by research on through-wall RFID signal manipulation\cite{wangThruwallEavesdroppingLoudspeakers2022} and the exploitation of aircraft landing systems\cite{sathayeWirelessAttacksAircraft2019}. At the protocol level, attacks leverage specific communication protocol mechanisms to achieve interference objectives.
    
    \paragraph{Work Review}
    
    Current research in this domain primarily focuses on two key areas: control system attacks and information system attacks.
    \subtitleone{(1) Control System Attacks.} The Brokenwire attack\cite{kohlerBrokenwireWirelessDisruption2023} demonstrates a novel approach to disrupting electric vehicle charging systems by exploiting the Carrier Sense Multiple Access with Collision Avoid (CSMA/CA) mechanism in the HomePlug Green PHY protocol. This attack achieves three significant advances: power efficiency three orders of magnitude higher than traditional noise interference, simultaneous disruption of multiple charging facilities, and implementation feasibility using standard radio hardware.
    \subtitleone{(2) Information System Attacks.} Several innovative attack vectors have emerged in this domain. Yan et al.\cite{yanSurfingattackInteractiveHidden2020} demonstrated voice assistant activation through solid-medium ultrasonic wave propagation, exploiting acoustic signal processing mechanisms. Liu et al.\cite{liuScreenGleaningScreen2021} developed an electromagnetic side-channel attack capable of reconstructing display content by intercepting screen-bound electromagnetic signals. Zhang et al.\cite{zhangInaudibleBackdoorAttack2024} engineered an inaudible backdoor attack on voice command systems by embedding sub-noise floor triggers in the audio spectrum.
    
    \paragraph{Technical Capabilities, Challenges and Development Trends}

    Current wireless signal attacks demonstrate several significant capabilities: \textit{(i) Non-contact} operation, enabling remote audio eavesdropping via mmEar\cite{xuMmEarPushLimit2024} and touchscreen control through Invisible Finger\cite{shanInvisibleFingerPractical2022}; \textit{(ii) Signal penetration} through physical barriers, as demonstrated by Wavesdropper's\cite{wangWavesdropperWallWord2022} through-wall speech eavesdropping; \textit{(iii)} \textit{Precise control} capabilities, exemplified by PuppetMouse's\cite{songPuppetMousePracticalContactless2024} millisecond-level mouse manipulation; and \textit{(iv)} high energy efficiency, with Brokenwire\cite{kohlerBrokenwireWirelessDisruption2023} achieving three orders of magnitude higher efficiency compared to traditional interference methods.
    
    Advancing these capabilities, particularly in channel manipulation attacks, presents two primary technical challenges:
    \subtitleone{(1) Protocol vulnerability exploitation.} Modern communication protocols incorporate multi-layered protection mechanisms, necessitating comprehensive analysis to identify exploitable weaknesses. Researchers have achieved breakthroughs through systematic protocol analysis, particularly at Media Access Control (MAC) and PHY, as demonstrated in the Brokenwire attack\cite{kohlerBrokenwireWirelessDisruption2023}.
    \subtitleone{(2) Signal synchronization precision.} Effective attacks, especially those requiring precise control, demand accurate timing coordination. This challenge has been addressed through adaptive timing tracking algorithms and real-time signal feedback, crucial for sophisticated operations like the through-wall screen attack\cite{liWaveSpyRemoteWall2020}.
    
    The field's evolution reveals two key technological trajectories: a transition from power-intensive interference to precise vulnerability-based control mechanisms, and the development of system-specific optimization strategies that leverage target characteristics rather than generic interference methods. These trends indicate a shift toward more sophisticated, energy-efficient, and targeted attack methodologies, particularly in channel manipulation research.

    \subsection{Passive Attacks Based on Wireless Signals} \label{sec:2.2}
    
    \subsubsection{Attack Framework Overview}
    
    Electronic devices inherently generate various signal emissions during operation, including electromagnetic radiation, acoustic signals, and thermal outputs, which can expose sensitive information. Unlike active attacks that require signal injection, passive attacks achieve their objectives by capturing and analyzing these naturally occurring emissions, resulting in enhanced attack stealth.
    
    Analysis of recent research reveals three primary categories of passive attacks, based on their underlying physical mechanisms: \subtitleone{(1) Based on device leakage signals}, which captures direct physical emissions such as electromagnetic radiation and heat during device operation and \subtitleone{(2) Based on network traffic}, which analyzes device-generated network traffic patterns to deduce user behavior and environmental information.
    
    \subsubsection{Information Theft Based on Device Leakage Signals}
    
\begin{figure}[!t]
    \centering
    \includegraphics[width=\linewidth]{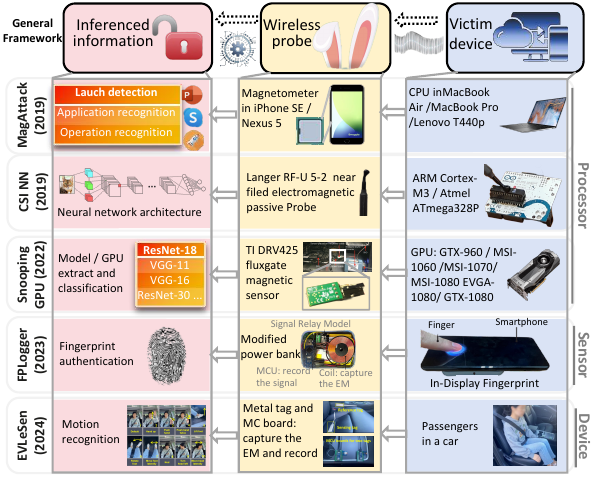}
    \caption{Examples of passive attacks based on electromagnetic leakage. Each attack system contains three key components: the attacked device (right), wireless detection device (middle), and inferred information (left).}
    \label{fig:passiveEMI}
    \end{figure}

    Modern electronic devices inevitably generate various physical signals during operation, often carrying information about the device's working state and processed data. Early research found that even ordinary hard disk drives produce exploitable acoustic and electromagnetic signals through their mechanical movements\cite{biedermannHardDriveSideChannel2015}.
    Fig.~\ref{fig:passiveEMI} illustrates how electromagnetic leakage attacks, from CPU activity detection to deep learning model theft and biometric recognition.
    
    \paragraph{Physical Principle} Current variations in device circuits generate electromagnetic radiation with intensity and spectral characteristics that correlate with operational states. For example, digital circuits produce distinct radiation patterns through clock signals and data transmission, enabling remote detection of CPU activities\cite{chenDOSEDetectingUserdriven2015}.
    
    \paragraph{Work Review}
    
    \subtitleone{(1) Processor Information Eavesdropping.} Fig.~\ref{fig:passiveEMI} demonstrates how electromagnetic leakage from processors enables computational activity eavesdropping. These attacks target two processor types: \subtitletwo{(1-1) CPU:} Studies using mobile device magnetometers have shown successful eavesdropping on computer activities\cite{chengMagAttackGuessingApplication2019,ningDeepMagSniffingMobile2018}. Batina et al.\cite{batinaCSINNReverse2019} demonstrated reverse-engineering of Advanced RISC Machine (ARM) Cortex-M3 neural network architectures through electromagnetic side channels, inferring network layers, neuron counts, and activation functions by analyzing computational radiation characteristics\cite{picekCurseClassImbalance2019}. Ji et al.\cite{jiNoSeeingAlso2023} developed an electromagnetic radiation-based covert channel achieving 98.6\% accuracy in CPU application behavior inference.
    \subtitletwo{(1-2) GPU:} Recent research has expanded to GPU information eavesdropping. Zhan et al.\cite{zhanGraphicsPeepingUnit2022} conducted the first systematic study of GPU electromagnetic leakage, identifying two vulnerabilities: DVFS-induced radiation leakage and fixed-frequency periodic signal leakage. Their analysis revealed strong correlations between leaked signals and GPU workload, enabling webpage fingerprinting with 95\% accuracy and keystroke timing inference from 2 meters. This research established the feasibility of remote physical side-channel attacks on non-shared GPUs and analyzed their limiting factors. Maia et al.\cite{maiaCanOneHear2022} explored neural network architecture inference through GPU electromagnetic leakage, developing a complete reconstruction system based on magnetic flux changes in GPU power lines. The MagTracer system by Xiao et al.\cite{xiaoMagTracerDetectingGPU2023} achieved over 98\% accuracy in detecting GPU cryptocurrency mining across 14 GPU models, with a false positive rate below 0.7\%.
    \subtitleone{(2) Sensor Information Eavesdropping.} \subtitletwo{(2-1) Fingerprint Sensors:} Ni et al.\cite{niRecoveringFingerprintsDisplay2023} demonstrated a fingerprint recovery attack exploiting electromagnetic side channels from under-display fingerprint sensors. Their research established the theoretical foundation of electromagnetic signal generation during fingerprint acquisition and developed a comprehensive signal processing and image reconstruction methodology. Testing on five mainstream smartphones demonstrated effective fingerprint image reconstruction with a 54\% successful unlocking rate within three attempts. \subtitletwo{(2-2) Microphones:} Ramesh et al. revealed dual aspects of microphone electromagnetic leakage. Their TickTock system\cite{rameshTickTockDetectingMicrophone2022} detected microphone operational states by analyzing electromagnetic radiation from clock signals, achieving over 90\% accuracy across 30 laptop models. Their subsequent research\cite{rameshYourMicLeaks2024} revealed potential recording content reconstruction through clock signal modulation analysis, introducing new device security challenges. Additionally, Xiao et al.\cite{xiaoKeystrokeRecognitionTapping2023} demonstrated accurate keystroke identification through smartphone microphone analysis of screen tapping sounds.
    \subtitleone{(3) Electromagnetic Leakage of Mobile Devices.} Mobile devices and electric vehicles present additional attack surfaces. Wang et al.\cite{wangMagnifiSenseInferringDevice2015} demonstrated smartphone electromagnetic radiation revealing device usage patterns. Baker et al.\cite{bakerLosingCarKeys2019} exposed Combined Charging System (CCS) vulnerabilities, showing 91.8\% communication data interception from adjacent parking spaces, including charging control and billing information. Cui et al.\cite{cuiEVLeSenVehicleSensing2024} identified electric vehicle electromagnetic leakage for inferring in-vehicle activities, while Liu et al.\cite{liuPrivacyLeakageWireless2024} revealed sensitive information leakage through electromagnetic induction during wireless charging.
    
    \paragraph{Key Challenges and Technical Evolution Analysis}
    
    These attacks face three main challenges:
    \subtitleone{(1) Signal Extraction Reliability:} Device leakage signals are weak and vulnerable to environmental interference. Detection techniques have evolved from basic magnetometer systems\cite{vaucelleCosteffectiveWearableSensor2009} to advanced high-sensitivity detection equipment\cite{longEMEyeCharacterizing2024}. Modern signal processing algorithms, particularly deep learning approaches, have enhanced signal reconstruction quality.
    \subtitleone{(2) Feature Recognition Accuracy:} Converting complex leakage signals into meaningful information requires advanced feature engineering. Some approaches utilized manual feature design, with time-frequency domain analysis for activity recognition serving as a representative example\cite{jiNoSeeingAlso2023}. Recent research has transitioned to end-to-end deep learning methods, achieving higher recognition accuracy.
    \subtitleone{(3) Attack Practicality:} Environmental interference and device heterogeneity present significant implementation challenges. Recent research has developed more robust solutions, such as cross-device detection~\cite{rameshYourMicLeaks2024} and coverting friction sound-based fingerprint theft technique~\cite{zhouPrintListenerUncoveringVulnerability2024}.
    
    \subsubsection{Information Inference Based on Protocol-Defined Signals}
    
\begin{figure}[!t]
    \centering
    \includegraphics[width=\linewidth]{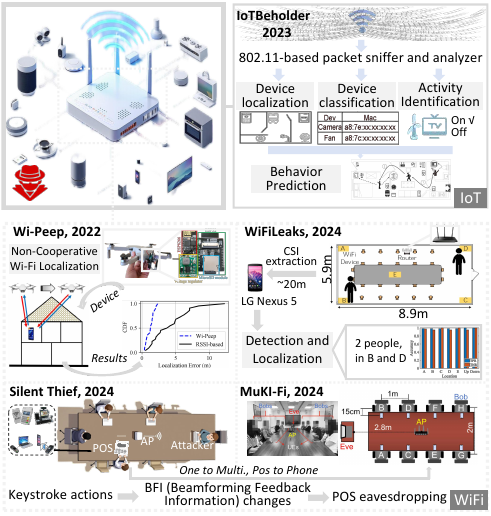}
    \caption{Examples of WiFi side-channel attacks based on protocol-defined signals.}
    \label{fig:trafficAnalysis}
    \end{figure}

    The widespread adoption of IoT and wireless communications has created new privacy vulnerabilities through device-generated network traffic and wireless signals. As illustrated in Fig.~\ref{fig:trafficAnalysis}, attackers can analyze these traffic patterns and signal characteristics to execute multi-level privacy breaches, ranging from device localization\cite{zhangWiCoRobustIndoor2023,wangLearningDomainInvariantModel2024} to user behavior inference. These passive attacks are particularly effective due to their stealthy nature, requiring only passive monitoring of open channels without direct device interaction.
    
    \paragraph{IoT Device Behavior Analysis}
    
    \subtitleone{Physical Principles:}
    Smart devices produce distinctive network traffic patterns during different operational states and tasks. These patterns manifest in three key aspects: (i) temporal characteristics, including packet size, interval, and burstiness; (ii) protocol features, encompassing protocol types and communication patterns; (iii) content features, such as encrypted traffic statistics. Analysis of these features enables inference of device states and user behavior.
    \subtitleone{Work Review:} IoTBeholder system\cite{zouIoTBeholderPrivacySnooping2023} (shown in Fig.~\ref{fig:trafficAnalysis}) demonstrates comprehensive WiFi traffic analysis for smart home privacy inference. The system introduces three key innovations: (1) an end-to-end attack framework operational on commercial devices without specialized hardware or privileges; (2) integrated modules for device localization, classification, and activity recognition that automatically process traffic data; (3) user automation rule inference through device temporal correlation analysis. Testing on 23 IoT devices, \cite{zouIoTBeholderPrivacySnooping2023} achieved accurate device identification, activity recognition, and user behavior prediction, highlighting significant privacy vulnerabilities in smart homes. Classen et al.\cite{classenAttacksWirelessCoexistence2022} identified novel security risks in wireless coexistence interfaces. Their research demonstrated how these interfaces, designed to coordinate Bluetooth, WiFi, and LTE technologies, enable cross-chip privilege escalation attacks. This vulnerability allows attackers to extract WiFi passwords or manipulate WiFi traffic through Bluetooth chips. The hardwired nature of coexistence interfaces in billions of deployed devices complicates security boundary enforcement between chips.
    
    \paragraph{WiFi Signal Analysis}
    
    \subtitleone{Physical Principles:}
    Environmental factors and human presence affect WiFi signal propagation, manifesting in variations of Received Signal Strength Indicator (RSSI), Channel State Information (CSI), and Beamforming Feedback Information (BFI). By analyzing changes in these characteristics, attackers can infer personnel activities and location information in the environment.
    \subtitleone{Work Review:} WiFi signal analysis attacks have evolved from coarse-grained location tracking to fine-grained behavior recognition. Zhu et al.\cite{zhuTuAlexaWhen2018} leveraged WiFi signal multipath effects to achieve covert motion tracking. Wi-Peep~\cite{abediNoncooperativeWifiLocalization2022} demonstrates a novel location leakage attack by exploiting 802.11 protocol vulnerabilities to trigger responses from unauthorized network devices. Using an innovative time-of-flight measurement scheme, the drone-mounted system accurately locates WiFi devices across multiple building floors. The WiFiLeaks system proposed by Gu et al.\cite{guWiFiLeaksExposingStationary2024} demonstrates how to detect personnel presence behind walls by analyzing CSI information captured by commercial mobile devices. Through innovative subcarrier correlation analysis methods, this system achieves 83.33\% detection accuracy in through-wall scenarios at 20 meters. The RFTrack system\cite{liRFTrackStealthyLocation2024} performs indoor position tracking through WiFi devices without requiring extensive device deployment or physical access, utilizing unlabeled RSSI time series for location inference. The Silent Thief system proposed by Chen et al.\cite{chenSilentThiefPassword2024} further expands the application range of WiFi signal analysis, demonstrating the possibility of inferring password inputs by analyzing the beamforming feedback information of POS terminals. Experiments show that this system can infer 6-digit POS passwords with 81\% accuracy within the first 100 attempts. MuKI-Fi~\cite{wangMuKIFiMultiPersonKeystroke2024} is the first to achieve keystroke inference in multi-person scenarios, naturally separating different users' keystroke behaviors using the near-field characteristics of BFI.
    
    \paragraph{Key Technical Challenges and Development Trends}
    
    Network traffic and WiFi signal-based information inference faces three primary challenges:
    \subtitleone{(1) Signal acquisition and feature extraction reliability.} Network traffic and wireless signals are susceptible to environmental noise, multipath effects, and user activity interference. Solutions have emerged at multiple levels: hardware-level optimizations in antenna design and signal processing circuits enhance acquisition quality; algorithm-level developments include adaptive processing methods, such as IoTBeholder's\cite{zouIoTBeholderPrivacySnooping2023} dynamic feature extraction algorithm for environmental adaptation.
    \subtitleone{(2) Time-varying data analysis accuracy and generalization.} Network traffic and wireless signal characteristics exhibit high temporal variability and user dependence, challenging traditional pattern recognition approaches. Advanced machine learning solutions address these issues. The Silent Thief system\cite{chenSilentThiefPassword2024} employs connected temporal classification to improve accuracy and cross-scenario generalization. Researchers are exploring unsupervised and transfer learning to reduce labeled data dependencies. RFTrack\cite{liRFTrackStealthyLocation2024} implements reinforcement learning for trajectory modeling to minimize environmental impacts.
    \subtitleone{(3) System real-time performance and resource efficiency.} Information inference systems must achieve real-time analysis within computational constraints, necessitating balance between algorithm complexity and inference accuracy. Researchers address this through model optimization and lightweight algorithms (e.g., Wi-Peep's\cite{abediNoncooperativeWifiLocalization2022} real-time localization capabilities on resource-constrained hardware).
    
    Development trends would focus on three directions: enhanced feature acquisition accuracy, refined feature mining techniques, and intelligent system integration. These advances target improvements in complex environment handling, multi-source data fusion, and adaptive learning capabilities. Additionally, balancing efficient information acquisition with privacy protection emerges as a critical consideration.

    \subsection{Defense Strategies Against Wireless Signal Attacks}
    
    This subsection systematically reviews defense strategies against wireless signal-based attacks.
    
    \subsubsection{Defense Framework Overview}
    Defense methods operate at two distinct levels.
    \subtitleone{(1) Physical layer defense} focuses on fundamental signal propagation characteristics, enhancing security through techniques such as signal path control, power allocation, and beamforming This bottom-level protection prevents information leakage at the source. \subtitleone{(2) Application layer defense} focuses on scenario-specific countermeasures and privacy protection mechanisms.
    
    \subsubsection{PHY Defense Strategies}
    
    Physical layer defense leverages modern wireless communication technology (e.g., RIS and cooperative transmission schemes) to establish fundamental security mechanisms.
    
    \paragraph{Beamforming and Intelligent Reflecting Surface Defense}
    
    \subtitleone{Physical Principles:}
    Beamforming utilizes multi-antenna arrays to create directional signal propagation by controlling phase and amplitude, enhancing desired signal paths while suppressing leakage in other directions. RIS employ programmable electromagnetic units to modulate incident signals, enabling active wireless channel control. The integration of these technologies significantly enhances physical layer security.
    \subtitleone{Work Review:}
    Staat et al.\cite{staatIRShieldCountermeasureAdversarial2022} introduced IRShield, a pioneering RIS-based defense system. This plug-and-play privacy protection scheme counters wireless sensing attacks through an innovative RIS configuration algorithm that dynamically adjusts reflection unit phases to confuse channel characteristics observed by eavesdroppers. IRShield reduced WiFi-based human motion detection attack success rates to below 5\%, establishing RIS technology's practical value in physical layer privacy protection.
    Asaad et al.\cite{asaadSecureActivePassive2022} investigated secure communication in RIS-assisted MIMO systems. Their research developed a joint optimization approach for downlink beamformers and RIS phase shifts to maximize weighted secrecy sum-rate. Results demonstrate that RIS integration in MIMO systems enhances both secrecy performance and system robustness against passive eavesdropping.
    
    \paragraph{Cooperative Transmission and Interference Injection Defense}
    
    \subtitleone{Physical Principles.}
    Cooperative transmission enables distributed beamforming through coordinated multi-node transmission, creating virtual antenna arrays. Interference injection enhances security by introducing artificial noise into communication signals, degrading channel quality for eavesdroppers while maintaining legitimate communication performance. These complementary techniques work together to minimize information leakage risk.
    \subtitleone{Work Review.}
    Li et al.\cite{liTwoWayAerialSecure2024} developed a distributed cooperative beamforming scheme for unmanned aerial vehicle communications. Their experimental results demonstrate effective eavesdropping resistance while maintaining communication quality.
    Jia et al.\cite{jiaSecureMultiantennaTransmission2022} addressed secure multi-antenna transmission against unknown eavesdroppers. They introduced two approaches: Optimal Adaptive Power Allocation (OAPA) and Suboptimal Fixed Power Allocation (SFPA), optimized for worst-case scenarios with unknown eavesdropper noise power. Their theoretical analysis and numerical results demonstrate security performance comparable to exhaustive search methods without requiring eavesdropper prior information.
    
    \paragraph{Key Technical Challenges and Development Trends}
    
    Physical layer defense technologies face two main technical challenges:
    \subtitleone{(1) Channel state uncertainty.} Practical wireless environments present incomplete or inaccurate channel state information, significantly complicating defense strategy design. Researchers have addressed this through robust optimization approaches. For instance, Jia et al.\cite{jiaSecureMultiantennaTransmission2022} developed the OAPA scheme optimized for worst-case scenarios, maintaining effectiveness despite unknown eavesdropper noise power.
    \subtitleone{(2) Real-time performance and computational complexity.} While physical layer defense requires rapid environmental adaptation, complex optimization algorithms often fail to meet real-time constraints. Recent research addresses this through algorithmic innovations. Asaad et al.\cite{asaadSecureActivePassive2022} developed low-complexity algorithms enabling rapid joint optimization of beamforming and RIS using fractional programming and alternating optimization. Li et al.\cite{liTwoWayAerialSecure2024} introduced an enhanced multi-objective swarm intelligence algorithm optimized for resource-constrained UAV platforms.
    
    The field is advancing through integration of conventional mechanisms (power control, beamforming) with emerging technologies like RIS \cite{asaadSecureActivePassive2022}, alongside machine learning algorithms. This evolution addresses the fundamental challenge: \textit{achieving reliable security with limited resources in dynamic environments.} Future development trends emphasize deeper integration of RIS and machine learning technologies.
    
    \subsubsection{Application Layer Defense Strategies}
    
    Application layer defense strategies design specialized protection mechanisms for specific application scenarios, effectively addressing security threats in particular contexts. These strategies primarily encompass perturbation and adversarial sample defense, as well as privacy protection mechanisms. It is noteworthy that these approaches have not been fully explored in wireless sensing. Therefore, here we introduce the application of these protection strategies in areas such as camera privacy protection and voice recognition protection, aiming to inspire subsequent research on wireless-based defense strategies.
    
    \paragraph{Physical Principle}
    These defense strategies are based on principles of signal processing, machine learning, information theory, and security protocols. Through carefully designed perturbations added to original signals or specialized algorithms and frameworks, they prevent attack models from correctly identifying or classifying target information while minimizing sensitive information leakage and ensuring functional implementation.

    \paragraph{Work Review}
    Recent research demonstrates the effectiveness of these strategies across various domains. Screen Perturbation~\cite{yeScreenPerturbationAdversarial2023} system addresses privacy protection for under-display cameras by adding imperceptible pixel perturbations to screen content, significantly reducing image classification and face recognition accuracy. In the audio domain, DARE-GP system~\cite{testaPrivacyRealTimeSpeech2023} uses genetic programming to generate universal adversarial perturbations that mask emotional features while preserving speech content, achieving robustness in real acoustic environments. Addressing privacy in augmented reality, BystandAR system~\cite{corbettBystandARProtectingBystander2023} utilizes AR-specific capabilities to protect bystander privacy with high accuracy and processing speed.
    
     \paragraph{Key Technical Challenges and Development Trends}
    
    The application of these strategies to wireless sensing faces challenges due to the high-dimensional complexity of wireless signals and the environmental dependence of signal propagation. However, the potential for adaptation is evident. The perturbation generation principles used in visual systems by Ye et al.\cite{yeScreenPerturbationAdversarial2023} could inspire similar approaches for wireless signals. The genetic programming method proposed by~\cite{testaPrivacyRealTimeSpeech2023} provides insights for handling time-varying characteristics of wireless signals.. As wireless signal processing technology and computational capabilities advance, transferring these mature application layer defense strategies to wireless signal attacks becomes increasingly feasible and valuable.

    \begin{table*}[!ht]
\centering
\caption{Summary of Wireless Signal-based Security Applications}
\label{tab:guardianSummary}
\resizebox{\textwidth}{!}{
\begin{tabular}{|c|p{3cm}|p{5cm}|p{4cm}|p{3.5cm}|p{4.5cm}|}
\hline
\textbf{Category} & \textbf{Subcategory} & \textbf{Representative Works} & \textbf{Core Principles} & \textbf{Main Challenges} & \textbf{Development Trends} \\
\hline
\multirow{4}{*}[-10em]{\rotatebox[origin=c]{90}{\textbf{\makebox[3em]{Biometric Authentication}}}} 
& Gait-based Authentication & 
\begin{itemize}[leftmargin=*]
\item 2D/3D-AoA based human characterization\cite{renPersonReidentification3D2023,heWiFiVisionSensing2020}
\item CSI and vision fusion recognition\cite{chenRFCamUncertaintyawareFusion2022,xuRadioBiometricsHuman2017}
\item Environment-independent recognition\cite{zhangWiPIGRPathIndependent2022,liangDCSGaitClassLevelDomain2024}
\item RFID-based gait recognition\cite{chenSensingHumanGait2024,huangAuIdAutomaticUser2019}
\end{itemize} & 
Identification using unique interaction patterns between human gait and wireless signals & 
Feature stability vs. environmental changes:
\begin{itemize}[leftmargin=*]
\item Environmental impact on features
\item Path dependency effects
\item Multi-person interference
\end{itemize} & 
\begin{itemize}[leftmargin=*]
\item Single to multi-dimensional features
\item Limited to open scenarios
\item Enhanced adaptability
\end{itemize} \\
\cline{2-6}
& Voice-based Authentication & 
\begin{itemize}[leftmargin=*]
\item Multimodal fusion recognition\cite{liuWavoIDRobustSecure2023,liuWavoiceNoiseresistantMultimodal2021}
\item Chest vibration authentication\cite{chenChestLiveFortifyingVoicebased2022}
\item Vocal cord vibration recognition\cite{liVocalPrintMmWaveBasedUnmediated2023,haoMmSafeVoiceSecurity2022}
\item Jaw motion authentication\cite{srivastavaJawthenticateMicrophonefreeSpeechbased2024}
\end{itemize} & 
Analysis of sound wave conduction and vocal organ movement characteristics & 
Security vs. usability:
\begin{itemize}[leftmargin=*]
\item Enhanced security needs
\item Feature collection complexity
\item Anti-spoofing challenges
\end{itemize} & 
\begin{itemize}[leftmargin=*]
\item Acoustic to physiological features
\item Enhanced anti-spoofing
\item User experience optimization
\end{itemize} \\
\cline{2-6}
& Face-based Authentication & 
\begin{itemize}[leftmargin=*]
\item Millimeter-wave penetration\cite{xuMaskDoesNot2022,zhaoAuthenticationMillimeterWaveBodyCentric2017}
\item Acoustic facial recognition\cite{zhangFaceRecognitionHarsh2024,shiFaceMicInferringLive2021}
\item Multimodal secure recognition\cite{kimScoresTellEverything2024,wuBioFace3DContinuous3d2021}
\end{itemize} & 
Wireless signal interaction with facial structures for contactless identification & 
Spatial resolution vs. detection range:
\begin{itemize}[leftmargin=*]
\item Wavelength limitations
\item Real-time constraints
\item Security performance balance
\end{itemize} & 
\begin{itemize}[leftmargin=*]
\item Enhanced resolution
\item Improved anti-spoofing
\item Expanded scenarios
\end{itemize} \\
\cline{2-6}
& Physiological Signal Authentication & 
\begin{itemize}[leftmargin=*]
\item Heartbeat-based recognition\cite{huangNFHeartNearfieldNoncontact2023,wangHeartPrintExploringHeartbeatBased2022}
\item Breathing authentication\cite{hanBreathSignTransparentContinuous2023,liuContinuousUserVerification2020}
\item Touch-based acoustic auth\cite{yangBioCasePrivacyProtection2023,wuEchoHandHighAccuracy2022}
\item Multi-person physiological auth\cite{wangSimultaneousAuthenticationMultiple2024}
\end{itemize} & 
Continuous authentication using physiological features & 
Feature uniqueness vs. stability:
\begin{itemize}[leftmargin=*]
\item Signal variability
\item Feature extraction reliability
\item Multi-user complexity
\end{itemize} & 
\begin{itemize}[leftmargin=*]
\item Multimodal fusion
\item Adaptive learning
\item Multi-user support
\end{itemize} \\
\hline
\multirow{2}{*}[-5em]{\rotatebox[origin=c]{90}{\textbf{\makebox[3em]{Device Authenticity}}}} & Device Fingerprinting & 
\begin{itemize}[leftmargin=*]
\item CPU radiation identification\cite{fengFingerprintingIoTDevices2023,chengDeMiCPUDeviceFingerprinting2019}
\item Physical layer authentication\cite{jooHoldDoorFingerprinting2020,jianRadioFrequencyFingerprinting2022}
\item Hardware feature recognition\cite{berdichSweepUnlockFingerprintingSmartphones2023}
\item Environmental radiation auth\cite{leeAEROKEYUsingAmbient2022,wangBCAuthPhysicalLayer2022}
\end{itemize} & 
Authentication using hardware and signal characteristics & 
Environmental dynamics vs. uniqueness:
\begin{itemize}[leftmargin=*]
\item Environmental impacts
\item Device scalability
\item Resource constraints
\end{itemize} & 
\begin{itemize}[leftmargin=*]
\item Environmental adaptation
\item Lightweight optimization
\item Multi-feature fusion
\end{itemize} \\
\cline{2-6}
& Media Authenticity & 
\begin{itemize}[leftmargin=*]
\item WiFi-vision verification\cite{fangNowhereHideDetecting2023,zhangFaceRecognitionHarsh2024}
\item Audio-pupil authentication\cite{zhuSoundLockNovelUser2023}
\item Multimodal anti-tampering\cite{sharmaLumosIdentifyingLocalizing2022}
\end{itemize} & 
Content verification through physical-digital consistency & 
Physical-digital consistency:
\begin{itemize}[leftmargin=*]
\item Signal capture
\item Feature alignment
\item Real-time verification
\end{itemize} & 
\begin{itemize}[leftmargin=*]
\item Improved mapping accuracy
\item Real-time processing
\item Anti-tampering enhancement
\end{itemize} \\
\hline
\multirow{2}{*}[-4em]{\rotatebox[origin=c]{90}{\textbf{\makebox[3em]{Privacy Protection}}}} 
 & Privacy Threat Detection & 
\begin{itemize}[leftmargin=*]
\item Radar-based detection\cite{qiuRadar2PassiveSpy2023}
\item WiFi drone detection\cite{dengDrDefenderProactive2024}
\item Multimodal threat recognition\cite{sharmaLumosIdentifyingLocalizing2022}
\end{itemize} & 
Active and passive wireless detection of privacy threats & 
Detection range vs. accuracy:
\begin{itemize}[leftmargin=*]
\item Coverage-accuracy tradeoff
\item Environmental interference
\item Multi-target complexity
\end{itemize} & 
\begin{itemize}[leftmargin=*]
\item Active-passive coordination
\item Algorithm optimization
\item Feature fusion
\end{itemize} \\
\cline{2-6}
& Privacy Data Protection & 
\begin{itemize}[leftmargin=*]
\item Multi-channel protection\cite{liuExploitingFineGrainedChannel2023,katsumataRevisitingFuzzySignatures2021}
\item Deep learning features\cite{kongPushLimitWiFibased2022,weiDualAdversarialRepresentationDisentanglement2024}
\item Physical layer security\cite{zhaoAuthenticationMillimeterWaveBodyCentric2017,maengPowerAllocationFingerprintBased2021}
\end{itemize} & 
Privacy protection through physical layer features & 
Security vs. practicality:
\begin{itemize}[leftmargin=*]
\item Resource consumption
\item System complexity
\item Implementation costs
\end{itemize} & 
\begin{itemize}[leftmargin=*]
\item Intelligent sensing
\item Lightweight security
\item Dynamic protection
\end{itemize} \\
\hline
\end{tabular}
}
\end{table*}
    
    \subsection{Summary and Insights}
    
    \subsubsection{Technical Characteristics of Attack Methods}
    
    Wireless signal-based attack methods exhibit three distinctive characteristics.
    \subtitleone{(1) Cross-domain Property.}
    These attack methods leverages diverse signal propagation mechanisms, demonstrated by acoustic extraction from optical signals (Lamphone\cite{nassiLamphonePassiveSound2022}) and electromagnetic leakage exploitation (GPU eavesdropping\cite{zhanGraphicsPeepingUnit2022}). 
    \subtitleone{(2) Stealthiness.} These attacks achieve stealth through passive listening\cite{rameshYourMicLeaks2024}, minimal power injection\cite{kohlerBrokenwireWirelessDisruption2023}, and behavior simulation\cite{chenSilentThiefPassword2024}. 
    \subtitleone{(3) Universality.} These attacks demonstrate universal applicability across devices and environments, evidenced by PPG-Hear's\cite{suPPGHearPracticalEavesdropping2024} cross-device compatibility and EchoLight's\cite{zhangEchoLightSoundEavesdropping2024} environmental adaptability.
        
    \subsubsection{Future Research Trends}
    
    The field evolves toward intelligent and systematic attacks, combining deep learning for adaptive parameter adjustment (demonstrated by mmEar\cite{xuMmEarPushLimit2024}) and coordinated exploitation of multiple physical characteristics (as in IoTBeholder\cite{zouIoTBeholderPrivacySnooping2023}). These advancements necessitate sophisticated defense strategies, including adaptive mechanisms like DARE-GP's\cite{testaPrivacyRealTimeSpeech2023} approach and comprehensive multi-dimensional protection frameworks, driving wireless sensing security toward systematic safeguards for IoT environments.

    \section{Wireless Signals as Guardians for Security Applications} \label{sec:roleGuardian}
    
    The rapid advancement of IoT technology introduces both opportunities and challenges for security applications. Traditional security methods, such as optical authentication and password-based verification, face inherent limitations, including line-of-sight requirements, reliance on user cooperation, and susceptibility to environmental factors\cite{renPersonReidentification3D2023}. In contrast, wireless signal-based security applications address these challenges by leveraging the unique physical properties of wireless signals.
    
    \noindent \textbf{Advantages of Wireless Signals.} Wireless signals offer several distinct advantages for security applications:
    \subtitleone{(1) Penetrability:} Wireless signals can operate in non-line-of-sight (NLOS) conditions, overcoming the limitations of traditional optical and acoustic methods. For example, mmWave radar penetrates clothing and obstacles via diffraction, enabling identity authentication based on body surface features\cite{wangRDGaitMmWaveBased2024} and even face recognition through face masks.
    \subtitleone{(2) Non-contact Operation:} Wireless signal-based systems enable security verification without physical interaction, facilitating continuous monitoring and authentication\cite{huangNFHeartNearfieldNoncontact2023}. For instance, WiFi-based systems can continuously verify user identity while allowing them to perform normal activities without interruption~\cite{yangHLocExploitingHeight2024,yangMultipleWiFiAccess2024}.
    \subtitleone{(3) Universality and Privacy:} Unlike optical sensors, wireless solutions do not collect visual data, thereby preserving user privacy while offering broader coverage. Additionally, the ubiquity of WiFi signals enhances the feasibility of large-scale implementation\cite{fangNowhereHideDetecting2023}.
    
    These characteristics make wireless signal-based security applications particularly well-suited for scenarios requiring continuous identity verification, authentication in complex environments, and privacy-preserving implementations.
    
    \noindent \textbf{Security Applications.}  
    The unique properties of wireless signals enable security applications across three complementary areas: biometric authentication, device authenticity verification, and privacy protection, as shown in Tab.~\ref{tab:guardianSummary}. These areas address distinct yet interconnected aspects of modern security needs.
    \subtitleone{(1) Biometric Authentication:} Wireless signals interact with the human body, reflecting unique physiological features like body shape and bioelectrical properties. These interactions allow contactless identity verification, even in challenging conditions such as through face masks or clothing.
    \subtitleone{(2) Device Authenticity Verification:} Manufacturing variations create unique RF fingerprints for wireless devices\cite{fengFingerprintingIoTDevices2023}, enabling counterfeit detection\cite{wangBCAuthPhysicalLayer2022} and lightweight authentication for resource-constrained IoT systems\cite{leeAEROKEYUsingAmbient2022}.
    \subtitleone{(3) Privacy Protection:} Wireless applications enhance privacy by detecting hidden surveillance devices\cite{qiuRadar2PassiveSpy2023}, identifying abnormal behaviors\cite{dengDrDefenderProactive2024}, and enabling identity anonymization\cite{weiDualAdversarialRepresentationDisentanglement2024}, ensuring secure systems without compromising user trust.
    
    This section examines these domains, highlighting principles, technical features, and representative works, followed by challenges and trends.

    \begin{table*}[htbp]
\caption{Comparison of Different Biometric Authentication Technologies}
\label{tab:authComparison} 
\centering
\begin{tabular*}{\textwidth}{@{}p{\dimexpr\textwidth/8-2\tabcolsep\relax}p{\dimexpr\textwidth/9-2\tabcolsep\relax}p{\dimexpr\textwidth/8-2\tabcolsep\relax}p{\dimexpr\textwidth/11-2\tabcolsep\relax}p{\dimexpr\textwidth/9-2\tabcolsep\relax}p{\dimexpr\textwidth/7-2\tabcolsep\relax}p{\dimexpr\textwidth/9-2\tabcolsep\relax}p{\dimexpr\textwidth/5-2\tabcolsep\relax}@{}}
\toprule
\rowcolor{gray!10} \textbf{Technology} & \textbf{Universality} & \textbf{Environmental Impact} & \textbf{Contactless} & \textbf{Spontaneous} & \textbf{Device Required} & \textbf{Anti-Replay} & \textbf{RF Signal Benefits}\\
\midrule
Fingerprint & Wide & Temperature & $\times$ & $\times$ & Dedicated & $\times$ & Unexplored\\
\rowcolor{red!0} Face\textsuperscript{\#} & Wide & Lighting & $\times$ & \checkmark & Camera & $\times$ & Anti-mask, Anti-puppet\\
Iris & Limited\textsuperscript{*} & Lighting & \checkmark & $\times$ & Dedicated & $\times$ & Unexplored\\
\rowcolor{red!0} Gait\textsuperscript{\#} & Limited\textsuperscript{†} & None & \checkmark & $\times$ & Dedicated & \checkmark & Low-light, Anti-occlusion\\
\rowcolor{red!0} Voice\textsuperscript{\#} & Limited\textsuperscript{‡} & Noise & \checkmark & $\times$ & Microphone & $\times$ & Anti-interference\\
\rowcolor{red!0} Vital sign\textsuperscript{\#} & Wide & None & \checkmark & \checkmark & WiFi & \checkmark & Penetration, Continuous\\
\bottomrule
\multicolumn{8}{l}{\textsuperscript{*}Not for blind\quad\textsuperscript{†}Not for mobility impaired\quad\textsuperscript{‡}Not for speech impaired}\\
\multicolumn{8}{l}{\textsuperscript{\#} Indicate authentication methods achievable through wireless sensing}\\
\end{tabular*}
\end{table*}
    \subsection{Human Identification and Authentication}

    Human Identification and Authentication (I\&A) plays an important role in security. While traditional optical-based biometric methods have proven effective, they face inherent limitations including environmental constraints and occlusion problems\cite{renPersonReidentification3D2023, xuMaskDoesNot2022}. Tab.~\ref{tab:authComparison} presents a detailed comparison of different biometric authentication technologies.
    

    Wireless sensing-based authentication demonstrates three key advantages over traditional methods:
    \subtitleone{(i) Environmental Adaptability.} Wireless signals maintain consistent performance across varying environmental conditions. Electromagnetic waves enable long-range identification through non-metallic materials\cite{chenRFCamUncertaintyawareFusion2022, wangContinuousUserAuthentication2019}, while sound waves excel in challenging environments\cite{hanAccuthSpoofingVoice2023, wuEchoHandHighAccuracy2022}. Multi-signal integration enhances system robustness, as demonstrated by authentication system in~\cite{yangOpenAuthHumanBodyBased2024} achieving 93.4\% accuracy through environment-independent feature extraction.
    \subtitleone{(ii) User-Friendliness.} Non-contact wireless authentication enables continuous identification without user intervention\cite{huangContinuousUserAuthentication2022}. Electromagnetic waves provide effective long-range verification\cite{zhaoAuthenticationMillimeterWaveBodyCentric2017}, while sound waves enable precise short-range biometric detection\cite{srivastavaJawthenticateMicrophonefreeSpeechbased2024, zouBiLockUserAuthentication2018}. Integration with existing WiFi infrastructure reduces deployment costs.
    \subtitleone{(iii) Security.} Wireless authentication leverages unique individual biometric modulation patterns that resist forgery\cite{katsumataRevisitingFuzzySignatures2021}. Electromagnetic waves capture comprehensive biometric characteristics\cite{avolaPersonReIdentificationWiFi2022, liuExploitingFineGrainedChannel2023}, while sound waves detect subtle physiological features like bone conduction properties\cite{hanBreathSignTransparentContinuous2023, wangLoweffortVRHeadset2023}. This physical-layer authentication provides inherent protection against replay attacks, enhancing IoT security\cite{dongSecureMmWaveRadarBasedSpeaker2021}. 

    These advantages have driven the successful application of wireless sensing in various I\&A methods, including gait-based, voiceprint-based, face-based, and vital signal-based approaches, as illustrated in Fig.~\ref{fig:humanIdentification}.

    \subsubsection{Gait-based I\&A}
    
\begin{figure*}[!t]
    \centering
    \includegraphics[width=0.9\linewidth]{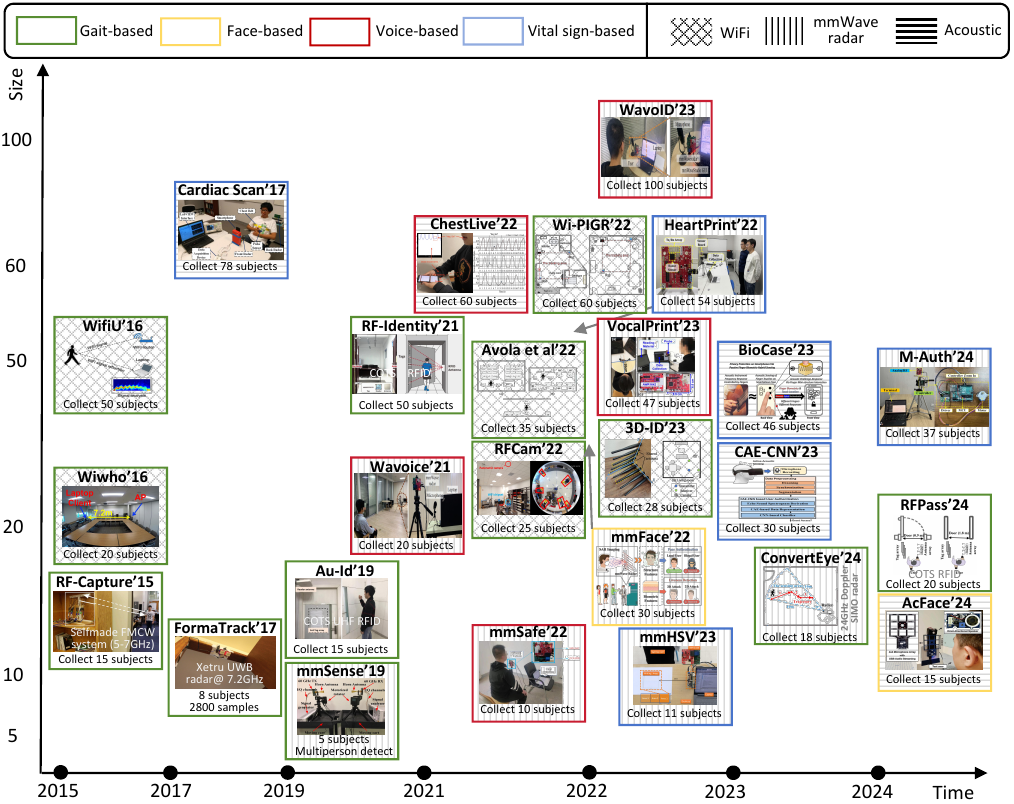}
    \caption{Evolution of wireless signal-based human I\&A technologies. Research in this field can be divided into four categories: gait-based (green box), face-based (yellow box), voiceprint-based (red box), and vital sign-based (blue box) identification methods.}
    \label{fig:humanIdentification}
    \end{figure*}

    Gait-based human identification, as illustrated in the green box in Fig.~\ref{fig:humanIdentification}, represents one of the earliest applications of wireless sensing recognition. 
    
    \paragraph{Physical Principle}
    
    Gait-based I\&A primarily utilizes three physical mechanisms. \subtitleone{(1) Electromagnetic wave interaction:} WiFi signals undergo reflection, scattering, and diffraction upon encountering human bodies, generating patterns linked to individual physiological characteristics\cite{gabrielDielectricPropertiesBiological1996, xuRadioBiometricsHuman2017}. These interactions manifest in Channel State Information (CSI) amplitude and phase variations\cite{avolaPersonReIdentificationWiFi2022}, Angle of Arrival (AoA) spatiotemporal distributions\cite{renPersonReidentification3D2023}, and Doppler effect frequency modulations\cite{wangRDGaitMmWaveBased2024}. \subtitleone{(2) Multi-modal signal fusion:} Different wireless signals provide complementary biometric information through distinct body interactions. \subtitleone{(3) Cross-modal mapping:} Establishing correlations between video and WiFi signals enables cross-modal identity recognition\cite{koranyXModalIDUsingWiFi2019}.
    
    \paragraph{Work Review}
    \subtitleone{(1) Gait Recognition.} Initial research established wireless signal-based gait recognition foundations. RF-Capture\cite{adibCapturingHumanFigure2015} demonstrated through-wall human figure capture and identification. Subsequent developments included WiWho\cite{zengWiWhoWiFiBasedPerson2016} and WiFi-ID\cite{zhangWiFiIDHumanIdentification2016} for WiFi-based recognition, with Wang et al.\cite{wangGaitRecognitionUsing2016} further advancing these techniques. FormaTrack\cite{kalyanaramanFormaTrackTracking2017} introduced radar-based body shape analysis, while mmSense\cite{guMmSenseMultiPersonDetection2019} achieved multi-person recognition using mmWave signals. Xu et al.\cite{xuAttentionBasedGaitRecognition2022} developed an attention mechanism-based system using RNN encoder-decoder architecture, achieving 89.77\%-97.32\% accuracy for groups of 4-10 people while simultaneously estimating walking directions. Wi-PIGR\cite{zhangWiPIGRPathIndependent2022} addressed path dependency through innovative receiver layouts and signal processing, reaching 77.15\% accuracy across 50 subjects. CovertEye\cite{sunCovertEyeGaitBasedHuman2024} enabled identification under weakly constrained trajectories, achieving 82.4\% accuracy. In RFID-based approaches, Au-Id\cite{huangAuIdAutomaticUser2019} demonstrated contactless identification during daily activities. RF-Identity\cite{fengRFIdentityNonIntrusivePerson2021} combined gait patterns and body features, achieving 94.2\% and 95.9\% accuracy for dynamic and static users respectively among 50 subjects. RFPass\cite{chenSensingHumanGait2024} proposed a method for gait authentication that is independent of environmental factors, utilizing commercial RFID devices. It leverages Doppler effect analysis and selects signals based on the direction of multi-path arrivals.
    \subtitleone{(2) Person Re-identification.} Person re-identification technology has been proposed in recent years, showing an evolution trend from single-modal to multi-modal fusion, and from static features to a combination of dynamic features. Early important work came from Avola et al.\cite{avolaPersonReIdentificationWiFi2022}, who systematically explored WiFi signal-based person Re-ID methods for the first time. This work designed a feature extraction framework based on siamese networks, achieving stable individual identification through CSI feature analysis, laying an important foundation for subsequent research. The RFCam system\cite{chenRFCamUncertaintyawareFusion2022} innovatively combined WiFi signals and visual information, significantly improving identification performance through uncertainty-aware feature fusion methods, achieving over 97\% identification accuracy in real-world setting. Ren et al.\cite{renPosterWiFiVisionbased2022,renPersonReidentification3D2023} achieved Re-ID based on WiFi vision\cite{heWiFiVisionSensing2020}, proposing a 2D-AoA-based human body visualization method, developing 3D body representation technology (3D-ID), and combining static body shape features with dynamic gait features, improving identification accuracy from 85\% to 92\%.
    
    \paragraph{Key Technical Challenges and Development Trends}
    
    The fundamental challenge in wireless signal-based gait recognition lies in mapping signal features to human biometric characteristics. Wireless signals undergo complex propagation effects including multipath, diffraction, and scattering when interacting with the human body, complicating stable feature extraction. Researchers have addressed this through multi-level approaches. At the signal collection level, multi-antenna spatial diversity techniques enhance signal stability, e.g., 3D-ID~\cite{renPersonReidentification3D2023} achieving 92\% identification accuracy through multi-dimensional feature fusion. At the feature extraction level, deep learning methods enable adaptive, environment-independent feature representations. Avola et al.\cite{avolaPersonReIdentificationWiFi2022} employed siamese networks while Chen et al.\cite{chenRFCamUncertaintyawareFusion2022} introduced uncertainty modeling to enhance feature robustness. For environmental adaptation, Wei et al.\cite{weiDualAdversarialRepresentationDisentanglement2024} developed a dual adversarial representation disentanglement model, and Wan et al.\cite{wanSelfSupervisedModalityAwareMultiple2023} proposed a self-supervised learning framework to achieve scene-invariant features. These advances indicate future trend: optimizing signal-feature mapping through deep learning while improving environmental adaptability via multi-modal fusion, marking a shift from experience-driven to theory-driven approaches and from single-feature to multi-dimensional feature analysis.

    \subsubsection{Voiceprint-based I\&A}
    
    Voiceprint recognition identifies individuals through their unique vocal characteristics. Modern approaches extend beyond traditional microphone-based methods by analyzing sound wave propagation in human tissues, offering enhanced security against forgery~\cite{dongSecureMmWaveRadarBasedSpeaker2021}.
    
    \paragraph{Physical Principle}
    
    Sound waves during speech propagate both through air and body tissues, with the latter process creating unique biometric features based on individual anatomical structures. The recognition mechanism relies on three key physical processes: (1) bone conduction through skull and skeletal structures~\cite{hanBreathSignTransparentContinuous2023, wangLoweffortVRHeadset2023}; (2) tissue vibration patterns from vocal cords, oral cavity, and facial tissues~\cite{srivastavaJawthenticateMicrophonefreeSpeechbased2024, liVocalPrintMmWaveBasedUnmediated2023}; and (3) distinctive sound wave reflections within the ear canal~\cite{gaoEarEchoUsingEar2019}.
    
    \paragraph{Work Review}
    Recent voiceprint recognition technology has evolved from single-feature to multi-modal fusion approaches. Wavoice~\cite{liuWavoiceNoiseresistantMultimodal2021} combined mmWave and audio signals for noise-resistant speech recognition, achieving a character recognition error rate below 1\% within 7 meters through attention-based feature fusion. ChestLive~\cite{chenChestLiveFortifyingVoicebased2022} system incorporated chest movement biometrics via acoustic sensing, enhancing anti-replay attack capabilities while maintaining 98.31\% authentication accuracy in complex environments. The mmSafe system\cite{haoMmSafeVoiceSecurity2022} applied mmWave technology to smart home scenarios, achieving 93.4\% speaker verification accuracy using weighted Mel-frequency cepstral coefficients (MFCCs). The VocalPrint system\cite{liVocalPrintMmWaveBasedUnmediated2023} utilized millimeter waves to capture vocal cord vibrations, demonstrating 96\% authentication accuracy with 41 participants. Recent research has emphasized practical multi-modal implementations. The Accuth~\cite{hanAccuthSpoofingVoice2023} captured speech vibration features via accelerometers, achieving over 90\% identification accuracy with 15 participants. The Jawthenticate~\cite{srivastavaJawthenticateMicrophonefreeSpeechbased2024} combined jaw movement and facial vibration features, reaching 97.07\% balanced accuracy across 41 speakers of different native languages. The state-of-the-art WavoID system~\cite{liuWavoIDRobustSecure2023} achieved over 98\% identification accuracy on 100 users by fusing mmWave sensed vocal cord vibrations with microphone-recorded signals.
    
    \paragraph{Key Technical Challenges and Development Trends}
    
    The primary challenge in voiceprint-based identification is balancing security with usability, as enhanced feature collection for security often compromises system accessibility. Researchers have developed multi-level solutions to address this challenge. At the signal acquisition level, methods like WavoID~\cite{liuWavoIDRobustSecure2023} achieve 98\% identification accuracy through mmWave and acoustic signal fusion. At the feature representation level, focus has shifted from air-propagated sound waves to tissue vibration features, exemplified by VocalPrint~\cite{liVocalPrintMmWaveBasedUnmediated2023} capturing vocal cord vibrations for improved anti-spoofing capabilities. Systems have strengthened anti-replay protection by analyzing comprehensive physiological responses during speech, such as chest cavity vibration analysis~\cite{chenChestLiveFortifyingVoicebased2022} and  jaw movement integration~\cite{srivastavaJawthenticateMicrophonefreeSpeechbased2024}. These developments reflect a clear trend toward strengthening security through multi-modal fusion of physiological features, while prioritizing minimal user interaction complexity.
    
    \subsubsection{Face-based I\&A}
    
    Wireless signal-based face recognition complements traditional optical methods by maintaining performance under challenging conditions like low light, smoke, and face mask occlusion\cite{xuMaskDoesNot2022}. Recent research has focused on facial feature reconstruction using mmWave radar and acoustic signals, along with multi-modal fusion techniques\cite{kimScoresTellEverything2024}.
    
    \paragraph{Physical Principle}
    The technology operates through three fundamental mechanisms: \subtitleone{(1) Electromagnetic wave imaging}, where mmWave reflections from facial surfaces create distinctive patterns based on contours and tissue structures\cite{luoImagingInteriorsImplicit2024, xuMaskDoesNot2022}; \subtitleone{(2) Acoustic reflection}, where sound wave interactions with facial surfaces generate unique spatial spectra corresponding to three-dimensional facial structures\cite{zhangFaceRecognitionHarsh2024}; and \subtitleone{(3) Physiological feature detection}, where wireless signals penetrate~\cite{zhangThroughwallHumanPose2024} surface tissues to capture deeper characteristics such as vascular distribution and tissue density\cite{zhaoAuthenticationMillimeterWaveBodyCentric2017}.
    
    \paragraph{Work Review}
    Wireless signal-based face recognition has seen significant advances recently. An et al.\cite{anImUPhysicalImpersonating2023} identified vulnerabilities in facial recognition systems, demonstrating how variations in facial styles can be exploited in attacks. Their findings underscore the critical need for more robust anti-spoofing mechanisms. The mmFace system\cite{xuMaskDoesNot2022} addressed these problems through mmWave-based face authentication, using mobile commercial mmWave radar to scan faces along specific trajectories. The system introduced distance-independent facial structure features to mitigate unstable face-to-device distances and implemented virtual registration through cross-modal photo-to-mmWave signal conversion.
    \textit{In terms of acoustic sensing,} Zhang et al.\cite{zhangFaceRecognitionHarsh2024} developed a system based on acoustic facial spectrum, which analyzes the distribution of sound wave reflections within 3D cubic units. Their novel multipath resolution algorithm accurately distinguishes signal reflections between units, maintaining over 95\% identification accuracy even with face mask occlusion.
    \textit{In multi-modal fusion,} the RFCam system\cite{chenRFCamUncertaintyawareFusion2022} combined wireless RF signals with visual information, extracting multi-dimensional features through multi-antenna WiFi devices and implementing uncertainty-aware fusion. Kim et al.\cite{kimScoresTellEverything2024} demonstrated the feasibility of high-quality face reconstruction using limited, non-adaptive queries, revealing new security challenges in this domain. While \cite{wuBioFace3DContinuous3d2021} and \cite{shiFaceMicInferringLive2021} achieved anti-spoofing recognition using single-ear biosensors and AR/VR device motion sensors respectively, wireless signal applications in this domain require further exploration.
    
    \paragraph{Key Technical Challenges and Development Trends}
    The fundamental challenge in wireless signal-based face recognition lies in achieving reliable identification with limited signal resolution. Researchers have addressed this through multi-level innovations: mmFace\cite{xuMaskDoesNot2022} enhanced spatial resolution through optimized antenna array configurations, while Zhang et al.\cite{zhangFaceRecognitionHarsh2024} overcame traditional limitations using the acoustic facial spectrum concept. The discovery of similarity score-based attack vulnerabilities by~\cite{kimScoresTellEverything2024} has driven research toward exploiting wireless signals' penetration capabilities for deep physiological feature-based anti-spoofing mechanisms. This field is evolving toward multi-physical quantity collaborative sensing, where electromagnetic and acoustic property fusion compensates for single-signal limitations.

    \subsubsection{Vital Signal-based I\&A}
    
    Vital signal-based authentication represents a cutting-edge approach in the realm of wireless sensing security. This method leverages the unique physiological characteristics of individuals, such as heartbeat and respiration patterns, to establish identity. Unlike traditional authentication methods that rely on explicit inputs like passwords or fingerprints, vital signal-based authentication offers a non-contact, continuous solution by analyzing the subtle interactions between wireless signals and the human body\cite{wangContinuousUserAuthentication2019}.
    
    \paragraph{Physical Principle}
    The fundamental principle underlying vital signal-based authentication is the ability of mmWave signals to detect minute vibrations on the body surface~\cite{liangContinualLearningRemote2025}. These surface vibrations are direct manifestations of internal physiological processes, primarily heart activity and respiration~\cite{xieRobustWiFiRespiration2024}. The key insight is that while these vital signals originate from the same physiological sources, they manifest as distinct representations on the body surface, which can be captured by high-frequency wireless signals.
    
    A notable advantage of vital signal-based authentication is its inherent resistance to forced authentication attempts. Unlike methods such as fingerprint scans, which can be coerced, vital signals are involuntary and difficult to manipulate. Any attempt at forced authentication would likely result in physiological stress responses, altering the vital signal patterns and triggering anomaly detection in the authentication system.
    
    \paragraph{Work Review}
    Recent research in vital signal-based authentication has advanced in three primary directions: 
    \subtitleone{(1) Respiration and Cardiac Activity-Based Authentication:}
    Cardiac Scan\cite{linCardiacScanNoncontact2017} pioneered remote cardiac motion sensing using continuous-wave radar, achieving 98.61\% Balanced Accuracy (BAC) and 4.42\% Equal Error Rate (EER) in tests with 47 participants. NF-Heart\cite{huangNFHeartNearfieldNoncontact2023} advanced this concept by integrating accelerometers into chairs for cardiac motion detection, achieving 96.45\% balanced accuracy with 105 subjects. MagSign\cite{caoMagSignHarnessingDynamic2024} utilized dynamic magnetic field characteristics through IoT device sensors for authentication, while BreathSign\cite{hanBreathSignTransparentContinuous2023} achieved 95.17\% accuracy using respiratory sound bone conduction within a single breathing cycle.
    \subtitleone{(2) Behavioral Biometrics and Device Interactions:}
    While not directly based on vital signs, these methods often complement physiological approaches in multi-modal systems. Wu et al.\cite{wuRobustDetectionPuppet2022} developed an anti-puppet authentication system to analyze fingertip touch behaviors. MotoPrint\cite{shenMotoPrintReconfigurableVibration2024} implemented reconfigurable vibration motor fingerprints through homologous signal learning. BioCase\cite{yangBioCasePrivacyProtection2023} combined physiological and behavioral features through acoustic perception-based fingerprint authentication, achieving 94\% identification accuracy at 5\% false positive rate in a 10-week study with 46 users.
    \subtitleone{(3) Multi-Feature Fusion Authentication:}
    Recognizing the potential of combining multiple biometric features, researchers have explored various fusion schemes. Wang et al.\cite{wangBiometricsAuthenticatedKeyExchange2021} integrated biometric features with cryptographic schemes in their key exchange framework. HeartPrint\cite{wangHeartPrintExploringHeartbeatBased2022} enabled multi-user authentication via mmWave radar, achieving 95\% accuracy with below 3\% attack success rate across 54 participants. The mmHSV system\cite{liMmHSVAirHandwritten2023} demonstrated 98.96\% AUC in mmWave-based air handwritten signature verification.
    \subtitleone{(4) Scenario-Specific Applications:}
    Recent research has also focused on adapting vital signal authentication to specific use cases. Wang et al.\cite{wangLoweffortVRHeadset2023} developed VR user authentication using head acoustic reflections, addressing echo suppression challenges. For multi-user environments, M-Auth\cite{wangSimultaneousAuthenticationMultiple2024} enabled simultaneous authentication through mmWave radar direction adjustment via a motor, achieving 96\% authentication accuracy and 95\% attack detection rate with 37 participants.
    
    \paragraph{Key Technical Challenges and Development Trends}
    
    The fundamental challenge in physiological signal-based authentication lies in balancing feature uniqueness with temporal stability. While these signals resist forgery, their time-varying nature complicates reliable authentication~\cite{zhangMonitoringLongtermCardiac2024}. Current research addresses this through two approaches: multi-modal fusion~\cite{yangBioCasePrivacyProtection2023}, and continuous authentication~\cite{islamRadarBasedNonContactContinuous2020}. Recent advances, such as multi-user authentication capability in~\cite{wangSimultaneousAuthenticationMultiple2024}, indicate a promising trend toward scenario-specific applications that optimize both security and practicality.

    \subsection{Device Authenticity Verification}
    
    The proliferation of IoT devices necessitates robust authenticity verification mechanisms. Traditional cryptographic methods face implementation challenges in resource-constrained IoT environments, whereas wireless signal-based verification offers a lightweight, forgery-resistant alternative leveraging hardware-level characteristics.
    
    This verification approach encompasses two primary directions: device fingerprint identification based on RF component characteristics, and media authenticity verification through wireless communication patterns. The underlying mechanism exploits inherent physical properties of wireless signals, where manufacturing variations create unique, device-specific signatures in signal transmission, modulation, and reception processes\cite{fengFingerprintingIoTDevices2023}, establishing a reliable foundation for authentication.

    \begin{figure}[!t]
    \centering
    \includegraphics[width=\linewidth]{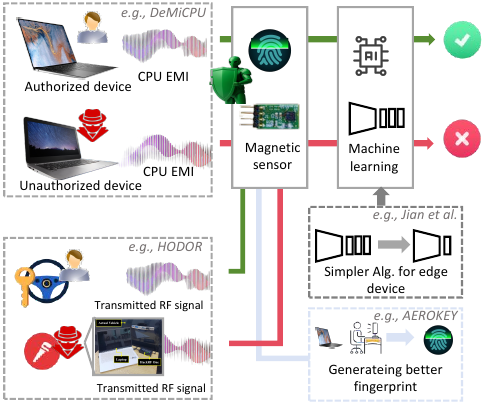}
    \caption{Wireless signal-based device fingerprint identification framework.}
    \label{fig:deviceIdentification}
    \end{figure}

    \subsubsection{Device Fingerprint Identification}

    \paragraph{Physical Principle}
    As shown in Fig.~\ref{fig:deviceIdentification}, device fingerprint identification relies fundamentally on distinct physical characteristics that make each device uniquely identifiable. These characteristics mainly originate from three levels: \subtitleone{(1) Hardware level}, where manufacturing deviations in RF front-end components create distinctive CPU electromagnetic radiation patterns\cite{fengFingerprintingIoTDevices2023}; \subtitleone{(2) Protocol level}, manifesting in unique behavioral patterns during wireless communication protocol implementation\cite{wangBCAuthPhysicalLayer2022}; and \subtitleone{(3) Environmental level}, characterized by specific device-environment electromagnetic interactions\cite{leeAEROKEYUsingAmbient2022}.
    
    \paragraph{Work Review}
    
    \subtitleone{Early important work} includes DeMiCPU~\cite{chengDeMiCPUDeviceFingerprinting2019}, which leverages CPU-generated magnetic induction signals for device fingerprinting. The system achieved 99.1\% precision across 90 mobile devices and 98.6\% accuracy among 30 identically configured devices by analyzing magnetic field characteristics from CPU modules and power circuits. Joo et al.\cite{jooHoldDoorFingerprinting2020} developed an RF fingerprint-based authentication system for keyless entry systems, effectively countering signal relay attacks through RF characteristic analysis.
    \subtitleone{Research then advanced toward complex fingerprint features.} Jian et al.\cite{jianRadioFrequencyFingerprinting2022} enhanced RF fingerprint identification on edge devices through structured pruning-based neural network optimization, achieving 27.2× convolutional layer compression while maintaining identification accuracy. Berdich et al.\cite{berdichSweepUnlockFingerprintingSmartphones2023} introduced speaker roll-off characteristics for smartphone fingerprinting, demonstrating robust performance across varying volumes and environments. The BCAuth system\cite{wangBCAuthPhysicalLayer2022} introduced enhanced physical layer authentication for backscatter communication, supporting both static and mobile device authentication while enabling multi-attacker identification and location tracking. Through spatial correlation analysis of scattering signals, the system implements preventive authentication mechanisms. AEROKEY\cite{leeAEROKEYUsingAmbient2022} pioneered ambient electromagnetic radiation-based authentication, generating unique keys for spatially proximate devices without additional hardware requirements, demonstrating effectiveness across diverse practical scenarios.
    \subtitleone{Recent research emphasizes practicality and security.} The Digitus system\cite{fengFingerprintingIoTDevices2023} introduced latent physical side channel-based fingerprinting, authenticating low-power IoT devices through processor clock electromagnetic radiation without supplementary transmission modules. The system maintains 95\% identification accuracy beyond 5-meter distances and in NLOS conditions. Shen et al.\cite{shenReceiverAgnosticCollaborativeRadio2024} developed receiver-agnostic collaborative RF fingerprint identification, employing adversarial training for feature extraction and collaborative inference for enhanced classification. CSI-RFF\cite{kongCSIRFFLeveragingMicroSignals2024} enables commercial WiFi device authentication through micro-CSI extraction, achieving nearly 99\% attack detection rates in mobile robot area access control applications.
    
    \paragraph{Key Technical Challenges and Development Trends}
    
    The fundamental challenge of device fingerprint identification lies in finding a balance between feature uniqueness and environmental dynamics. While hardware-specific characteristics provide unique device signatures, environmental variables including temperature fluctuations and electromagnetic interference can significantly impact authentication reliability, posing obstacles for widespread deployment. Recent research has addressed this challenge through a multi-tiered approach: at the hardware level, the Digitus system~\cite{fengFingerprintingIoTDevices2023} employs feature decomposition techniques to enhance signal stability; at the protocol level, BCAuth~\cite{wangBCAuthPhysicalLayer2022} implements dynamic authentication leveraging spatial information; and at the system level, AEROKEY~\cite{leeAEROKEYUsingAmbient2022} converts environmental variability into authentication assets. These hierarchical solutions indicate an emerging paradigm where multi-dimensional feature integration transforms environmental dynamics from authentication constraints into security-enhancing mechanisms.
    
    \subsubsection{Media Authenticity Verification}
    
    With the rapid rise of deepfake technologies, the demand for effective media authentication has become more urgent than ever. Physical-layer verification using wireless signals presents a promising solution by examining the intrinsic properties of media during its creation and transmission. Unlike traditional content-based analysis, this approach offers robust tamper-resistance, relying on the physical characteristics of signal patterns to detect manipulation.
    
    \paragraph{Physical Principle}
    
    Media authenticity verification leverages two fundamental physical mechanisms: signal propagation characteristics, which exhibit unique spatiotemporal patterns in real-world scenarios\cite{fangNowhereHideDetecting2023}, and device response characteristics, where media devices such as cameras, microphones, and speakers demonstrate distinctive signal response signatures\cite{zhuSoundLockNovelUser2023}. These physical properties resist precise simulation, establishing a reliable foundation for authenticity verification.
    
    \paragraph{Work Review}
    
    WiSil\cite{fangNowhereHideDetecting2023} introduced video authenticity verification through visual-WiFi contour correspondence. The system constructs object wavefront models in monitored areas, reconstructs object contours via deep learning, and performs authenticity verification using siamese networks for semantic feature extraction. Experimental results demonstrated 95\% accuracy in detecting tampered frames with robust performance across environmental and personnel variations. SoundLock\cite{zhuSoundLockNovelUser2023} developed an authentication framework for VR devices based on auditory-pupillary response analysis, achieving FAR and FRR of 0.76\% and 0.91\% respectively, validating the effectiveness of multi-modal physiological responses in authentication.
    
    \paragraph{Key Technical Challenges and Development Trends}
    
    The fundamental challenge in media authenticity verification lies in establishing reliable correspondence between physical and digital domains. This requires both precise capture of physical spatiotemporal characteristics and verification of their consistency with digital content. Multi-modal fusion has emerged as a promising solution: WiSil\cite{fangNowhereHideDetecting2023} integrates WiFi-visual contours, while SoundLock\cite{zhuSoundLockNovelUser2023} combines auditory-pupillary responses. This fusion approach, incorporating both physical features (visual/auditory) and physiological responses (WiFi/pupillary), represents an emerging paradigm in authentication system design: leveraging multi-dimensional physical coupling relationships to create tamper-resistant verification mechanisms.
    
    \subsection{Privacy Protection}
    
    Wireless signals provide unique advantages for privacy protection through their non-contact monitoring and environmental sensing capabilities. These signals enable two primary protection mechanisms: \textit{privacy threat detection} for identifying unauthorized devices and suspicious behaviors, and \textit{data privacy protection} for securing wireless communications through physical-layer security approaches.

    \begin{figure}[!t]
    \centering
    \includegraphics[width=\linewidth]{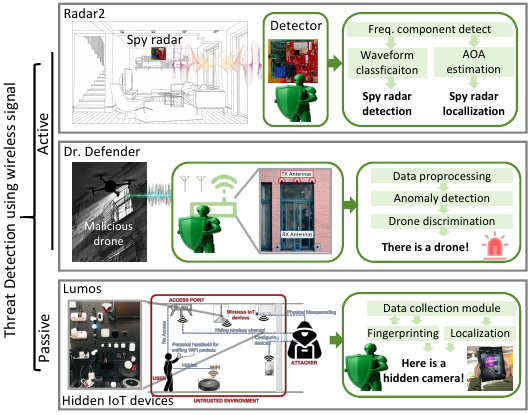}
    \caption{Classification of wireless signal-based privacy threat detection systems. Each system contains two key modules: detection hardware and signal processing workflow.}
    \label{fig:ThreatDetection}
    \end{figure}

    \subsubsection{Privacy Threat Detection}
    
    Privacy threat detection employs wireless signal analysis to identify potential privacy threats such as concealed cameras and eavesdropping devices. As illustrated in Fig.~\ref{fig:ThreatDetection}, detection can be implemented through active or passive approaches using specialized detection equipment and signal processing algorithms. This physical-layer detection methodology offers enhanced detection capabilities compared to traditional software or network traffic analysis approaches.
    
    \paragraph{Physical Principle}
    Three fundamental physical mechanisms enable threat detection: \textit{electromagnetic radiation patterns} generated during device operation, \textit{spatial distribution characteristics} enabling device localization, and \textit{behavioral signatures} reflected in dynamic signal patterns, as shown in Fig.~\ref{fig:ThreatDetection}. These physical properties provide robust foundations for threat identification and localization.
    
    \paragraph{Work Review}
    As illustrated in Fig.~\ref{fig:ThreatDetection}, privacy threat detection encompasses two main approaches: \subtitleone{(1) Active detection} utilizes deliberately transmitted wireless signals. The Radar2 system\cite{qiuRadar2PassiveSpy2023} implements commercial mmWave radar for environmental scanning, achieving hidden radar localization through spectral analysis and triangulation, with detection rates exceeding 96\% and localization error below 0.3 meters. The Dr. Defender system\cite{dengDrDefenderProactive2024} extends detection to aerial threats by analyzing WiFi CSI features for drone identification within 10-meter ranges. \subtitleone{(2) Passive detection} leverages inherent device emissions. The Lumos system\cite{sharmaLumosIdentifyingLocalizing2022} integrates wireless signal characteristics with device behavior analysis, achieving 1.5-meter localization accuracy while scanning 1000 square feet within 30 minutes, effectively combining passive detection's energy efficiency with active detection's precision.
    
    \paragraph{Key Technical Challenges and Development Trends}
    
    The fundamental challenge in privacy threat detection lies in balancing detection range and accuracy. This trade-off stems from physical constraints: extending detection range requires increased transmission power or reception sensitivity, which introduces environmental interference and multipath effects that compromise localization accuracy. Research addresses this challenge through parallel approaches: active detection systems like Radar2\cite{qiuRadar2PassiveSpy2023} enhance performance through optimized signal processing, achieving sub-0.3-meter accuracy, while passive detection systems like Lumos\cite{sharmaLumosIdentifyingLocalizing2022} improve adaptability through multi-feature fusion. This dual-path development strategy indicates an emerging trend toward achieving optimal range-accuracy trade-offs across diverse scenarios, particularly in complex environments such as aerial threat detection demonstrated by Dr. Defender\cite{dengDrDefenderProactive2024}, through the integration of active and passive detection technologies.
    
    \subsubsection{Privacy Data Protection}
    
    The widespread adoption of IoT devices has elevated the importance of data protection. Physical-layer security mechanisms can leverage wireless signal characteristics to provide theoretical security guarantees with reduced computational overhead.
    
    \paragraph{Physical Principles}
    
    Privacy data protection leveraging wireless signals relies on three fundamental physical mechanisms:
    \subtitleone{(1) Channel reciprocity and randomness} leverages the inherent randomness of wireless channels, allowing legitimate parties to generate consistent keys while ensuring spatial separation prevents attackers from accessing identical channel information.
    \subtitleone{(2) Spatial selectivity} extends protection through directional propagation, employing beamforming and spatial filtering to optimize signal quality along intended paths while naturally degrading unauthorized reception.
    \subtitleone{(3) Artificial noise injection} actively strengthens protection by strategically degrading potential eavesdroppers' signal quality while maintaining legitimate communication performance.
    
    \paragraph{Work Review}
    
    Research on privacy data protection based on wireless signals has gone through three development stages:
    \subtitleone{(1) Basic physical feature protection} explored inherent wireless signal characteristics for secure communication. Zhao et al.\cite{zhaoAuthenticationMillimeterWaveBodyCentric2017} pioneered security enhancement in mmWave communication through human body scattering characteristics.
    \subtitleone{(2) Active environmental control protection} evolved toward active communication environment regulation. Katsumata et al.\cite{katsumataRevisitingFuzzySignatures2021} developed a channel-adaptive fuzzy signature scheme achieving 112-bit security with high efficiency. Kong et al.\cite{kongPushLimitWiFibased2022} enhanced system performance in dynamic environments through deep learning-based environment-invariant feature extraction.
    \subtitleone{(3) Intelligent comprehensive protection} emphasized multi-dimensional protection strategies. Liu et al.\cite{liuExploitingFineGrainedChannel2023} developed an integrated security framework combining multiple physical features including arrival angle, channel gain, and phase noise to enhance anti-eavesdropping capabilities.
    
    \paragraph{Key Technical Challenges and Development Trends}
    
    The fundamental challenge in privacy data protection lies in balancing security and practicality, as enhanced security measures typically demand increased signal processing complexity and system resources, conflicting with IoT device constraints. The field has evolved from passive utilization to active control strategies: early approaches like \cite{zhaoAuthenticationMillimeterWaveBodyCentric2017} relied on natural channel characteristics, while recent research explores active protection mechanisms, exemplified by \cite{liuExploitingFineGrainedChannel2023} multi-dimensional physical feature integration. Modern approaches demonstrate improved environmental adaptability, with Kong et al.\cite{kongPushLimitWiFibased2022} leveraging deep learning and the AEROKEY system\cite{leeAEROKEYUsingAmbient2022} transforming environmental dynamics into security enhancement mechanisms. This progression toward intelligent environmental perception and control indicates the trend toward achieving robust security with minimal system overhead.
    
    \subsection{Summary and Insights}
    
    Recent years have witnessed rapid advancement in wireless signal-based security mechanisms. Through systematic review, this field demonstrates clear technological progression across biometric authentication, device verification, and privacy protection. Wireless sensing technology offers distinct advantages: penetration capability overcomes limitations of optical solutions\cite{renPersonReidentification3D2023}, non-contact sensing enables natural user interaction\cite{huangNFHeartNearfieldNoncontact2023}, and universal deployment facilitates comprehensive security coverage\cite{fangNowhereHideDetecting2023}.
    
    The evolution of wireless security technologies exhibits three significant trends: \subtitleone{(1) Multi-modal fusion}, progressing from single-modal to multi-modal approaches, exemplified by the RFCam system\cite{chenRFCamUncertaintyawareFusion2022} which enhances identification accuracy through WiFi and visual information fusion. \subtitleone{(2) Intelligence integration}, where deep learning implementation enables enhanced environmental adaptability, as demonstrated in \cite{wanSelfSupervisedModalityAwareMultiple2023}. \subtitleone{(3) Active protection mechanisms}, illustrated by the innovative power allocation scheme proposed in~\cite{maengPowerAllocationFingerprintBased2021} .
    
    These technological advances signify a fundamental paradigm shift in security applications, transitioning from conventional explicit security mechanisms to ubiquitous physical-layer protection. Notably, wireless signal-based privacy protection introduces an efficient security approach, achieving robust protection through physical-layer characteristics while avoiding the computational complexity of traditional cryptographic methods\cite{katsumataRevisitingFuzzySignatures2021}.
    
    \section{Statistical Analysis of Existing Work} \label{sec:roleStat}
    
    \begin{figure}[!t]
    \centering
    \includegraphics[width=0.9\linewidth]{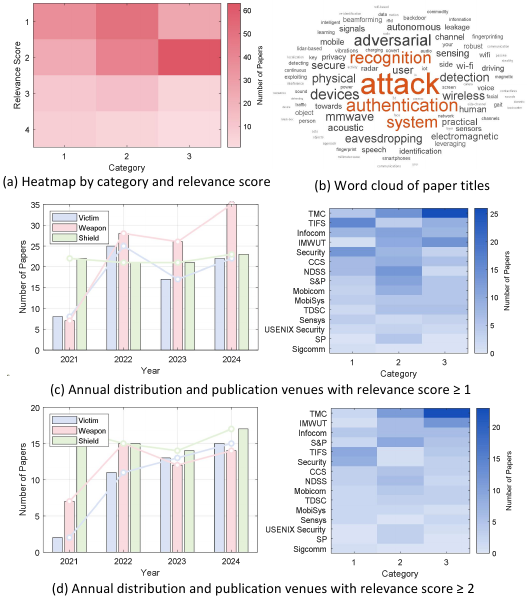}
    \caption{Statistical analysis results of the database.}
    \label{fig:statisticalAnalysis}
    \end{figure}
    
    \subsection{Awesome-WS-Security}
    
    To achieve a systematic understanding of the wireless sensing security field, we manually constructed the first literature collection focused on wireless sensing security: Awesome-WS-Security. This database contains over 250 papers, covering most important research works between 2020-2024. The database records five key attributes for each entry: title, year, source, category, and relevance, with the following distinctive features:
    
    \subtitleone{Category Labels:} Papers are categorized into three types: wireless systems as attack targets (1), wireless signals as attack tools (2), and wireless signals as guardians (3).
    
    \subtitleone{Relevance Labels:} Each paper receives a relevance score (1-4): score 1 indicates sensing/communication security papers with potential inspiration; score 2 represents wireless security papers with clear relevance; scores $\geq$3 indicate direct wireless sensing security research.
    
    \subtitleone{Open Source:} The database and analysis code are publicly available on GitHub, enabling researchers to conduct custom analyses across multiple dimensions.
    
    \subsection{Visualization and Analysis}
    
    Analysis of the database, as shown in Fig.~\ref{fig:statisticalAnalysis}, reveals several key insights:
    
    \subtitleone{Distribution:} The heat map (Fig.~\ref{fig:statisticalAnalysis}(a)) shows a predominance of papers with relevance scores 1-2, indicating the field's early developmental stage with significant growth potential.
    
    \subtitleone{Focus:} The word cloud analysis highlights frequent occurrences of terms like ``attack,'' ``authentication,'' ``identification,'' and ``adversarial,'' reflecting core research themes.
    
    \subtitleone{Temporal Trends:} Papers using wireless signals as attack tools show significant growth (relevance $\geq$1), while research on attacks against wireless systems dominates at relevance $\geq$2 (Fig.~\ref{fig:statisticalAnalysis}(c,d)).
    
    \subtitleone{Publication Venues:} TMC, TIFS, and Infocom lead publications at relevance $\geq$1, while TMC, IMWUT, and Infocom dominate at relevance $\geq$2, highlighting strong interest from mobile and ubiquitous computing communities.
    
    More in-depth statistics and findings await further exploration by researchers through Awesome-WS-Security. We hope that the Awesome-WS-Security database will facilitate researchers in quickly locating wireless sensing security-related works among the vast array of mobile computing and security articles.

	\section{Challenges and Future Directions}  \label{sec:discussion}

    \begin{table*}[!t]
\caption{Key Challenges and Future Directions in Wireless Sensing Security Research}
\label{tab:challenges}
\renewcommand{\arraystretch}{1.1}
\begin{tabular}{|p{2cm}|p{6cm}|p{9cm}|}
\hline
\textbf{Fundamental Challenges} & \textbf{Specific Manifestations} & \textbf{Future Directions} \\
\hline
\multirow{3}{*}{\begin{tabular}[c]{@{}l@{}}Physical-Digital\\Interaction\\Complexity\end{tabular}} & 
• Micro: Random noise causing signal randomness\cite{chenRFCamUncertaintyawareFusion2022}\newline
• Macro: Environmental changes causing channel instability\cite{basakMmSpySpyingPhone2022,hanBreathSignTransparentContinuous2023}\newline
• System: Unstable sensing results\cite{liIFNetImagingFocusing2024} & 
• Cross-domain attack models: Unified multi-physical domain framework\cite{yangMultipleWiFiAccess2024,weiMetasurfaceenabledSmartWireless2023}\newline
• Foundation models: Intelligent signal processing algorithms\cite{songRFURLUnsupervisedRepresentation2022,guanTalk2RadarBridgingNatural2024,wengLargeModelSmall2024} \\
\hline
\multirow{3}{*}{\begin{tabular}[c]{@{}l@{}}Security\\Mechanism\\Lag\end{tabular}} & 
• Delayed attack detection\cite{huntMadRadarBlackBoxPhysical2024}\newline
• Insufficient defense adaptability\cite{staatIRShieldCountermeasureAdversarial2022}\newline
• Slow response to new threats\cite{basakMmSpySpyingPhone2022} & 
• Multi-layer defense: Complete defense chain\cite{asaadSecureActivePassive2022,liuExploitingFineGrainedChannel2023}\newline
• Intelligent defense\cite{weiDualAdversarialRepresentationDisentanglement2024,wanSelfSupervisedModalityAwareMultiple2023}\newline
• Proactive defense: Early threat identification\cite{sharmaLumosIdentifyingLocalizing2022,dengDrDefenderProactive2024} \\
\hline
\multirow{3}{*}{\begin{tabular}[c]{@{}l@{}}Application\\Scenario\\Diversity\end{tabular}} & 
• Varied scenario requirements\cite{reddyvennamMmSpoofResilientSpoofing2023}\newline
• Rapid scenario evolution\cite{haoMmSafeVoiceSecurity2022}\newline
• Multi-objective optimization challenges\cite{hanBreathSignTransparentContinuous2023} & 
Cognitive Security Example:\newline
• Traditional limitations: Specialized equipment\cite{wangNewMethodEEGBased2013}, active cooperation\cite{weissLookingGoodLying2006}\newline
• RF advantages: Contactless, continuous monitoring, easy deployment\cite{tsiamyrtzisImagingFacialPhysiology2007,harmerAutomaticBlushDetection2010,daiDetectingMentalDisorders2023,alhanaiDetectingDepressionAudio2018}\\
\hline
\end{tabular}
\end{table*}
    
    \subsection{Challenges Faced by Wireless Sensing Security}
    
    Beyond the specific technical challenges discussed previously, wireless sensing security faces three fundamental challenges stemming from its essential characteristics (Tab.~\ref{tab:challenges}).
    
    \subsubsection{Complexity of Wireless Signal Physical-Digital Interaction}
    
    The physical-digital interaction presents a fundamental challenge in wireless sensing security. Unlike traditional digital security, wireless sensing systems must contend with physical world uncertainties at multiple levels: randomness in signal characteristics at the physical layer, instability in wireless channels due to environmental dynamics, and uncertainty in information extraction at the sensing layer.

    This uncertainty affects wireless sensing security in three contexts. As attack targets, signals become vulnerable when attackers exploit environmental changes to mask malicious activities, as demonstrated by the RFCam system\cite{chenRFCamUncertaintyawareFusion2022}. As attack tools, environmental factors can compromise attack precision and reliability, as seen in the mmSpy system\cite{basakMmSpySpyingPhone2022}. As security mechanisms, this uncertainty directly impacts reliability, requiring complex signal processing solutions like those implemented in the BreathSign system\cite{hanBreathSignTransparentContinuous2023}.
    
    Addressing these challenges requires advances in wireless signal processing and deep learning technologies, including synthetic aperture\cite{liIFNetDeepImaging2024,bianLargescaleSingleshotMillimeterwave2024}, motion compensation\cite{manzoniMotionEstimationCompensation2023,alvarez-narciandiShorttimeCoherentIncoherent2024}, autofocusing\cite{chenCoSenseExploitingCooperative2024,zhangSinglePointMultiPointReflection2024,changMSenseBoostingWireless2024,zhangMobi2Sense2022,maMobi2SenseEnablingWireless2022}, super-resolution\cite{yangIMIAInterferenceMitigation2023,yangMultipleWiFiAccess2024,wang2021high}, sparse aperture\cite{huSparseArrayOptimization2024,songFastFactorizedKirchhoff2024,wang2023cvae}, multi-base cooperation\cite{tagliaferriCooperativeCoherentMultistatic2024}, supervised learning\cite{liIFNetImagingFocusing2024,zhangDenseAccurateRadar2024,liDomainIndependentRealTimeGesture2023}, self-supervised learning\cite{songRFURLUnsupervisedRepresentation2022}, modal alignment\cite{guanTalk2RadarBridgingNatural2024,wengLargeModelSmall2024}, and large model development for wireless signals.
    
    \subsubsection{Lag in Security Protection Mechanisms}
    
    The lag in security protection mechanisms represents a critical challenge in wireless sensing security. Unlike traditional network security, wireless sensing security must address concurrent threats from both physical and digital domains, making the design and deployment of protection mechanisms highly complex.
    
    Protection lag manifests in three scenarios: \subtitleone{(1) Signals as attack targets:} Conventional defenses show significant delays in responding to emerging threats, (e.g., mmWave radar spoofing attack\cite{huntMadRadarBlackBoxPhysical2024}), while even dynamic solutions like IRShield\cite{staatIRShieldCountermeasureAdversarial2022} remain constrained by predefined parameters; \subtitleone{(2) Signals as attack vectors:} Current protection mechanisms struggle with novel attack methodologies, as demonstrated by AEROKEY's\cite{leeAEROKEYUsingAmbient2022} vulnerability to targeted attacks despite its adaptive design, and mmSpy's\cite{basakMmSpySpyingPhone2022} successful circumvention of conventional security measures; \subtitleone{(3) Signals as security mechanisms:} System effectiveness is compromised by protection lag, evident in BreathSign's\cite{hanBreathSignTransparentContinuous2023} continuous feature model update requirements and Lumos's\cite{sharmaLumosIdentifyingLocalizing2022} insufficient adaptability to emerging threats.
    
    This limitation fundamentally stems from the linear ``problem identification, cause analysis, countermeasure design, and defense deployment'' sequence in current frameworks. As attack methodologies advance, particularly with emerging foundation models, \textit{future research should transcend traditional passive defense limitations through active, intelligent, and collaborative mechanisms}.
    
    \subsubsection{Diversity and Rapid Development of Application Scenarios}
    
    The diversity and rapid evolution of application scenarios present a significant challenge in wireless sensing security.
    
    Scenario diversity drives distinct requirements: \subtitleone{(1) Signals as attack targets:} Security demands vary dramatically between contexts - autonomous driving requires millisecond-level responses in high-mobility environments~\cite{reddyvennamMmSpoofResilientSpoofing2023}, while smart home applications prioritize long-term stability and energy efficiency; \subtitleone{(2) Signals as attack tools:} Attack effectiveness varies significantly across environments, with Wu et al.\cite{wuDeceptionDetectionVideos2017} demonstrating how indoor-optimized signal eavesdropping techniques require substantial adaptation for outdoor settings, creating a scenario-dependent complexity that can be leveraged defensively; \subtitleone{(3) Signals as security guardians:} Systems must balance multiple competing objectives, exemplified by Lumos's\cite{sharmaLumosIdentifyingLocalizing2022} need to maintain privacy without compromising communication functionality, and BreathSign's\cite{hanBreathSignTransparentContinuous2023} requirements to optimize between authentication accuracy, latency, and usability.
    
    Besides, emerging technologies such as 5G/6G and IoT create novel application contexts. For instance, the metaverse introduces unprecedented mixed reality security requirements, while quantum computing developments may fundamentally alter cryptographic paradigms. As demonstrated by the mmSafe system\cite{haoMmSafeVoiceSecurity2022}, future security frameworks must adapt to dynamic scenario evolution.
    
    Future wireless sensing security research should extend beyond traditional application boundaries. Emerging fields such as cognitive security and biometric authentication present opportunities for transformative advances. Security solution design should address current requirements while maintaining extensibility for future scenario evolution.
    
    \subsection{Future Outlook: Breakthroughs in Wireless Signal Processing and Sensing Algorithms}
    
    Addressing physical-digital interaction complexity in wireless sensing security requires fundamental breakthroughs in signal processing and sensing algorithms. These advances are anticipated in two primary directions: cross-domain attack models and foundation model-enabled intelligent attacks.
    
    \subsubsection{Cross-Domain Attack Models}
    Current wireless sensing algorithms predominantly operate within single signal domains, creating vulnerabilities to cross-domain attacks. For example, Gao et al.\cite{gaoKITEExploringPractical2023} demonstrated acoustic signals interfering with inertial sensor operations, circumventing traditional signal domain defenses. Li et al.\cite{liSpiralspyExploringStealthy2022} established covert channels through mechanical vibrations, overcoming electromagnetic domain isolation. These findings highlight critical coupling relationships between physical domains that create novel attack vectors.
    
    Future signal processing algorithms require a unified theoretical framework spanning multiple physical domains. Developing coupling models across sound, light, electromagnetic, and other physical quantities enables systems to predict and understand cross-domain attack patterns. This theoretical advancement provides foundations for novel defense algorithms. \textit{Multi-modal} approaches, such as LiDAR-supervised wireless sensing, demonstrate enhanced sensing capabilities\cite{gengDREAMPCDDeepReconstruction2024,laiEnablingVisualRecognition2024,chengNovelRadarPoint2022}.
    Particularly noteworthy is that with the development of programmable \textit{metasurface} technologies like RIS, the coupling between physical domains becomes more complex and controllable\cite{weiMetasurfaceenabledSmartWireless2023}.
    
    \subsubsection{Intelligent Attacks Empowered by Foundation Models}
    
    The advancement of deep learning, particularly foundation models, presents increasingly sophisticated attack threats to wireless sensing systems. Traditional rule-based and simple machine learning approaches show limitations against intelligent attacks. For example, Xie et al.\cite{xieUniversalTargetedAdversarial2023} demonstrated foundation models' capabilities in generating covert attack signals.
    
    Future signal processing algorithms should leverage foundation model capabilities. Unsupervised representation learning frameworks show superior promise in enhancing attack pattern recognition through pre-trained universal signal representations~\cite{songRFURLUnsupervisedRepresentation2022,fangPRISMPretrainingRF2024,songUnleashingPotentialSelfSupervised2024}. Recent developments, such as large model-assisted radar signal interpretation method in~\cite{guanTalk2RadarBridgingNatural2024}, highlight foundation models' potential in enhancing system interpretability.
    
    The evolution of wireless signal processing and sensing algorithms exhibits a clear trend toward intelligence. The integration of traditional signal processing with AI technologies enables robust algorithmic frameworks, particularly crucial for multi-physical domain environments. This advancement necessitates a balanced focus on interpretability and reliability, establishing foundations for trustworthy wireless sensing security systems.
    
    \subsection{Future Outlook: Building a Stronger Defense System for Wireless Sensing Systems}
    
    Addressing the fundamental challenge of security protection mechanism lag requires constructing next-generation wireless sensing defense systems. These systems must rapidly perceive and respond to emerging threats while maintaining robust protection against known attacks. Future defense systems will advance in three critical directions: multi-level collaborative defense, intelligent defense strategies, and preventive active defense mechanisms.
    
    \subsubsection{Multi-Level Collaborative Defense}
    
    Single-layer defense strategies demonstrate significant limitations against complex wireless attacks. Physical layer interference suppression alone proves insufficient against intelligent spoofing attacks, while application layer protection remains vulnerable to physical layer exploitation. Staat et al.\cite{staatIRShieldCountermeasureAdversarial2022} demonstrated that intelligent reflecting surface defense systems remain vulnerable to carefully designed adversarial samples without upper-layer protocol cooperation.
    
    Multi-level collaborative defense requires establishing comprehensive defense chains spanning physical to application layers. At the physical layer, programmable devices like RIS enable active channel characteristic control, as demonstrated by MIMO-RIS joint optimization scheme in~\cite{asaadSecureActivePassive2022}. Fine-grained channel feature authentication mechanism in~\cite{liuExploitingFineGrainedChannel2023} provides cross-layer security verification at the protocol layer.
    
    \subsubsection{Intelligent Defense Strategies}
    
    Contemporary intelligent attacks exceed traditional rule-based defense capabilities. Modern defense systems require advanced learning and reasoning capabilities for complex attack pattern comprehension and response. The dual adversarial representation disentanglement model in~\cite{weiDualAdversarialRepresentationDisentanglement2024} demonstrates deep learning-based attack feature pattern capture.
    
    Adaptive defense represents a crucial advancement. The self-supervised learning framework proposed by~\cite{wanSelfSupervisedModalityAwareMultiple2023} enhances security cognition through multi-modal data analysis, enabling both known attack pattern identification and continuous learning adaptation. Similarly, the uncertainty-aware fusion method developed by~\cite{chenRFCamUncertaintyawareFusion2022} demonstrates improved defense decision reliability through precise uncertainty quantification, establishing foundations for trustworthy intelligent defense systems.
    
    \subsubsection{Preventive Active Defense Mechanisms}
    
    Despite existing active defense strategies, most current defense methods remain passive, responding only post-attack with limited effectiveness. Advanced defense systems require proactive threat identification and prevention capabilities, as demonstrated by Lumos system\cite{sharmaLumosIdentifyingLocalizing2022}  through active environmental scanning for hidden listening devices. The effectiveness of such preventive defense approaches relies heavily on accurate threat prediction models. The AEROKEY system\cite{leeAEROKEYUsingAmbient2022} validates physics-based prediction through environmental electromagnetic characteristic analysis, while the fuzzy signature scheme proposed by~\cite{katsumataRevisitingFuzzySignatures2021} provides theoretical security guarantees. Further advancing this proactive paradigm, the defense system developed by~\cite{dengDrDefenderProactive2024} prevents potential eavesdropping through wireless channel characteristic adjustment.
    
    \subsection{Future Outlook: Exploring New Security Applications}
    
    The convergence of IoT and AI technologies expands wireless sensing applications across traditional physical security, information security, and emerging domains such as cognitive security. This section examines cognitive security as an exemplar of wireless sensing technology's potential in novel security applications.
    
    \subsubsection{The Definition and Significance of Cognitive Security}
    
    Cognitive security (CogSec) is an emerging branch of cybersecurity that has evolved to address the human cognitive dimension of cyber threats. As a multidisciplinary research field, it leverages knowledge from social science, psychology, cognition science, neuroscience, AI and computer science to study human-content interaction patterns and their impacts on human cognition \cite{guoMassFakeNews2020}. CogSec encompasses understanding both the intrinsic cognitive mechanisms of human information processing and the external manifestations of cognitive states through physiological and behavioral patterns. This extension of traditional cybersecurity reflects the evolution of cyber threats beyond technical systems to target human cognitive processes, characterized by its dual focus on: (1) protecting human cognitive processes from malicious information influence, and (2) monitoring cognitive states through objective measurements for early risk detection.
    
    Cognitive security research demonstrates significant practical implications. At the individual level, it enables mental health issue identification and prevention. Han et al.\cite{hanDeepLearningMobile2021} established that early cognitive abnormality detection aids in preventing depression and anxiety disorders. At the organizational level, cognitive state monitoring enhances decision reliability. Research by Wu et al.\cite{wuDeceptionDetectionVideos2017} and Ding et al.\cite{dingFaceFocusedCrossStreamNetwork2019} demonstrates effective deceptive behavior identification through cognitive feature analysis.
    
    \subsubsection{Challenges Faced by Cognitive Security}
    
    Current cognitive security research faces three challenges:
    
    \textit{(i) Accuracy of cognitive assessment.} Traditional questionnaire surveys and clinical interviews have drawbacks such as high subjectivity and poor real-time performance. Although researchers have developed various objective assessment methods, such as depression detection based on eye movement features by Alghowinem et al.\cite{alghowinemEyeMovementAnalysis2013} and assessment schemes combining reaction times by Pan et al.\cite{panDepressionDetectionBased2019}, these methods still struggle to achieve high-precision real-time monitoring.
    
    \textit{(ii) Feasibility of continuous monitoring.} Most existing methods require specialized equipment and specific scenarios. For example, the EEG-based method proposed by Wang et al.\cite{wangNewMethodEEGBased2013}, although achieving 95\% accuracy, is difficult to apply in daily environments. Even thermal imaging solutions developed by Pavlidis et al.\cite{pavlidisThermalImageAnalysis2002} and Rajoub et al.\cite{rajoubThermalFacialAnalysis2014} are limited by equipment and environmental conditions.
    
    \textit{(iii) User acceptance issues.} Most assessment methods require active user cooperation. From the findings of \cite{weissLookingGoodLying2006} to the advancements of the latest TrueWatch system\cite{alperTrueWatchPolygraphExamination2024}, researches consistently show that user engagement significantly impacts assessment effectiveness. How to reduce user burden while ensuring assessment effectiveness has become a key challenge. In practical applications of cognitive security, cognitive safety assessments should be conducted without users' awareness.
    
    \subsubsection{Advantages and Potential of Wireless Signal-Based Cognitive Security}
    
    Wireless sensing technology offers distinct advantages in addressing these challenges:
    
    \textit{(i) Non-contact acquisition of multi-dimensional physiological features.} Wireless signals capture multiple cognitive state-related physiological indicators simultaneously, including cardiac and respiratory characteristics\cite{hanDeepLearningMobile2021}. Research by Tsiamyrtzis et al.\cite{tsiamyrtzisImagingFacialPhysiology2007} and Harmer et al.\cite{harmerAutomaticBlushDetection2010} demonstrates subtle physiological changes' reflection of cognitive state variations.
    
    \textit{(ii) Feasibility of continuous monitoring.} Wireless sensing eliminates device wear and active user cooperation requirements. The large-scale study of 9,000 participants by Dai et al.\cite{daiDetectingMentalDisorders2023}  validates long-term activity-based psychological abnormality detection effectiveness. Lin et al.\cite{linDepressionDetectionCombining2021} and Shen et al.\cite{shenDepressionDetectionAnalysing2021} demonstrate enhanced detection accuracy through multi-feature integration.
    
    \textit{(iii) Deployment convenience.} With the development of ISAC, wireless sensing can leverage existing communication infrastructure and smart devices for large-scale deployment. Applications range from airport security lie detection\cite{warmelinkThermalImagingLie2011} to facial emotion analysis\cite{shusterLieMyFace2021}. Al Hanai et al.\cite{alhanaiDetectingDepressionAudio2018} demonstrate effective mental state assessment through voice analysis in daily scenarios.
    
    Cognitive security demonstrates wireless sensing technology's broader security applications. Its non-contact and beyond-line-of-sight capabilities enable advances in physiological feature recognition\cite{liSBRFFineGrainedRadar2024}, behavioral pattern analysis\cite{liRealtimeFallDetection2022}, emotional state assessment\cite{xuContactlessGSRSensing2022}, and NLOS eavesdropping\cite{liuImagingPrivacyThreats2024}. Continued advancement in sensing technology and AI algorithms will expand wireless sensing-based security applications, contributing to safer and more intelligent future societies.
    
    \section{Conclusion} \label{sec:conclusion}
    
    This paper presented the first systematic review of wireless sensing security, introducing an innovative role-based classification framework that categorizes research according to wireless signals' functions as victims, weapons, and shields. This framework not only provided an intuitive and logical organization of the field but also revealed fundamental patterns across diverse security applications. Through comprehensive analysis of over 200 publications mainly from 2020-2024, we further identified key technological challenges and emerging trends. This systematic review established a foundation for understanding wireless sensing security, offering both newcomers and established researchers clear perspectives on the field.
    
	\ifCLASSOPTIONcaptionsoff
	\newpage
	\fi

    \footnotesize
    \section*{List of Abbreviations}

    \begin{IEEEdescription}[\IEEEsetlabelwidth{$\alpha\omega\pi\theta\mu\mu \mu$}\IEEEusemathlabelsep] 
    \item[AI] Artificial Intelligence
    \item[AoA] Angle of Arrival
    \item[ARM] Advanced RISC Machine
    \item[ASR] Automatic Speech Recognition
    \item[BFI] Beamforming Feedback Information
    \item[BAC] Balanced Accuracy
    \item[CCS] Combined Charging System
    \item[CDAE] Conditional Denoising Autoencoder
    \item[CSI] Channel State Information
    \item[CSMA/CA] Carrier Sense Multiple Access with Collision Avoid
    \item[CogSec] Cognitive Security
    \item[CPU] Central Processing Unit
    \item[DE] Differential Evolution
    \item[DL] Deep Learning
    \item[DNN] Deep Neural Network
    \item[DoA] Direction of Arrival
    \item[EER] Equal Error Rate
    \item[EMI] Electromagnetic Interference
    \item[FL] Fingerprinting Localization
    \item[FMCW] Frequency-Modulated Continuous Wave
    \item[GPS] Global Positioning System
    \item[GPU] Graphics Processing Unit
    \item[HAR] Human Activity Recognition
    \item[HRRP] High-Resolution Range Profile
    \item[I\&A] Identification and Authentication
    \item[IDS] Intrusion Detection Systems
    \item[ILS] Intelligent Reflecting Surfaces
    \item[IMU] Inertial Measurement Unit
    \item[IoT] Internet of Things
    \item[ISAC] Integrated Sensing and Communication
    \item[ISAR] Inverse Synthetic Aperture Radar
    \item[LiDAR] Light Detection and Ranging
    \item[LIME] Local Interpretable Model-agnostic Explanations
    \item[LOS] Line-of-Sight
    \item[MAC] Media Access Control / Medium Access Control
    \item[MEMS] Micro-Electro-Mechanical Systems
    \item[mmWave] Millimeter Wave / millimeter wave
    \item[MIMO] Multiple Input Multiple Output / Multiple-Input Multiple-Output
    \item[NLOS] Non-Line-of-Sight
    \item[NES] Natural Evolution Strategies
    \item[OAPA] Optimal Adaptive Power Allocation
    \item[PHY] Physical Layer
    \item[RF] Radio Frequency
    \item[RFID] Radio-Frequency Identification
    \item[RIS] Reconfigurable Intelligent Surface
    \item[RMBN] Robust Mean-Based Normalization
    \item[RSSI] Received Signal Strength Indicator
    \item[SAR] Synthetic Aperture Radar
    \item[SBR] Shooting and Bouncing Rays
    \item[SFPA] Suboptimal Fixed Power Allocation
    \item[SNR] Signal-to-Noise Ratio
    \item[SRS] Speech Recognition Systems
    \item[SVM] Support Vector Machine
    \item[UAV] Unmanned Aerial Vehicle
    \item[UWB] Ultra-Wideband
    \item[WBR] WiFi-Based Behavior Recognition
    \item[WLAN] Wireless Local Area Network
    \item[WiFi] Wireless Fidelity
    \end{IEEEdescription}
    
	\bibliographystyle{IEEEtran}
	\bibliography{bib1,bib2}

\end{document}